 \documentclass[twocolumn,showpacs,nofootinbib]{revtex4}

\usepackage[latin1]{inputenc}
\usepackage{graphicx}%
\usepackage{dcolumn}
\usepackage{amsmath}

\makeatletter
\def\btt#1{\texttt{\@backslashchar#1}}%
\DeclareRobustCommand\bblash{\btt{\@backslashchar}}%
\makeatother

\begin{document}

 \title{Vehicle Overacceleration -- \\ A Fundamental Microscopic Mechanism for Traffic Breakdown Control Using Automated Vehicles and AI}

\author{Boris S. Kerner $^1$}

 \affiliation{$^1$
Faculty of Physics, University of Duisburg-Essen,
47048 Duisburg, Germany}


\begin{abstract}
This review article addresses a fundamental controversial question in traffic theory:
  Is the nucleation character of traffic breakdown at a bottleneck governed by vehicle overdeceleration (overbraking) or by 
	discontinuous vehicle acceleration, referred to as vehicle overacceleration.
	
A widely accepted view in traffic science attributes traffic breakdown to traffic instability induced by vehicle braking, in which the speed of a braking vehicle becomes lower than that of its preceding vehicle. Within most standard traffic theories, such overdeceleration is assumed to generate traffic instability, which in turn is regarded as the primary mechanism of traffic breakdown. Based on this assumption, various traffic-management approaches, including jam absorption driving, have been proposed to suppress traffic instability and thereby prevent congestion.
		In contrast, three-phase traffic theory proposes a fundamentally different mechanism. It suggests that traffic breakdown is not caused by braking-induced overdeceleration leading to traffic instability, but rather by  vehicle overacceleration. Because the prevailing view has long associated traffic breakdown with braking-induced instability, the hypothesis that overacceleration governs the nucleation of traffic breakdown remains controversial.
For this reason, in this review we analyze whether the empirical  nucleation nature of traffic breakdown at a bottleneck is caused by   vehicle  overdeceleration  leading to traffic instability or by   vehicle  overacceleration. This question is of particular importance in the context of automated vehicles and AI, whose individual dynamic behavior can enable reliable strategies for traffic breakdown control in the future.
 We show that, in both human-driven and automated traffic flow, traffic breakdown is governed by vehicle overacceleration rather than vehicle overdeceleration.
With this objective, in microscopic modeling
we  separate  traffic breakdown caused
 by   overacceleration  from traffic instabilities caused by overdeceleration due to braking behavior,
  while following recent papers
[B.S. Kerner, Phys. Rev. E  108 (2023) 014302; B.S. Kerner, 108  (2023) 064305; B.S. Kerner and S.L. Klenov,  112 (2025)  034309].

	To emphasize the fundamental features of  overacceleration, we utilize single-leader car-following models, where the follower's
	motion depends \textit{solely} on the immediate preceding vehicle. We find that  overacceleration occurs in these \textit{elementary} models when: (i) \textit{No} additional information from vehicles further ahead of the preceding vehicle is used.  (ii) \textit{No} driver psychological speed thresholds 
	and \textit{no} non-linear bounded rationality are applied.  (iii)
		There is  \textit{no} stronger tendency to accelerate than to decelerate among drivers or automated systems in response to upstream speed disturbances,
		\textit{no} heightened preference for free-flow conditions over safe spacing,
	and   \textit{no} other model extensions---such as heterogeneous driving behavior, randomness, or multi-phase velocity-spacing relationships---are incorporated.

To emphasize the fundamental importance of  overacceleration in traffic theory, we demonstrate that it occurs naturally in  the single-leader car-following models \textit{without} being explicitly introduced. This reveals that  overacceleration is an inherent model property: it becomes evident when \textit{no} vehicle overdeceleration under free-flow conditions is incorporated into the model. 
\end{abstract}
  
\maketitle
 \tableofcontents

\section{Introduction 
\label{Introduction}}

\subsection{Importance of Traffic Breakdown Control}

Traffic breakdown is a transition from free flow  to congested traffic. Traffic breakdown occurs mostly at
a bottleneck. During traffic breakdown the average speed decreases and vehicle density increases.
Travel time and energy consumption in congested traffic increase, whereas traffic safety decreases.
For these reasons, there have been developed different theories for the explanation of the nature of traffic breakdown
as well as  a huge number of  various   macroscopic and microscopic  modeling   approaches to 
traffic breakdown simulation and control  at the bottleneck 
(see, e.g., papers~\cite{MayAthol1963A,May1964A,Drake-Shofer1967,HallAllen1986A,HallGunter1986A,Hall1987A,HallBarrow1988A,HallPersaud1989A,HallAgyemang-Duah1991A,HallHurdle1992A,HallPushkar1993A,HallBrilon1994A,PersaudHurdle1988A,PersaudHall1989A,PersaudHall1990A,PersaudYagar1998A,PersaudYagar2001A,Banks1989A10,Cassidy-Skabardonis1989,Banks1990B10,Banks1991A10,Banks1991B10,Chin-May1991,Banks1993A10,HsuBanks1993A,Acha-Daza_Hall1994A,ElefteriadouRoess1995A,Brilon2005A,KondyliElefteriadou2013A,ElefteriadouKondyli2014A} and
 reviews~\cite{May,Manual2010,Gartner2,ElefteriadouBook2014_Int1,Da,Sch,Brockfeld2003,Bellomo,Ferrara2018A,Leu,Mahnke,MahnkeKLub2009A,Wh2,Schadschneider2011,Saifuzzaman2015A,Pa1983,Newell1963,Prigogine1971,New,Nagel2003A,Nagatani_R,Ashton,Drew,Gerlough,Gazis,Barcelo2010,Treiber-Kesting,DaihengNi,MakridisZhang,Kessels,Schadschneider,Chowdhury,Helbing,Mannering1998,Brackstone1999,Shvetsov2003,Maerivoet2005,Rakha2009,Piccoli2009,Roess2014,Hegyi2017,Seo2017,MATSim_Nagel2016,KernerBook1}). Examples of macroscopic approaches to traffic breakdown control
 are on-ramp metering (see, e.g.,  references in~\cite{Papageorgiou2008,Papageorgiou_1990A,Papageorgiou_1991B,Papageorgiou_1997B,Kerner_Control,KernerBook2}) and speed limit control
(see, e.g.,~\cite{Kerner_TRB2007,KernerBook2,WANG_ZHANG2012,YangZhai2018,HanHegyi2022,HegyiHoogendoorn2010,ZhangZhang2024,KhondakerKattan2015}).
Microscopic approaches  to traffic breakdown control are based on features of 
microscopic vehicle characteristics that   control can prevent  traffic breakdown.
 It was found that real
traffic breakdown at the bottleneck exhibits the empirical nucleation nature. Therefore, modeling approaches for traffic breakdown control should be able to
simulate the empirical nucleation nature of traffic breakdown~\footnote{We do not consider in this review
classical Lighthill-Whitham-Richards (LWR) model of traffic breakdown~\cite{LWR,LWR2}  as well as its
 applications like famous cell-transmission model of Daganzo~\cite{LWR3,LWR4} 
because the LWR model cannot explain the empirical nucleation nature of traffic breakdown~\cite{KernerBook2}.}.

In the age of automated vehicles and artificial intelligence (AI), the importance of investigating which microscopic vehicle features can be effectively used  for  traffic breakdown control has been increased considerably.
It is assumed that through an appropriate change in microscopic
behavior of automated vehicles, traffic breakdown could be avoided. Conversely, it is also assumed
that AI can be used to determine
when and how this change in the microscopic behavior of automated vehicles 
should be applied to achieve the greatest possible effect in maintaining the free flow of traffic at the bottleneck.
For this reason, in this review we limit the consideration on {\it microscopic}  vehicle characteristics that   control through automated vehicles and AI can prevent  traffic breakdown
  at the bottleneck.

	\subsection{Purpose of the Review}
 
This review article addresses a fundamental, long-standing point of contention in traffic theory:
\begin{itemize}
\item[--] Is the nucleation character of traffic breakdown at a bottleneck governed by vehicle overdeceleration (overbraking) or by discontinuous vehicle acceleration -- termed    $\lq\lq$vehicle  overacceleration"?
\end{itemize}

\subsection{Vehicle Overdeceleration (Overbraking), Traffic Instabilities, and  Jam Absorption Driving  
\label{Overdeceleration_S}}

Traffic instability was
  discovered in microscopic traffic simulations by Herman, Gazis,  Montroll, Potts, and  Rothery~\cite{GM_Com1,GM_Com2,GM_Com3} as well as
Kometani and  Sasaki~\cite{KS,KS1,KS2,KS4}   at the end of 1950s. 
Traffic instability occurs due to  driver's delays, in particular,  a driver reaction time  
 that can  cause {\it  vehicle
overdeceleration}. Vehicle
overdeceleration is as follows: The speed of the braking vehicle becomes lower that the speed of the preceding vehicle. If overdeceleration is realized for the following  vehicles, traffic instability, i.e., a growing
 speed wave of a local speed decrease~\footnote{For clarity, the term ``local speed decrease'' is used throughout
 this work as a synonym for the more common terms ``local speed drop'' and ``local speed reduction''.}
propagating upstream occurs in traffic flow.  Note that
vehicle  overdeceleration is also called vehicle overbraking.

The idea to traffic instability that should explain traffic breakdown is used in 
the most microscopic classical traffic flow models,
in particular, well-known and widely used models by Newell~\cite{Newell}, Gipps~\cite{Gipps1981,Gipps1986},
Wiedemann~\cite{Wiedemann},
macroscopic models of Payne~\cite{Payne_1,Payne_2} and  Aw-Rascle~\cite{Aw-Raschle}, cellular automation (CA) model of Nagel and Schreckenberg~\cite{Nagel_S},
 optimal velocity model (OVM) of Bando et al.~\cite{Bando_1,Bando,Bando_2,Bando_3},
lattice traffic flow model of Nagatani~\cite{Nagatani_1,Nagatani_2},
intelligent driver model (IDM)  of Treiber~\cite{Treiber}, 
stochastic microscopic   model of Krau{\ss}~\cite{Krauss,Kra_PhD}, and full velocity difference model (FVDM)   of Jiang et al.~\cite{Jiang2001}.
There are a huge number of other microscopic and macroscopic
traffic flow models in which traffic breakdown is simulated through
the classical traffic instability caused by vehicle overdeceleration  
(see, e.g., papers~\cite{Barlovic,Chen2012A,Chen2012B,Chen2014,Ngoduy2013A,Ngoduy2015A,Ngoduy2019A} 
and reviews~\cite{Sch,Brockfeld2003,Wh2,Schadschneider2011,Saifuzzaman2015A,Nagatani_R,Gazis,Barcelo2010,DaihengNi,Treiber-Kesting,Schadschneider,Chowdhury,Helbing,Brackstone1999,Shvetsov2003,Maerivoet2005}). 

 In 1993, it was found~\cite{KK1993} that the development of classical traffic instability leads to the formation of a moving jam (J) within the initially free   flow (F), which is referred to as the  F$\rightarrow$J transition.
 The moving jam is an upstream propagating congestion pattern spatially bounded by two jam fronts. Within the moving jam, vehicle density is high, while speed, and thus the  flow rate, can drop to zero. Sequences of moving jams
 (called also stop-and-go traffic, or jammed flow, or jams  or else traffic oscillations) have been observed in real traffic almost since the beginning of 
traffic research (see classical papers by Koshi, Edie, Treiterer et al.~\cite{Koshi2,Edie19582,Edie1960A2,Edie1961A2,Edie1980A2,Treiterer1974A2,Treiterer19752})
 (see empirical examples of sequences of moving jams, e.g., in Figs.~2.13 
and~2.14  of Sec.~2.6 of~\cite{KernerBook2}).
 In simulations of the   standard traffic models~\footnote{In the review,
the term {\it standard traffic models} means
  the models where traffic breakdown at the                bottleneck is explained by
traffic instability resulting from overdeceleration. This is because in accordance with the review focus (see Sec.~\ref{Focus_S})
the discussion of
other standard traffic models is out of the scope of the review.},
 where traffic breakdown at the                bottleneck is explained by
traffic instability, the formation of moving jams during and after   traffic breakdown
 at the  bottleneck exhibits complex spatiotemporal    traffic flow dynamics~\cite{Nagel2003A,Nagatani_R,Treiber-Kesting,Helbing,Bando_2,KK1994,KK1995A,Helbing1999A,Helbing2001A,Helbing2002A}.

It is assumed that future vehicular traffic is mixed traffic that consists of human-driving and automated vehicles
(e.g.,~\cite{Ioannou,Ioannou1996,Ioannou2006,IoannouChien2002A,Levine1966A_Aut,Liang1999A_Aut,Liang2000A_Aut,Meyer2014A,Bengler2014A,Swaroop1996A_Aut,Swaroop2001A_Aut,Varaiya1993A,Lin2009A,Martinez2007A,Brummelen2018A,fail_Shladover1995A,Rajamani2012A_Aut,Davis2004B9,Davis2014C,Davis2015A_Int1,Davis2006A,Davis2006B,Davis2007A,Davis2008A,Dharba1999A,Marsden2001A,VanderWerf2001A,VanderWerf2002A,TreiberH2001A,Shrivastava2002A,Kukuchi2003A,BoseIoannou2003A,Suzuki2003A,Zhou2005A,vanArem2006,Kesting2007A,Kesting2008A,Kesting2010A,Shladover2012A,Ngoduy2012A,Ngoduy2013B,Papageorgiou2015B,Papageorgiou2015A,Papageorgiou2015C,Perraki2018A,Talebpour2016A,Wang2017A,Mamouei2018A,Sharon2017A,HanAhn2018A,ChenAhn2018A,Zhou2017B,Klawtanong2020A,ZhangZhang2019A,ZhuZhou2020A,ZhouZhu2020A,ZhouAhn2020A,ChenSrivastava2020A,ZhongLee2020A,ZhengLiu2020A,JinSun2020A,Zhou-Zhu2020A,Yao-Hu2019A,YeYamamoto2019A,Zhu-Zhang2018B,Wen-XingLi-Dong2018A,YeYamamoto2018A,YeYamamoto2018B,Xin-Chang2020A}).
Traffic instability   caused by vehicle overdeceleration can also occur in traffic of automated vehicles.
 This traffic instability called {\it string instability} can occur in automated vehicle platoons even when
a reaction time of the automated vehicle can be considered negligible
(e.g.,~\cite{Ioannou,Ioannou1996,Ioannou2006,IoannouChien2002A,Levine1966A_Aut,Liang1999A_Aut,Liang2000A_Aut,fail_Shladover1995A,Davis2004B9,Davis2014C,Davis2015A_Int1,Davis2006A,Davis2006B,Davis2007A,Davis2008A}).
In this case, the choice of dynamic coefficients in the control of the automated vehicle determines whether  the vehicle overdecelerates and, therefore, string instability occurs or not. It should be emphasized that for simulations of mixed traffic usually standard traffic models for human-driving vehicles
are used  (e.g.,~\cite{Dharba1999A,Marsden2001A,VanderWerf2001A,VanderWerf2002A,TreiberH2001A,Shrivastava2002A,Kukuchi2003A,BoseIoannou2003A,Suzuki2003A,Zhou2005A,vanArem2006,Kesting2007A,Kesting2008A,Kesting2010A,Shladover2012A,Ngoduy2012A,Ngoduy2013B,Papageorgiou2015B,Papageorgiou2015A,Papageorgiou2015C,Perraki2018A,Talebpour2016A,Wang2017A,Mamouei2018A,Sharon2017A,HanAhn2018A,ChenAhn2018A,Zhou2017B,Klawtanong2020A,ZhangZhang2019A,ZhuZhou2020A,ZhouZhu2020A,ZhouAhn2020A,ChenSrivastava2020A,ZhongLee2020A,ZhengLiu2020A,JinSun2020A,Zhou-Zhu2020A,Yao-Hu2019A,YeYamamoto2019A,Zhu-Zhang2018B,Wen-XingLi-Dong2018A,YeYamamoto2018A,YeYamamoto2018B,Xin-Chang2020A}).

 Based on the above-mentioned classical works to the theory of traffic instabilities leading to moving jam formation,
 microscopic approaches to {\it jam absorption driving}
(also referred to as stop-and-go wave dissipation, stop-and-go wave suppression, mitigation of
traffic oscillations, or shock wave damping) have been developed
(see, e.g., papers~\cite{SternCui2018,WangJin2023,HanWang2021,NishiTomoeda2013,TaniguchiNishi2015,Nishi2020A,HeZheng2017,LiYanagisawa2024,LiuZheng2025,ZhengZhang2020,WangLi2022,SuzukiNishi2026}); a history of the jam absorption driving can be found in a recent review of He   et al.~\cite{ZhengbingHe}.  It is assumed in these studies that the cause of traffic breakdown is traffic instability  resulting from
vehicle overdeceleration.

\subsection{Why Cannot Traffic Instability   
Be  The Cause of Empirical Traffic Breakdown? \label{No_OD_Br_Sec}}

Contrary to theoretical predictions of 
standard traffic models of Sec.~\ref{Overdeceleration_S}~\cite{Sch,Brockfeld2003,Wh2,Schadschneider2011,Saifuzzaman2015A,Nagatani_R,Gazis,Barcelo2010,DaihengNi,Treiber-Kesting,Schadschneider,Chowdhury,Helbing,Brackstone1999,Shvetsov2003,Maerivoet2005,Bando_2,KK1994,Helbing1999A,Helbing2001A,Helbing2002A,SternCui2018,WangJin2023,HanWang2021,NishiTomoeda2013,TaniguchiNishi2015,Nishi2020A,HeZheng2017,LiYanagisawa2024,LiuZheng2025,ZhengZhang2020,WangLi2022,SuzukiNishi2026,ZhengbingHe}, as shown in   books~\cite{KernerBook1,KernerBook2,KernerBook3,KernerBook4}, the empirical
traffic breakdown {\it observed in real field traffic data} is a transition from free flow (F) to synchronized flow (F) (F$\rightarrow$S  transition), {\it not}
moving jam emergence.
Synchronized flow is a new traffic phase of congested traffic
 introduced in the three-phase traffic theory~\cite{KernerBook1,KernerBook2,KernerBook3,KernerBook4}. 
  Traffic breakdown (F$\rightarrow$S  transition) at the bottleneck
 exhibits the nucleation nature.

\begin{figure*}
\begin{center}
\includegraphics[width = 10.5 cm]{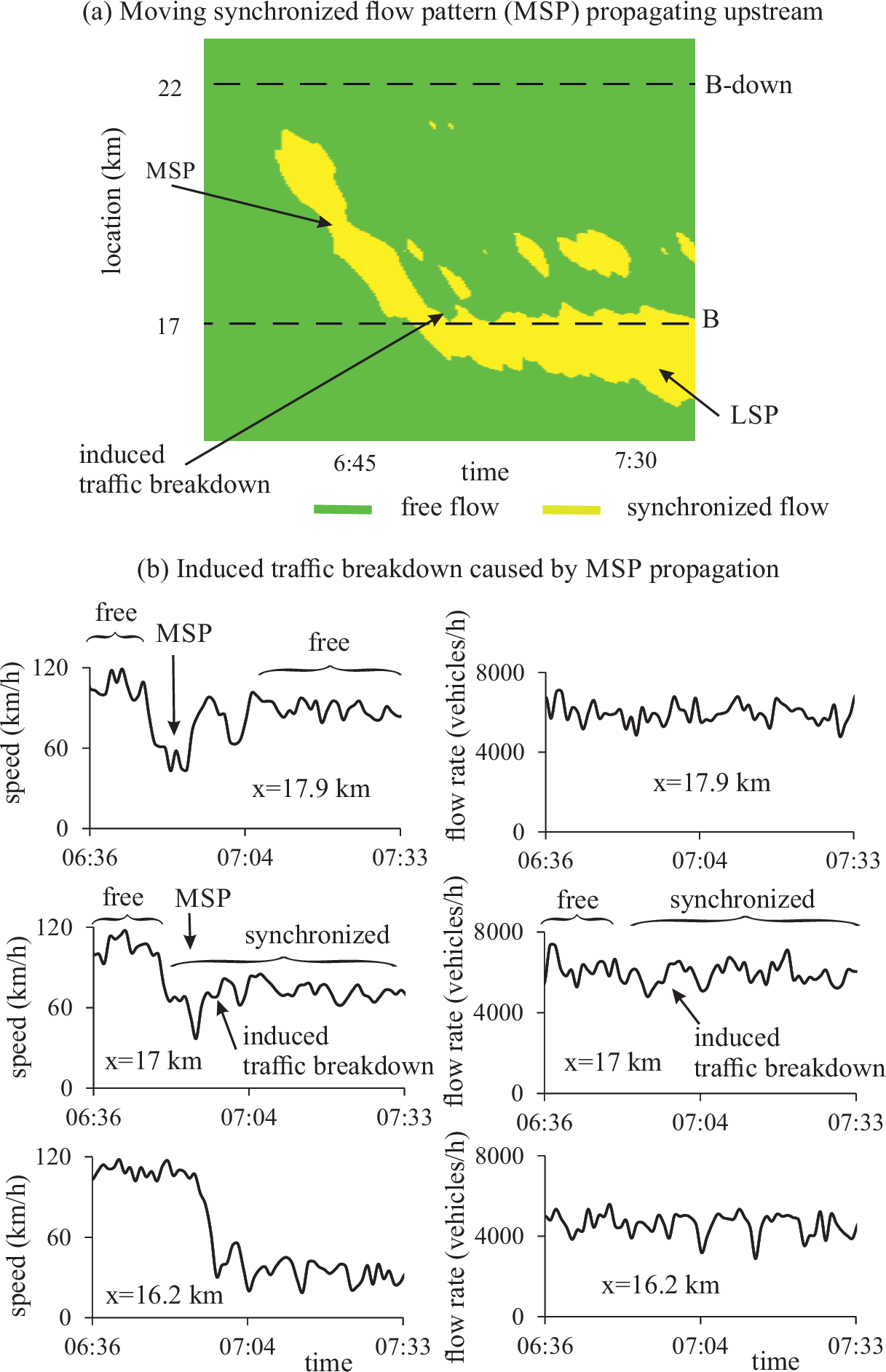}
\end{center}
\caption[]{Empirical nucleation nature of  traffic breakdown (F$\rightarrow$S transition) at bottleneck.
(a) Empirical   speed data  measured through road detectors installed along   road, which are  presented  
in space and time with averaging method described in Sec.~C.2 of Ref.~\cite{TomTom}.
(b) 1-min averaged speed (left) and total flow rate 
across the road (right) related to (a) measured at three road locations; $\lq\lq$free" and $\lq\lq$synchronized" are free flow and synchronized flow, respectively.
Adapted 
from~\cite{KernerBook1,KernerBook2,KernerBook3,KernerBook4}. 
}
\label{20041998_MSP}
\end{figure*}

 But what is the empirical nucleation nature 
of traffic breakdown (F$\rightarrow$S  transition) at the bottleneck?
 One   empirical example
of the empirical nucleation nature of traffic breakdown 
 is presented in Fig.~\ref{20041998_MSP}(a): 
A  moving synchronized flow pattern (MSP) occurs at a downstream off-ramp bottleneck B-down and  it propagates further  upstream. 
 When the MSP reaches the upstream on-ramp bottleneck B, the MSP   induces traffic breakdown   (F$\rightarrow$S  transition) 
at bottleneck B (Fig.~\ref{20041998_MSP}(b), $x=$ 17.0 km). Due to the breakdown a localized synchronized flow pattern (LSP) is realized
at bottleneck B  (Fig.~\ref{20041998_MSP}(b), $x=$ 17.0 and 16.2 km).

Contrary to a moving jam, where   the flow rate is significantly lower than outside the jam,
the flow rate within   empirical synchronized flow  
is nearly    identical to the flow rate in free flow outside the MSP and LSP.   In other words:
 {\it No}   moving jams emerge within the MSP
(Fig.~\ref{20041998_MSP}(b), $x=$ 17.9 km) and
{\it no}   moving jams occur within the LSP (Fig.~\ref{20041998_MSP}(b), $x=$ 16.2 km). Therefore,
 empirical traffic breakdown at a bottleneck is indeed the F$\rightarrow$S  transition that exhibits
the empirical nucleation nature  (Fig.~\ref{20041998_MSP}(a))\footnote{See other empirical examples of the empirical nucleation nature of traffic breakdown (F$\rightarrow$S  transition) in 
books~\cite{KernerBook1,KernerBook2,KernerBook3,KernerBook4,RehbornBook}.}.
As shown in   books~\cite{KernerBook1,KernerBook2,KernerBook3,KernerBook4}  (see, e.g., Chap.~10 of~\cite{KernerBook2}),
{\it none } of  
  the standard traffic theories (see, e.g., papers~\cite{GM_Com1,GM_Com2,GM_Com3,KS,KS1,KS2,KS4,Newell,Gipps1981,Gipps1986,Wiedemann,Payne_1,Payne_2,Aw-Raschle,Nagel_S,Bando_1,Bando,Bando_2,Bando_3,Nagatani_1,Nagatani_2,Treiber,Krauss,Kra_PhD,Jiang2001,Barlovic,Chen2012A,Chen2012B,Chen2014,KK1993,KK1994,KKK1997,Helbing1999A,Helbing2001A,Helbing2002A}
	and reviews~\cite{May,Manual2010,Gartner2,ElefteriadouBook2014_Int1,Da,Sch,Brockfeld2003,Bellomo,Ferrara2018A,Leu,Mahnke,MahnkeKLub2009A,Wh2,Schadschneider2011,Saifuzzaman2015A,Pa1983,Newell1963,Prigogine1971,New,Nagel2003A,Nagatani_R,Ashton,Drew,Gerlough,Gazis,Barcelo2010,Treiber-Kesting,DaihengNi,MakridisZhang,Kessels,Schadschneider,Chowdhury,Helbing,Mannering1998,Brackstone1999,Shvetsov2003,Maerivoet2005,Rakha2009,Piccoli2009,Roess2014,Hegyi2017,Seo2017,MATSim_Nagel2016}) can explain the empirical nucleation nature of traffic breakdown
	(F$\rightarrow$S  transition) at the bottleneck
	(Fig.~\ref{20041998_MSP}).

Over the past decades,
	there are substantial advancements by further developments of classical models (e.g., IDM, FVDM,  OVM, and Wiedemann-model), in which numerous modern modifications  
	have successfully incorporated driver psychological thresholds, acceleration asymmetry,   non-linear bounded rationality
	as well as other features of driver' behavior (see, e.g., papers by Panwai and   Dia~\cite{Panwai-Dia2005A},  Treiber, Kesting, 
	and Helbing~\cite{Kesting2010A,Treiber_Kesting2006A,Treiber_Kesting2013B}, Gong et al.~\cite{Gong-Liu2008A},
	Tang et al.~\cite{Tang-Caccetta2012,Tang-Huang2015,Tang-Huang2017}, Bifulco et al.~\cite{Bifulco-Pariota2013}, Yu and Shi~\cite{Yu-Shi2013A}, 
	and  papers~\cite{Saifuzzaman2015A,Pariota-Bifulco2015,HamdarTreiber2015,Durrani-Lee2019A,Jafaripournimchahi2020,Tang-Luo2016A,Wageningen-Kessels2015A,Md-Shafiul-Azam2026A,Yifan-Zhang2022A,XiaoyunChen2020A,YeeunKim2024A,SamerHamdar2020A,Md-Zakir-Hosen2026A,Zihao-Wang2020A,Hossain-Tanimoto2022,SroczyskiCz2023A,Fabian-Hart2024A}  
as well as references there).

	However, these modifications do not change the fact that traffic breakdown is explained in these models by traffic instability caused by overdeceleration.
	For this reason, these model advancements cannot also explain the empirical nucleation nature of traffic breakdown
	(F$\rightarrow$S  transition) at the bottleneck
	(Fig.~\ref{20041998_MSP}). 
	This is the origin of the statement that vehicle overdeceleration and the resulting traffic instability
cannot be  the cause of empirical traffic breakdown.
A detailed presentation of the standard theory of traffic flow dynamics,  in which traffic instability is assumed to be
the origin of  traffic breakdown,
 can be found in the book by
 Treiber and Kesting~\cite{Treiber-Kesting}. It must be emphasized that a large amount of field data on road traffic with human-driven vehicles, collected over many years in various countries, supports the  critical conclusion regarding the standard theory of traffic flow dynamics~\footnote{In contrast to traffic consisting of human-driven vehicles,
there is hardly any field data available for mixed traffic. Therefore, it could be a very interesting task for future empirical studies to investigate, using   field data on mixed traffic, whether overdeceleration of automated vehicles and the resulting traffic instability could be the cause of empirically observed traffic breakdown at a bottleneck in mixed traffic.}. Therefore, we can draw the following conclusion:
		 \begin{itemize}
\item [--]   Vehicle overdeceleration and the resulting traffic instability {\it cannot}
be  the cause of empirical traffic breakdown (F$\rightarrow$S transition) at the bottleneck.  
\end{itemize}

\subsection{How Does  Three-Phase Traffic Theory Explain The Cause of Empirical Traffic Breakdown? \label{3_OA_Br_Sec}}

The cause
for the introduction   of
   three-phase traffic theory~\cite{Kerner1999B,Kerner1999A,Kerner1999C,Kerner1998C,Kerner1998B,Kerner2000} was the empirical nucleation nature of
traffic breakdown (F$\rightarrow$S transition).
The three-phase traffic theory 
	  is the framework for
		the description of empirical vehicular traffic in the three traffic phases~\footnote{{\bf The definition
		of the     traffic phase $\lq\lq$wide moving jam" is as follows}:
		The   wide moving jam traffic phase   
 is a moving jam that exhibits the jam characteristic feature to propagate through any states of
	 free flow and synchronized flow as well as through 
	any bottleneck
	while maintaining the mean velocity of the downstream jam front (see Chap.~11 of~\cite{KernerBook1})~\cite{KK1994,KK1995A,KR1996,KKK1997}.   Microscopically,
	the characteristic feature of the wide moving jam is explained by the existence of at least one flow interruption interval within a wide moving jam: Traffic flow is interrupted within the wide moving jam.
	The existence of a flow interruption interval(s) (criterion for the flow interruption interval has been considered in Sec.~2.6.1 of~\cite{KernerBook2})
		can be considered as a {\it microscopic criterion of the wide moving jam traffic phase}~\cite{KernerBook2,KKHR2,KKHR3}.
	It is worth emphasizing that, in the  literature,   the wide moving jam traffic phase within the three-phase traffic theory is frequently described using terms such as $\lq\lq$stop-and-go traffic" or $\lq\lq$jammed flow" or else $\lq\lq$jams" 
(see, e.g., Secs.~12.6.3  and~14.3.2  of~\cite{Treiber-Kesting}), which are related to a sequence of moving jams.
However, if within the sequence of moving jams of congested traffic no flow interruption intervals  is observed, then the jam sequence belongs
to the synchronized flow traffic phase (see, e.g., Fig.~2.14 
  of~\cite{KernerBook2}), rather than to the wide moving jam traffic phase. Furthermore, in most cases, the empirically observed   sequence of moving jams constitutes a sequence comprising two or three distinct traffic phases  (see, e.g., Fig.~2.13 
  of~\cite{KernerBook2}). Consequently, terms such as $\lq\lq$jammed flow" or $\lq\lq$jams" cannot adequately describe the wide moving jam traffic phase.},\footnote{{\bf The definition
		of the     traffic phase $\lq\lq$synchronized flow" is as follows}: In contrast  to the wide moving jam phase,
there are no flow interruption intervals in synchronized flow. 
The  
  downstream front of synchronized flow can  be localized at a highway bottleneck.
	The synchronized flow traffic phase  ensures
the nucleation nature of the F$\rightarrow$S transition at the
bottleneck.  In the three-phase traffic theory,
the  distinguishing 
between  
  the synchronized flow and wide moving jam  traffic phases is made as follows: If in a set of empirical traffic data, we have identified congested traffic states associated with the wide moving jam  phase, then all remaining congested states in the empirical data set are related to the synchronized flow 
	  phase (page 21 of~\cite{KernerBook2} and page 20 of~\cite{KernerBook4}).}:
 \begin{itemize}
		\item [1.] Free flow (F). 
		\item [2.] Synchronized flow (S). 
	\item [3.] Wide moving jam (J).
		\end{itemize}
		Three-phase traffic theory
		follows {\it exclusively}   from the analysis of empirical spatiotemporal
 traffic data, not from
simulations of   mathematical traffic models.
\begin{itemize}
\item [--]  Through the analysis of real traffic data collected over many years in various countries, the three-phase traffic theory has established that there are
{\it common} empirical spatiotemporal traffic phenomena that do not require confirmation through mathematical traffic models.  
\end{itemize}
One of the most important of these empirical spatiotemporal traffic phenomena that does
 not require confirmation through mathematical traffic models is the following (see Chap.~5 of~\cite{KernerBook1} and   Fig.~\ref{20041998_MSP}(a)):
\begin{itemize} 
\item [--]  Empirical traffic breakdown at a bottleneck is the F$\rightarrow$S  transition that exhibits
the empirical nucleation nature.
\end{itemize}

The three-phase traffic theory explains the common empirical spatiotemporal traffic phenomena through a series of theoretical assumptions (hypotheses) that
	should be confirmed by mathematical traffic models~\cite{Kerner1999B,Kerner1999A,Kerner1999C,Kerner1998C,Kerner1998B,Kerner2000}. 
	One of the most important of the theoretical assumptions of the three-phase traffic theory is the following:
	 \begin{itemize}
\item [--]  The empirical nucleation of
traffic breakdown (F$\rightarrow$S transition) at the bottleneck  is caused by  
a {\it discontinuous character}
 of vehicle {\it    acceleration}  behaviors called by  vehicle  overacceleration. 
\end{itemize}

		\subsection{Overdeceleration versus Overacceleration -- \newline
		The Fundamental Contradiction between
		Standard and Three-Phase Traffic Theories}

		There is a widespread belief among traffic scientists that traffic breakdown should be explained primarily by features of vehicle braking leading to vehicle overdeceleration and, therefore, to traffic instability  (Sec.~\ref{No_OD_Br_Sec}).  
 In contrast to this conclusion of the standard traffic theories
(e.g.,~\cite{Gartner2,ElefteriadouBook2014_Int1,Sch,Brockfeld2003,Bellomo,Leu,Mahnke,MahnkeKLub2009A,Schadschneider2011,Saifuzzaman2015A,Pa1983,Newell1963,Nagel2003A,Nagatani_R,Ashton,Drew,Gerlough,Gazis,Barcelo2010,Treiber-Kesting,DaihengNi,MakridisZhang,Kessels,Schadschneider,Chowdhury,Helbing,Mannering1998,Brackstone1999,Shvetsov2003,Maerivoet2005,Hegyi2017,MATSim_Nagel2016}), 
three-phase traffic theory  proposes a fundamentally different
 mechanism~\cite{KernerBook1,KernerBook2,KernerBook3,KernerBook4,KernerReviews}:
\begin{itemize}
\item [--]
 Three-phase traffic theory assumes that traffic breakdown is caused not by braking behavior   leading to vehicle overdeceleration, but by a {\it discontinuous character of vehicle acceleration}: The probability of vehicle acceleration drops
when due to traffic breakdown free flow transforms to synchronized flow. In three-phase traffic theory, this type of vehicle  acceleration behavior   is referred to as {\it overacceleration}. 
\end{itemize}
Because of the long-standing belief that traffic breakdown is driven by traffic instability due to overdeceleration, the idea that overacceleration could be the underlying cause for real traffic breakdown is difficult for many researchers to accept.

  The concept of $\lq\lq$discontinous overacceleration," intended to explain the empirical origins of traffic breakdown
 (F$\rightarrow$S transition), was first    introduced in 1999 
 within the framework of three-phase traffic theory~\cite{Kerner1999B,Kerner1999A,Kerner1999C}.
At that time, instead of the term $\lq\lq$probability of overacceleration,"  the term $\lq\lq$probability of overtaking"
 (Fig.~4(b) in~\cite{Kerner1999C}) 
  was used. That is to say, in the terminology used now, an overacceleration mechanism through lane changes was assumed
	(the overacceleration mechanism through lane changes will be considered in Sec.~\ref{Lane_OA_sec}).
	 
	The first   mathematical  model of overacceleration was introduced in 2002 in the stochastic microscopic
	three-phase traffic model by Kerner and Klenov~\cite{KKl}. 
 Different  mathematical formulations of discontinuous overacceleration  
 has been further   applied in   other there-phase traffic flow models
(e.g.,~\cite{KKl,KKW,KKl2003A,KKl2004A,KKl2006A,KKl2009A,KKHS2013_Int1}). Traffic flow models  
 in the framework of three-phase traffic theory have been further developed and
used in many   traffic simulations, in particular in papers by Wang et al.~\cite{WANG_ZHANG2012}, Yang et al.~\cite{YangZhai2018,YangZhao2024A}, Wu et al.~\cite{Wu2008},  He et al.~\cite{HeGuan2010A},   
Hu et al.~\cite{HuWang2012A,HuLiu2021A,HuZhang2019A,HuHao2020A,HuQiao2022A,HuYu2025A,HuLinHao2025A},
Qian et al.~\cite{QianFeng2017A}, Lyu et al.~\cite{LyuHu2022A}, and Zeng et al.~\cite{ZengQian2019A,ZengQiao2021A} 
(see also papers~\cite{ZhaoLin2020A,ChenZhu2024A,Lubashevsky2019A,Chechina2026A,Haoyu-Fang2026A,Haopeng-Deng2026A,Jianpeng-Sun2025A}
and references there).
 
Mathematical formulations of the concept of discontinous overacceleration based
 on {\it stochastic} three-phase traffic models~\cite{KKl,KKW,KKl2003A,KKl2004A,KKl2009A,KKHS2013_Int1}
  have already been reviewed 
in~\cite{KernerBook1,KernerBook2,KernerBook3,KernerBook4,KernerReviews}.
 However, in the stochastic models both overacceleration and the opposite effect of overdeceleration are described 
on average through   model fluctuations~\cite{KKl,KKW,KKl2003A,KKl2004A,KKl2009A,KKHS2013_Int1}.
 It seems that, based on simulations of the   stochastic three-phase traffic models, 
 it is difficult to show that the empirical nucleation nature of traffic breakdown can be explained {\it exclusively} by discontinuous overacceleration, i.e.,
 {\it without any effect of vehicle overdeceleration}. 

	\subsection{Overdeceleration versus Overacceleration -- \newline
		Behavioral Origin \label{OD-OA_Beh_Or_S}}
 
\begin{itemize}
\item [--]
The behavioral origin of vehicle
  overdeceleration is related to the wish of the vehicle to avoid vehicle
collisions.
\end{itemize}
\begin{itemize}
\item [--] The behavioral origin of vehicle overacceleration
is related to the vehicle's desire to move in free
flow.
\end{itemize}

  \subsection{Methodology  and Central Question to Be Answered \label{Meth_Int_S}}

Contrary to reviews~\cite{KernerBook1,KernerBook2,KernerBook3,KernerBook4,KernerReviews}, we use in this review a different methodology: We consider  {\it deterministic} three-phase traffic flow models for
	both  human-driving and automated vehicles~\cite{Kerner2023A,Kerner2023B,KKl2025A}.
	We choose model parameters in the deterministic three-phase traffic flow models~\cite{Kerner2023A,Kerner2023B,KKl2025A} at which
	{\it no} vehicle overdeceleration is realized. This means that no classical traffic instability and no string instability
	is realized in the models. 
	The central question that we would like to answer in this review is as follows:
\begin{itemize}
\item [--] Can mathematical three-phase traffic models, in which neither    vehicle overdeceleration nor traffic instability occur,   nevertheless simulate   the empirical nucleation of
traffic breakdown (F$\rightarrow$S transition) at the bottleneck {\it exclusively}   through
vehicle {\it  acceleration}  behaviors?
\end{itemize}
	The absence of vehicle overdeceleration and, as a result, the absence of traffic instabilities in the models  allows us
  to explain   that and why overacceleration is
  a fundamental microscopic mechanism in traffic breakdown control both for human-driving and automated vehicles.
	
	The main contribution of this review is that with the use of 
microscopic {\it deterministic} models of Ref.~\cite{Kerner2023A,Kerner2023B,KKl2025A} we   show  
that the   control of vehicle overacceleration, which can be made through individual control of  vehicle motion,  is a key element of traffic breakdown control.  
 For this reason, vehicle overacceleration is a fundamental microscopic characteristic for controlling traffic breakdown.

 Using the methodology described above, we show that,
even if no vehicle overbraking effect and no traffic instability occur in the traffic flow, the traffic breakdown, as observed in real traffic data, still exhibits the nucleation characteristic caused by discontinuous overacceleration.   
This challenges the widespread assumption that effective management of traffic breakdown is possible through
jam absorption driving. However, we discuss qualitatively    that 
 jam absorption driving in cooperation with overacceleration management
can be efficient  for traffic congestion mitigation (Sec.~\ref{S-C-F_sec}).

	\subsection{The Focus of the Review \label{Focus_S}}
	
  There have been developed a   number of   different 
	modeling approaches to interpret traffic breakdown mechanisms.  In particular,
	to explain the empirical F$\rightarrow$S transition and resulting synchronized flow, model
	extensions like  heterogeneous driving behavior, randomness, or multi-phase velocity-spacing relationships
	have been developed in models of Jiang and Wu~\cite{Jiang-Wu2004}, model of Tian et al.~\cite{TianYuan2012} as well as in
	several CA models called  \textit{generalized Nagel-Schreckengerg CA models}~\cite{TianJiang2019,Tian2020A}, 
	like models of Gao et al.~\cite{Gao2007A,Gao2009A}, model of Zhao et al.~\cite{Bo-Han-Zhao2009},
	model of Jia et al.~\cite{Bin-Jia2011A}, and two-state models of Tian et al.~\cite{Tian-Treiber2015A,Tian-Li2016}.
	To describe synchronized flow,  comfortable driving model of Knospe et al.~\cite{Knospe2000A}
	has been introduced that incorporates anticipation effects, reduced acceleration capabilities and an enhanced interaction horizon for braking (brake light model); further model modifications and applications of the comfortable driving model
	  have been done by Knospe et al.~\cite{Knospe2001A,Knospe2002A}, Jiang and Wu~\cite{Jiang-Wu2003A}, Tian et al.~\cite{Tian-Jia2009A,Tian-Jia2014,Tian-Jia2017},
and Wang et al.~\cite{Wang-Ma2019} (see reviews of Tian et al.~\cite{TianJiang2019,Tian2020A}).
To explain synchronized flow, the model by Lee et al.~\cite{Lee_Sch2004A} incorporates situation-dependent vehicle deceleration safety margins by applying two discrete behavioral states (optimistic and pessimistic) combined with collision-free braking.
 The model of Lee et al. has have been
	further developed and used by Pottmeier et al~\cite{Pottmeier2007},
	Li et al~\cite{Li-Gao2006}, Kokubo et al~\cite{Kokubo2011}, Tian et al~\cite{Tian-Li2016B},   L{\'a}rraga and   Alvarez-Icaza~\cite{Larraga2010}
	as well as Habel and   Schreckenberg~\cite{Habel-Sch2016}.
  As mentioned in Sec.~\ref{No_OD_Br_Sec},
	there are many other
	substantial advancements in traffic theory made by the broader traffic flow community over the past decades; in particular, numerous modern modifications to classical models have successfully incorporated driver psychological speed thresholds   and non-linear bounded rationality. 
	However, the discussion of all these    modern modeling approaches   is out of the scope of  this review.
	
	This is because the focus of the review is to 
 address  a fundamental point of contention in traffic theory formulated above:
 \textit{Is the nucleation character of traffic breakdown at the bottleneck governed by vehicle overdeceleration   or 
by discontinuous vehicle acceleration   (overacceleration)?}

		To achieve this goal and mathematically model the nucleation nature of traffic breakdown, Secs.~\ref{Safety_OA_sec} and~\ref{TPACC_St_S} utilize single-leader deterministic car-following traffic flow models. These \textit{elementary} models include (i) no reliance on look-ahead traffic data and no situation-dependent safety margins, (ii) no driver or vehicle psychological speed thresholds, (iii) no non-linear bounded rationality, (iv)  no  stronger tendency to accelerate than to decelerate among drivers or automated systems in response to upstream speed disturbances, 
		(v) no greater preference for free-flow conditions over compliance with safe spacing, 
		and (vi) no other model extensions, such as heterogeneous driving behavior, randomness, or multi-phase velocity-spacing relationships.

To emphasize the basic importance of discontinuous overacceleration for traffic theory, in 
the deterministic models of Secs.~\ref{Safety_OA_sec} and~\ref{TPACC_St_S}  {\it no} overacceleration mechanism
has been explicitly introduced. 
  \begin{itemize}	
 \item[--] 
Discontinuous overacceleration does occur in the single-leader deterministic car-following traffic flow models {\it without} explicitly introducing
it in the model.
 \item[--] The overacceleration mechanism  
becomes  evident when vehicle overdeceleration is not incorporated into the model
 \textit{under free-flow conditions}.  
\item[--] The  overacceleration mechanism is {\it inherent model property}: it  results
 directly from mathematical single-leader deterministic car-following traffic flow models for both human-driven and automated vehicles.
\end{itemize}

To focus the review on   the explanation   of the  
 nucleation nature of traffic breakdown (F$\rightarrow$S transition) through discontinuous overacceleration and on the discussion
 of overacceleration  versus overdeceleration,
we do not   review other important features of overacceleration, in particular, we do not consider
the theory of the S$\rightarrow$F instability occurring in synchronized flow due to discontinuous 
overacceleration~\cite{Kerner2015C}~\footnote{The exception is Sec.~\ref{S-F1_sec} in which an idea about the application of overacceleration management for
the initiating of the  S$\rightarrow$F instability leading to free flow recovering is qualitatively discussed.}.
 We do not also  review a spatiotemporal competition of the 
   S$\rightarrow$F instability caused by overacceleration with the S$\rightarrow$J instability
	caused by overdeceleration~\cite{Kerner2019AA}. 

 \subsection{Structure}
 
The review is organized as follows: 
In section ~\ref{Def_OA_sec} we give a qualitative explanation 
of the concept of overacceleration. 

In Secs.~\ref{Safety_OA_sec}--\ref{AV_MSP_S}, taking into account various traffic conditions,
for a traffic flow consisting of either automated vehicles or human-driven vehicles,
a mathematical confirmation of the following fundamental statement of the study is provided:
\begin{itemize}
\item [--] Even if no vehicle overdeceleration effect and no traffic instability occur in the traffic flow, the traffic breakdown, as observed in real traffic data, still exhibits the nucleation characteristic caused by discontinuous overacceleration.
\end{itemize}

In Sec.~\ref{OA_Not_OD_Sec}, we consider a microscopic model for human-driving vehicles that integrates both overacceleration and overdeceleration mechanisms. The goal   of Sec.~\ref{OA_Not_OD_Sec} is as follows:
We show
that the nucleation characteristic of traffic breakdown (F$\rightarrow$S transition) is caused exclusively
 by discontinuous overacceleration and not by overdeceleration, which determines the characteristics of the formation of moving jams in synchronized traffic flow after  
traffic breakdown.

Sec.~\ref{Dis_S}  discusses the following topics: (i) that and why
discontinuous overacceleration is an inherent   property of
   single-leader car-following  models 
	(Sec.~\ref{Dis_OA_Feature_S}), (ii) inconsistent interpretations of the concept ``overacceleration'' 
	(Sec.~\ref{Term_OA_OA_S}),   (iii) why
discontinuous overacceleration can be considered a critical self-organized spatiotemporal traffic phenomenon
(Sec.~\ref{Dis_OA_Sp-Tem_S}), (iv)
  controversial views regarding   theoretical highway capacities   (Sec.~\ref{OA-OD_Cap_S}),
	(v) about possible contribution of  overdeceleration   to the nucleation of traffic breakdown (F$\rightarrow$S transition)
	at   bottlenecks in real-world traffic (Sec.~\ref{OA-OD_Real-World_S}) and in  
	 mathematical modeling   (Sec.~\ref{OA-OD_Modeling_S}),  
	(vi) commonality and individuality of different overacceleration mechanisms
	(Sec.~\ref{OA-Commonality_S}),
and (vii) differing views regarding the empirical validation  of traffic models    (Sec.~\ref{Valid_OA_Model_S}).

Sec.~\ref{S-F_gen_sec}  considers some ideas to microscopic  overacceleration management  through     automated vehicles  and AI:
(i) about strategies for traffic breakdown prevention 
(Sec.~\ref{Ad_Abs_Sec}),
(ii) how microscopic  overacceleration management can be organized 
(Sec.~\ref{S-F_sec}), (iii) when and how
a cooperation between discontinuous overacceleration and jam absorption  driving to restore free
 traffic flow at the bottleneck possible (Sec.~\ref{S-C-F_sec}),
(iv)  a possibility of
 dissolution of  synchronized flow at the bottleneck   initiated by a single automated vehicle through 
the nucleation  of an S$\rightarrow$F transition caused by the discontinuous nature of the overacceleration
(Sec.~\ref{S-F1_sec}).  

Conclusions are formulated in Sec.~\ref{Conl_S}.

\section{Definition and Physics of Vehicle Overacceleration  
\label{Def_OA_sec}}

\subsection{Speed Adaptation within Indifferent Zone for Car-Following 
\label{Cause_Adap_sec}}

Before we define the term {\it overacceleration}, we explain 
  why empirical traffic breakdown at a bottleneck is not caused by traffic instability, i.e., no overdeceleration
occurs in free flow at the bottleneck. 
 
 We consider a car-following scenario occurring in a hypothetical steady state of traffic flow, in which identical vehicles move at a given speed
$v$ satisfying the condition $0 <v < v_{\rm free}$, where $v_{\rm free}$ denotes a maximum free-flow speed~\footnote{In the limiting
 case of the steady state with $v= v_{\rm free}$, any sufficiently large time-headway to the preceding vehicle is possible.}.
In the most standard traffic models (e.g.,~\cite{GM_Com1,GM_Com2,GM_Com3,KS,KS1,KS2,KS4,Newell,Gipps1981,Gipps1986,Wiedemann,Payne_1,Payne_2,Nagel_S,Bando_1,Bando,Bando_2,Bando_3,Nagatani_1,Nagatani_2,Treiber,Krauss,Kra_PhD,Jiang2001,Chen2012A,Chen2012B,Chen2014,KK1993,KK1994}),  it is assumed that
in the steady state each of the vehicles should maintain a fixed time-headway to the preceding vehicle.
A multitude of such steady states of traffic flow lie on the fundamental diagram, i.e., on 1D-curve in the flow--density (or, equivalently, space-gap--speed)
plane.

\begin{figure} 
\begin{center}
\includegraphics[width = 8 cm]{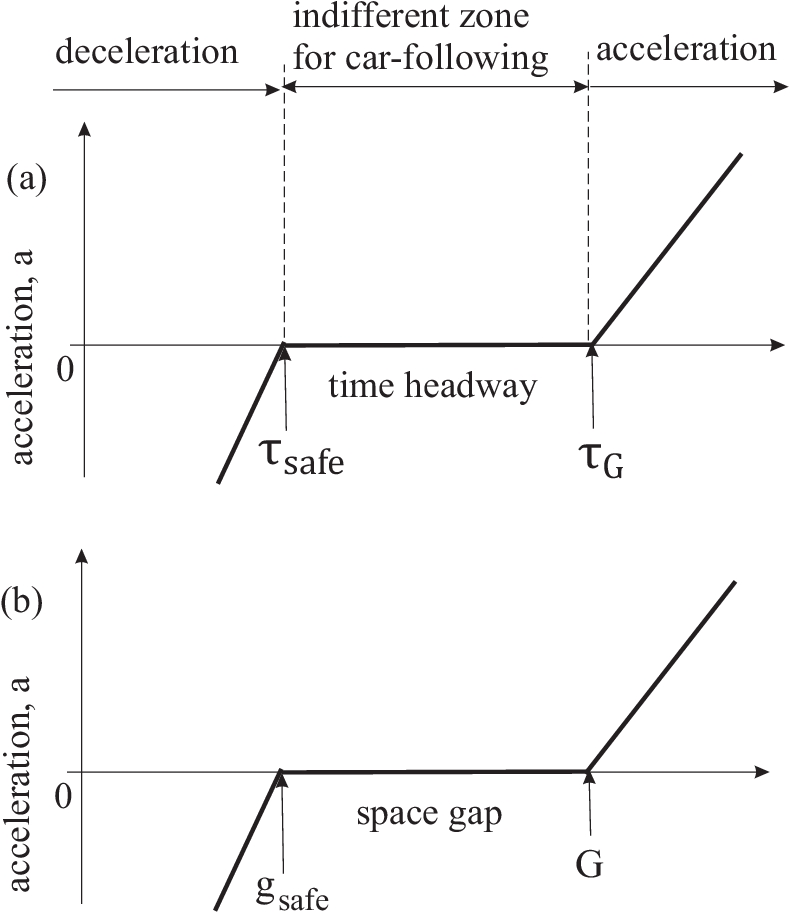}
\caption{Qualitative  presentation of indifferent zone for car-following of three-phase traffic theory~\cite{KernerBook1,KernerBook2,KernerBook3,KernerBook4}:
 Vehicle acceleration (deceleration) as functions of time-headway  (a) and space-gap (b). 
More detailed explanations of the indifferent zone for car-following can be found in Appendix~A of the book~\cite{KernerBook4}.
We assume in these figures that within the indifferent zone $g_{\rm safe} \leq g \leq     G$ the speed difference $\Delta v= v_{\ell} -v=0$; 
$v$ is the vehicle speed,
$v_{\ell}$ is the speed of the preceding vehicle. 
}
\label{IZ_tau_g}
\end{center}
\end{figure}

Rather than the fixed time headway  between vehicles, it is assumed
 in the three-phase traffic theory  that 
  there is an indifferent region for car-following  behavior
\begin{equation}
\tau_{\rm safe} \leq \tau \leq \tau_{\rm G},
\label{indif_tau}
\end{equation}
where 
\begin{equation}
\tau_{\rm G} > \tau_{\rm safe},
\label{indif_tau_more}
\end{equation}
 $\tau$ is time-headway of the vehicle to the preceding vehicle,
 $\tau_{\rm G}$ is a synchronization time-headway, and $\tau_{\rm safe}$ is a safe time-headway.
Under condition (\ref{indif_tau}) a vehicle moves independent of the time-headway to
the preceding vehicle. This explains the term $\lq\lq$indifferent zone for car-following".
When the time-headway $\tau$ becomes less than the safe time-headway $\tau_{\rm safe}$, the vehicle decelerates (Fig.~\ref{IZ_tau_g}(a)).
Respectively, when the time-headway $\tau$ becomes longer than the synchronization
 time-headway $\tau_{\rm safe}$, the vehicle accelerates (Fig.~\ref{IZ_tau_g}(a)).

Space gap between vehicles $g$ is determined   through the obvious formula
\begin{equation}
 g=v\tau,
\label{tau_g}
\end{equation}
where we assume   through the paper~\footnote{The exception is Sec.~\ref{OA_Not_OD_Sec}.} that the vehicle speed $v>0$;
  $g= x_{\ell} - x - d$, $x$ and $x_{\ell}$
	are, respectively, the coordinates of the vehicle and the
	 preceding   vehicle, $d$ is the vehicle length. 
Therefore, formula (\ref{indif_tau}) for the indifferent zone in time-headway for car-following is equivalent to
formula
\begin{equation}
g_{\rm safe} \leq g \leq     G,
\label{indif_g}
\end{equation}
for  the indifferent zone in space-gap between vehicles.
In (\ref{indif_g}),
\begin{equation}
G > g_{\rm safe},
\label{indif_g_more}
\end{equation}
 $G$ is a synchronization space-gap and $g_{\rm safe}$ is a safe space-gap for which we get
\begin{equation}
 G=v\tau_{\rm G} \ {\rm and} \   g_{\rm safe}=v\tau_{\rm safe}.
\label{G_g-safe}
\end{equation}

\begin{figure} 
\begin{center}
\includegraphics[width = 8 cm]{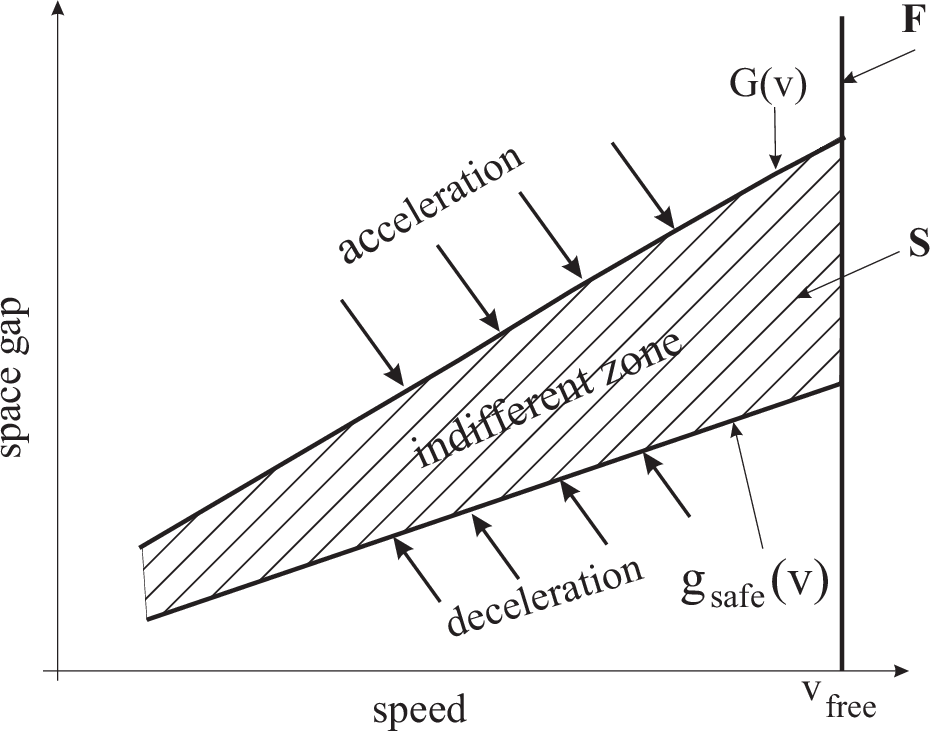}
\caption{Qualitative explanations of the origin of the indifferent zone for car-following
 and asymmetric deceleration--acceleration driver behavior~\cite{KernerBook1,KernerBook2,KernerBook3,KernerBook4}:
A part of the steady states of traffic flow (dashed 2D-region)   shown in the space-gap--speed plane;
 $g_{\rm safe}(v)$ is the speed-dependence of the safe space gap $g_{\rm safe}$, $G(v)$ is the speed-dependence of the synchronization space gap $G$, $v_{\rm free}$ the maximum speed in free flow; F --   free flow, S --   synchronized flow.
}
\label{IndZone}
\end{center}
\end{figure}

The hypothesis about the existence of an
 2D-region of  steady states of traffic flow 
 introduced qualitatively at the end  1990s~\cite{Kerner1999B,Kerner1999A,Kerner1999C,Kerner1998C,Kerner1998B}~\footnote{In the literature, 
the hypothesis of three-phase traffic theory about 2D-region  of steady states of traffic flow
	  is often erroneously interpreted as a 2D-region of traffic states
resulting from the dynamics of moving jams and/or stochastic effects. In fact, both the dynamics of moving jams
 and stochastic effects, which cause spatiotemporally time-dependent
and spatially inhomogeneous
traffic dynamics, lead to a 2D-region of  traffic states in the flow--density plane.
The hypothesis of   the 2D-region of   states of traffic flow in the three-phase traffic theory, however, refers to
{\it hypothetical} steady   states of traffic flow, not to the spatiotemporal, time-dependent
and spatially inhomogeneous
dynamics of traffic.} 
is {\it the origin} of    
  the indifferent region of car-following behavior and the asymmetric acceleration--deceleration behavior of 
	drivers~\footnote{Asymmetric acceleration--deceleration behavior of drivers was probably firstly observed by Treiterer in  empirical observations of
	``jam loops'' resulting from vehicles propagation through a wide moving jam~\cite{Treiterer19752}:
	at the upstream jam front---i.e., where the vehicles slow down---the mean time-headway 
	or space-gap between vehicles is significantly shorter than at the downstream jam front, 
	i.e., where vehicles accelerate upon leaving the   jam
(see theoretical 
 explanations of  Treiterer's ``jam loops'' made in Sec.~IV.C of~\cite{KKHR3} or in Sec.~B.3 of~\cite{KernerBook4}). }
		(\ref{indif_tau}) (or, equivalently, (\ref{indif_g}))
(Fig.~\ref{IZ_tau_g})  as follows: (i) The existence of the 2D-region of steady states of traffic flow 
    (Fig.~\ref{IndZone}) is equivalent to the existence of
the indifferent zone of car-following behavior within which at  $v=v_{\ell}$  vehicle motion is indifferent to the space-gap value, i.e., 
 vehicle	acceleration/deceleration $a=0$  under condition (\ref{indif_g}); (ii)
The existence 
of the boundaries $g=g_{\rm safe}(v)$ and $g=G(v)$   of   the 2D-region of steady states of traffic flow  (Fig.~\ref{IndZone})
implies the asymmetric acceleration--deceleration behavior of drivers---i.e.,  
  a vehicle accelerates ($a>0$) when the space-gap satisfies $g>G$,
	 whereas the vehicle decelerates ($a<0$) under condition $g<g_{\rm safe}$ 
(Fig.~\ref{IndZone}).

	 The first mathematical implementation
of the indifferent region of car-following behavior and the asymmetric acceleration--deceleration behavior of drivers
 was carried out in 2002
in the Kerner-Klenov model~\cite{KKl} as well as in the KKW CA model~\cite{KKW}.
Thus, the hypothesis  regarding  the
 2D-region of  steady states of traffic flow of the three-phase traffic theory constitutes the origin
of drivers' asymmetric acceleration and deceleration behavior, as well as the indifference zone of car-following behavior~\footnote{See more detailed
 explanations in Appendix~A of~\cite{KernerBook4}.}; both are important mathematical features of  the stochastic as well as the deterministic 
 Kerner-Klenov   models~\cite{KKl,KKl2003A,KKl2006A}~\footnote{For a review, see  Chap.~11 of~\cite{KernerBook2}.}.

 \begin{figure} 
		\begin{center}
\includegraphics[width = 8 cm]{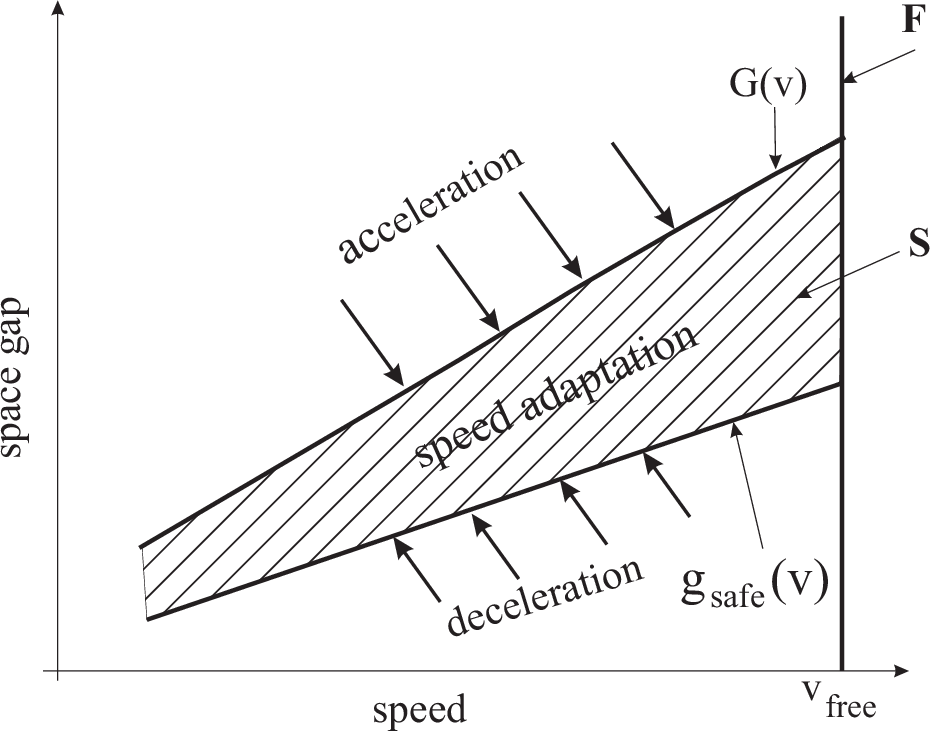}
\caption{Qualitative explanation of vehicle speed adaptation within the indifferent zone for car-following 
 in the
three-phase traffic theory~\cite{KernerBook1,KernerBook2,KernerBook3,KernerBook4}.
 2D-dashed region of steady states in the speed--space-gap plane is  taken from Fig.~\ref{IndZone}.    
}
\label{IZ_Asym2}
\end{center}
\end{figure}

Rather than steady states of traffic flow, dynamic states are realized in real traffic.   However, the above consideration of
 the indifferent zone of car-following behavior related to the steady states of traffic flow is important
  for the dynamic states of traffic flow. In particular, in the three-phase traffic flow models~\cite{KKl,KKW,KKl2003A,KKl2004A,KKl2006A} it is assumed 
  that
acceleration/deceleration $a$ of a vehicle within the indifferent zone  
 satisfies equation
\begin{equation}
a =  K_{\Delta v}\Delta v   \ \textrm{at $g_{\rm safe} \leq g \leq G$},
\label{SA_indif}
\end{equation}
where $K_{\Delta v}$ is a positive dynamic coefficient,
\begin{equation}
 \Delta v= v_{\ell} -v,
\label{delta-v}
\end{equation}
 $v_{\ell}$ is the speed of the preceding vehicle.

Eq.~(\ref{SA_indif}) describes speed adaptation within the indifferent zone of car-following. 
From (\ref{SA_indif})
  we can draw the following conclusion: 
\begin{itemize}
\item[--]
 As long as condition (\ref{indif_g}) is satisfied, there is {\it no}
driver overreaction and, therefore, no overdeceleration of the vehicle
during {\it speed adaptation}.
\end{itemize}

Speed adaptation effect within indifferent zone for car-following  (\ref{indif_g}) (Figs.~\ref{IZ_tau_g} and~\ref{IZ_Asym2}) explains why   in {\it free flow} 
at the bottleneck
 no vehicle overdeceleration and, as a result, no
  traffic instability with moving jam emergence should necessarily appear. 
	In other words, speed adaptation within the indifferent zone for car-following
explains why  in real field traffic data a spontaneous F$\rightarrow$J transition, i.e.,
  moving jam emergence in free flow
 is not observed~\cite{KernerBook1,KernerBook2,KernerBook3,KernerBook4}.
	
	However,   speed adaptation   within indifferent zone for car-following  (\ref{indif_g}) (Figs.~\ref{IZ_tau_g} and~\ref{IZ_Asym2}) does not explain
 the empirical nucleation nature of traffic breakdown (F$\rightarrow$S transition) at the bottleneck. 
The nucleation nature of traffic breakdown   at the bottleneck is
the cause of the introduction of the concept of vehicle overacceleration~\cite{KernerBook1,KernerBook2,KernerBook3,KernerBook4}.

We should emphasized that speed adaptation within the indifferent zone for car-following discuss above is only one
of the examples of speed adaptation. In general,
speed adaptation is a vehicle behavior that tries to equalize
the vehicle speed to the speed of the preceding vehicle. In particular, speed adaptation occurs due to the
  braking of the vehicle
that approaches the bottleneck at which the average speed is lower than the average speed is in free flow
upstream of the bottleneck. Due to speed adaptation a speed wave of vehicle deceleration upstream  of the bottleneck
can occur.

\subsection{Definition of Vehicle Overacceleration  
\label{Definition_OA_sec}}
 
 For human-driving and automated-driving vehicles we apply the following definition of
overacceleration:
\begin{itemize}
\item[--] Vehicle acceleration behavior that causes the free flow
metastability with respect to the F$\rightarrow$S transition at the
bottleneck is called vehicle overacceleration.
\end{itemize}
The definition of overacceleration, which can be considered a
\textit{sufficient} condition for overacceleration, is made through the
macroscopic effect of overacceleration on vehicular traffic:
When free flow is in a metastable state with respect to the
F$\rightarrow$S transition at the bottleneck, the overacceleration does
occur in vehicular traffic.

Through the analysis of whether an F$\rightarrow$S transition at the
bottleneck exhibits nucleation nature, the macroscopic definition
of overacceleration is applicable for a clear distinguishing
whether overacceleration occurs or not in both real field traffic
data and mathematical modeling: If the F$\rightarrow$S transition at
the bottleneck exhibits the nucleation nature, then overacceleration
does occur; otherwise, no overacceleration occurs. 

To explain this macroscopic overacceleration definition, firstly we should  emphasize that overacceleration is in a spatiotemporal
competition with the opposite effect of   speed adaptation~\cite{KernerBook1,KernerBook2,KernerBook3}. 
\begin{itemize}
\item[--] Overacceleration  causes a tendency toward free
flow at the bottleneck.
\item[--] Contrary
to overacceleration, speed adaptation causes a tendency toward synchronized
flow. 
\end{itemize} 
This means that free flow at the bottleneck is maintained only if the acceleration occurring within the local speed decrease 
at the bottleneck is, on average, stronger than the opposing effect of speed adaptation. Otherwise, 
speed adaptation   causes the wave of reduced speed to propagate upstream from the bottleneck---ultimately leading to a transition from free flow to synchronized traffic (F$\rightarrow$S transition) at the bottleneck.

The definition of discontinuous overacceleration presented above reflects a core assumption of the three-phase traffic theory: that the nucleation of traffic breakdown at bottlenecks in real-world human-driven traffic is governed by a spatiotemporal competition between discontinuous overacceleration and the opposing effect of speed adaptation.

	To justify this assumption in this review, we apply mathematical modeling to single-leader deterministic car-following traffic flow models
	(Secs.~\ref{Safety_OA_sec} and~\ref{TPACC_St_S}). These models do not account for information from vehicles further ahead of the preceding vehicle or situation-dependent safety margins. Furthermore, they do not incorporate driver/vehicle psychological speed thresholds, non-linear bounded rationality, and  a stronger tendency to accelerate than to decelerate among drivers or automated systems in response to upstream speed disturbances. The models also do not consider a preference for free-flow conditions over safe spacing, nor do they include extensions like heterogeneous driving behavior, randomness, or multi-phase velocity-spacing relationships.
	
	In Secs.~\ref{Safety_OA_sec} and~\ref{TPACC_St_S} as well as in the following discussion
	of the results of the mathematical modeling of the single-leader deterministic car-following traffic flow models
		made in Secs.~\ref{Dis_OA_Feature_S} and~\ref{Dis_OA_Sp-Tem_S}, we demonstrate that when the overdeceleration mechanism is absent in these  models under free-flow conditions, it becomes evident that the interplay between discontinuous overacceleration and speed adaptation drives the nucleation-like nature of traffic breakdown at a bottleneck. We show that the overacceleration mechanism is an inherent model property resulting directly from single-leader deterministic car-following traffic flow models for both human-driven and automated vehicles.  
 
In the   microscopic mathematical
modeling  made in the review (Secs.~\ref{Safety_OA_sec}--\ref{Lane_OA_sec}),
after we have recognized the occurrence
of overacceleration through the nucleation nature of
traffic breakdown,  
 we study overacceleration mechanism(s) on microlevel.
As a way for a quantitative mathematical analysis of distinguishing
whether overacceleration occurs or not, we use  
simulations of a spatiotemporal dynamic behavior of an addition
time-limited local speed decrease in free flow: The F$\rightarrow$S transition
exhibits the nucleation nature, i.e., overacceleration occurs, when a
small enough local speed decrease decays over time, whereas a large
enough local speed decrease whose amplitude exceeds some critical
amplitude (critical nucleus) grows over time leading to the F$\rightarrow$S
transition. For the mathematical simulation of the time-limited local
speed decrease in free flow, we use usually either a short-time
breaking of one of the vehicles or a short-time addition increase in
the on-ramp inflow rate at the bottleneck~\footnote{A study of other possible
mathematical ways for the analysis whether the F$\rightarrow$S transition at
the bottleneck exhibits the nucleation nature or not and, therefore,
overacceleration occurs or not (like the use of a noise term in the
model) that is out of the scope of this review paper can be an interesting task
for further traffic research.}.

 \subsection{Physics of Vehicle Overacceleration  
\label{Cause_OA_sec}}

 To explain the physics of discontinuous overacceleration, we consider firstly
 empirical data shown in Fig.~\ref{20041998_MSP}. We can made the following conclusions: 

(i) Before the MSP reaches upstream bottleneck B, free flow has been at bottleneck B ($x=$ 17 km in Fig.~\ref{20041998_MSP}(b)). This means that 
  vehicle speed adaptation to slower moving
 vehicles at the bottleneck does not cause   traffic breakdown. We can assume that this is only possible when  the vehicle acceleration
at bottleneck B is on average stronger
than the speed adaptation. 

(ii) Later the MSP reaches  bottleneck B. During the time-interval when 
 the MSP is at the bottleneck location ($x=$ 17 km in Fig.~\ref{20041998_MSP}(b)), synchronized flow is at bottleneck B. We see that
 the vehicle acceleration is not able
to return free flow at   bottleneck B.  
We can assume that this is only possible when  the vehicle acceleration
at bottleneck B is on average weaker
than the speed adaptation within the synchronized flow of the MSP. 

(iii) After this time-interval ($t> $ 07:04   in Fig.~\ref{20041998_MSP}(b)), 
 free flow returns downstream of upstream bottleneck B ($x=$ 17.9 km in Fig.~\ref{20041998_MSP}(b)). Nevertheless,
the vehicle acceleration cannot also
 return free flow at   bottleneck B: A permanent synchronized flow   remains, i.e.,
the MSP induces an F$\rightarrow$S transition ($x=$ 17 km in Fig.~\ref{20041998_MSP}(b)) at   bottleneck B. As in item (ii),
we can assume that the vehicle acceleration
at bottleneck B becomes on average weaker
than the speed adaptation when   synchronized flow is at bottleneck B. 

\begin{figure}
\begin{center}
\includegraphics[width = 8 cm]{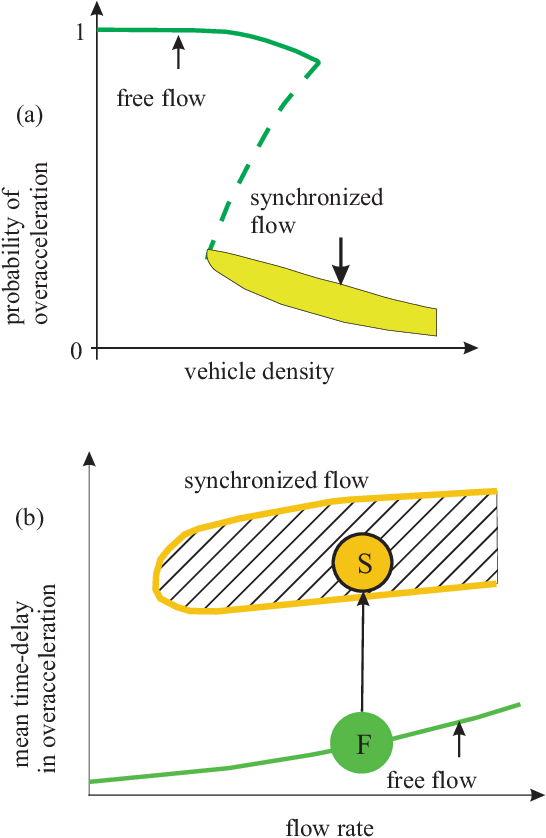}
\end{center}
\caption[]{Hypothesis of three-phase traffic theory about the 
discontinuous character of driver overacceleration~\cite{Kerner1999B,Kerner1999A,Kerner1999C}.
(a) Qualitative vehicle density-dependence of driver overacceleration
probability per a   time interval; dashed curve shows     the probability of the occurrence of the critical
 ``nucleus'' for the F$\rightarrow$S transition   at the bottleneck. (b) Equivalent presentation of (a)
as  a
 discontinuous flow-rate dependence of the mean time-delay in
 overacceleration. In (b), the up-arrow shows qualitatively
 the jump in   the mean time-delay in
 overacceleration that occurs when the F$\rightarrow$S transition is realized;  F and S are states of free flow and
synchronized flow, respectively. 
}
\label{Hyp-OA}
\end{figure}

Thus, we can assume that during the F$\rightarrow$S transition the probability (per time-unit) of
vehicle acceleration to free flow exhibits a discontinuity (Fig.~\ref{Hyp-OA}(a)): According to the definition of overacceleration
(Sec.~\ref{Definition_OA_sec}), it is therefore assumed that in free   flow the probability of overacceleration
 from a lower speed to a higher speed is significantly greater than the probability of overacceleration in synchronized  flow.

 Empirical spatiotemporal traffic data collected at various bottlenecks suggest that
the  local speed decrease   in free flow at the bottleneck, within which the minimum speed is low enough to initiate traffic breakdown,
 can be regarded as a 
  ``nucleus'' for synchronized flow   at the bottleneck. Indeed,
when the minimum speed within the nucleus is higher than some critical speed, 
the probability of vehicle acceleration within
the   local speed decrease is high enough. As a result, no traffic breakdown  
occurs at the bottleneck. In the opposite case, when 
the minimum speed within the nucleus  is less than the critical speed, 
the probability of vehicle acceleration within
the   local speed decrease is  low. In this case, the growth of the   nucleus
 results in  traffic breakdown (F$\rightarrow$S transition)
  at the bottleneck.

We will show below (Secs.~\ref{Safety_OA_sec}--\ref{Adv_TPACC_model_sec}) that the discontinuous character of overacceleration (Fig.~\ref{Hyp-OA}) is a fundamental
microscopic vehicle characteristic
 for both human-driving and automated-driving vehicles
moving in traffic flow.

We note that on the microlevel of vehicle motion there is a finite
time-delay in vehicle overacceleration.  In general, the mean
time-delay in vehicle overacceleration can considerably depend
on traffic situation.
In synchronized  flow, the mean time-delay in overacceleration should be significantly longer than in free   flow. At given initial traffic parameters
(e.g., at given flow rates at traffic network boundaries), 
during the F$\rightarrow$S phase transition the probability of overacceleration drops  abruptly whereas the mean time-delay in
 overacceleration jumps abruptly.  This is related to the discontinuous overacceleration assumed in the three-phase traffic theory  (Fig.~\ref{Hyp-OA}(b)).

Although the discontinuities in the overacceleration probability and the mean overacceleration time-delay (Fig.~\ref{Hyp-OA}) 
help explain the underlying physics of this discontinuity, they are difficult to measure in real-world traffic. Consequently, instead of calculations
of the nucleation characteristics shown in Fig.~\ref{Hyp-OA}, we model 
the impact of discontinuous overacceleration on traffic flow by simulating speed--flow characteristics (Secs.~\ref{Safety_OA_sec}--\ref{Adv_TPACC_model_sec}), which are highly useful for engineering purposes. Note that the initial calculations of discontinuous speed--flow characteristics were performed using the Kerner-Klenov stochastic model (see Fig.~3(a) of~\cite{KKl2003A}).

For a qualitative consideration of discontinuous  overacceleration as the cause for range of stochastic highway capacities
discussed in next Sec.~\ref{Z_highway_C_Sec},
 common results of the mathematical analysis of discontinuous speed--flow characteristics
are presented in Fig.~\ref{Breakdown_C}.

 \subsection{Discontinuous  Overacceleration as the Cause for Range of Stochastic Highway Capacities \label{Z_highway_C_Sec}}

\begin{figure} 
\begin{center}
\includegraphics[width = 8 cm]{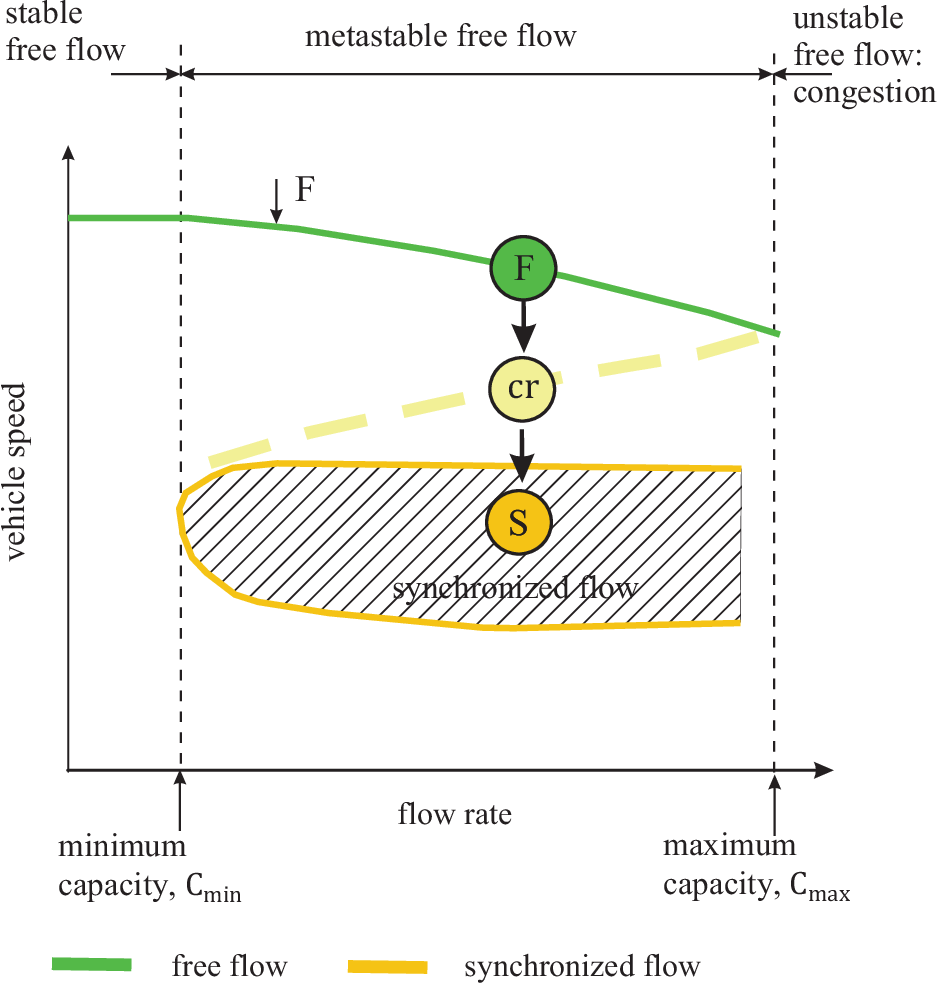}
\caption{Qualitative explanation of the range of highway capacities  (\ref{range_C}) and
the nucleation nature
of traffic breakdown at the bottleneck~\cite{KernerBook1,KernerBook2,KernerBook3,KernerBook4}.
Dashed  curve is related to critical nucleus for  the F$\rightarrow$S transition.
Down-arrow from a state of free flow $\lq\lq$F"
	to a state $\lq\lq$cr" on   curve  for critical nucleus
	  shows symbolically
	the emergence of a nucleus for an F$\rightarrow$S transition, whereas  down-arrow from  the state
	$\lq\lq$cr" to
	 a state $\lq\lq$S" of synchronized flow shows symbolically
  the development of the F$\rightarrow$S transition (traffic breakdown) at the bottleneck.
}
\label{Breakdown_C}
\end{center}
\end{figure}

The empirical nucleation nature of the F$\rightarrow$S transition at the bottleneck
means that there is a range of the flow rates in free flow within which free flow is in a metastable state
 with respect to the F$\rightarrow$S   transition. Accordingly,
 the three-phase traffic theory~\cite{KernerBook1,KernerBook2,KernerBook3,KernerBook4}  postulates that, at any given time $t$,
the capacity $C(t)$ of free traffic flow at the bottleneck lies within a specific capacity range  (Fig.~\ref{Breakdown_C}):
\begin{equation}
C_{\rm min}(t)\leq C(t) \leq C_{\rm max}(t),
\label{range_C}
\end{equation}
where $C_{\rm min}$ and $C_{\rm max}$ are, respectively, the minimum and maximum highway capacities.
Both $C_{\rm min}$ and $C_{\rm max}$  are stochastic values. At each given time instant $t$, we assume that
in (\ref{range_C})
  $C_{\rm max}(t)>C_{\rm min}(t)$.

The hypothesis about the range of highway capacities in free flow at the bottleneck (\ref{range_C})
 postulated in the three-phase traffic theory
is based
on the empirical nucleation nature of the F$\rightarrow$S transition (traffic breakdown) at the
bottleneck.
In turn, the range of highway capacities appears through the effect of the discontinuous character of overacceleration.
Indeed, speed adaptation describes the tendency to synchronized flow. Contrarily, overacceleration
describes the tendency to free flow. 

First we assume that free flow is at the bottleneck (curve F in Fig.~\ref{Breakdown_C}).
When the flow rate in free flow increases, within a local speed decrease    at the bottleneck the average speed  
due to speed adaptation effect decreases. However, as long as
 overacceleration at the bottleneck is on average stronger
than speed adaptation, free flow is maintained at the bottleneck. This explains the maximum capacity $C_{\rm max}$
as the maximum flow rate at which free flow can be still maintained at the bottleneck:
At a larger flow rate, the F$\rightarrow$S transition (traffic breakdown)   occurs spontaneously
at the bottleneck. When the flow rate is less than $C_{\rm max}$ but it is larger than
  $C_{\rm min}$, there is
a critical nucleus for the F$\rightarrow$S transition (dashed curve in  Fig.~\ref{Breakdown_C}). This means that within a given time interval traffic breakdown occurs with some probability. This is realized if a random local speed decrease appears at the bottleneck whose
 minimum speed   is less
than the speed within the critical nucleus.

Now we assume that    synchronized flow   is at the bottleneck ($\lq\lq$synchronized flow" in Fig.~\ref{Breakdown_C}).
When the flow rate in free flow downstream of the synchronized flow pattern (SP) at the bottleneck decreases,  then we can assume that  within synchronized flow the density decreases, whereas the speed 
increases. As a result, overacceleration effect within synchronized flow becomes more probable.
However, as long as
 overacceleration at the bottleneck is on average weaker
than speed adaptation, synchronized flow is maintained at the bottleneck. This explains the minimum capacity $C_{\rm min}$
as the minimum flow rate at which
synchronized flow can be still maintained at the bottleneck: At a smaller flow rate,
 overacceleration at the bottleneck becomes on average stronger
than speed adaptation in synchronized flow.  Consequently,
a return S$\rightarrow$F transition    occurs, the SP dissolves, and  free flow  is re-established  at the bottleneck.
\begin{itemize}
\item [--] The maximum and minimum 
highway capacities $C_{\rm min}$ and $C_{\rm max}$ of free flow at the bottleneck are determined    
by a competition between discontinuous vehicle 
  overacceleration  and speed adaptation.
\end{itemize}

\subsection{Justification for the Term ``Overacceleration'' \label{Term_OA_S}}

Given that discontinuous overacceleration is an emergent spatiotemporal phenomenon resulting
 from the competition between vehicle overacceleration and speed adaptation---rather than an isolated event of independent vehicle acceleration---a fundamental question arises: Why was the term ``overacceleration'' chosen to describe this complex traffic phenomenon, considering it does not explicitly mention this underlying spatiotemporal competition?

The selection of the term ``overacceleration'' is justified based on the following three core theoretical arguments:
\begin{enumerate}
\item[1.] In the spatiotemporal competition between vehicle overacceleration and speed adaptation, the \textit{discontinuity} in overacceleration fundamentally dictates the nucleation characteristics of the F$\rightarrow$S transition. Consequently, the acceleration component must be explicitly emphasized to properly characterize the driving mechanism of this spatiotemporal competition.
\item[2.] Discontinuous overacceleration is conceptually complementary to overdeceleration, representing its exact \textit{opposite effect} regarding localized speed variations. Specifically, discontinuous overacceleration can 
generate an amplifying wave of speed \textit{increase} within synchronized flow (an S$\rightarrow$F instability, see Fig.~\ref{SF-instability_2} below). 
Conversely, overdeceleration generates an amplifying wave of localized speed \textit{decrease} within synchronized flow
(the S$\rightarrow$J instability). Thus,
 the prefix in ``over\textbf{ac}celeration'' intentionally mirrors and contrasts with ``over\textbf{de}celeration'' to highlight their opposing dynamical impacts on synchronized flow.
\item[3.] Among the numerous variations of vehicle acceleration behaviors, only those exhibiting a distinct discontinuity during the transition from free flow to synchronized flow are classified as overacceleration. Therefore, the term ``overacceleration'' serves as a precise taxonomical descriptor to distinguish acceleration behavior involving such discontinuities from standard, continuous acceleration processes.
\end{enumerate}

	Taken literally, the term  {\it overacceleration} suggests excessive acceleration, which often causes ambiguity.  
	Inconsistent interpretations of the concept ``overacceleration'' will be analyzed in Sec.~\ref{Term_OA_OA_S}.

\subsection{Explanation of Behavioral Origin of Overacceleration \label{Bev_Origion_S}}

As mentioned in Sec.~\ref{OD-OA_Beh_Or_S}, the behavioral origin of vehicle overacceleration
is related to the vehicle's desire to move in free
flow.
To understand the  behavioral origin of the discontinuous character
of vehicle overacceleration, we   consider  drivers that reach a local speed decrease
at a bottleneck. 
  First, we assume that the minimum speed within this local speed decrease
 is only slightly lower than the free-flow speed upstream and downstream of the bottleneck. Consequently, the drivers approaching this minor speed decrease
 are motivated to accelerate strongly within the bottleneck to escape it as quickly as possible. Under these conditions, the mean acceleration time-delay can be sufficiently short (Fig.~\ref{Hyp-OA}(b)), and the probability of vehicle overacceleration within the local speed decrease is great (Fig.~\ref{Hyp-OA}(a)). Accordingly, no upstream propagation of congested traffic occurs at the bottleneck; instead, the local speed decrease remains localized. The same dynamic behavior can be assumed for automated vehicles approaching the local speed decrease at the bottleneck. 
Discussions on mathematical modeling confirming this qualitative assumption can be found in Sec.~\ref{Dis_OA_Feature_S}.

Conversely, when the minimum speed within the local speed decrease
 at the bottleneck is low enough, drivers must execute acceleration cautiously to avoid collisions with the preceding vehicle. This cautious acceleration behavior increases the mean time-delay in acceleration (Fig.~\ref{Hyp-OA}(b)). In this case, the probability of vehicle overacceleration within the local speed decrease is sufficiently low (Fig.~\ref{Hyp-OA}(a)). For this reason, the upstream propagation of synchronized flow (congested traffic) occurs at the bottleneck, leading to a traffic breakdown. The same acceleration behavior within the local speed decrease at the bottleneck can be assumed for automated vehicles.

\section{Overacceleration through Safety Acceleration in Helly's Model of Adaptive
Cruise Control (ACC)  in Automated Vehicles
\label{Safety_OA_sec}}  

  \subsection{Helly's Model for ACC}
	
\begin{figure}
\begin{center}
\includegraphics[width = 8 cm]{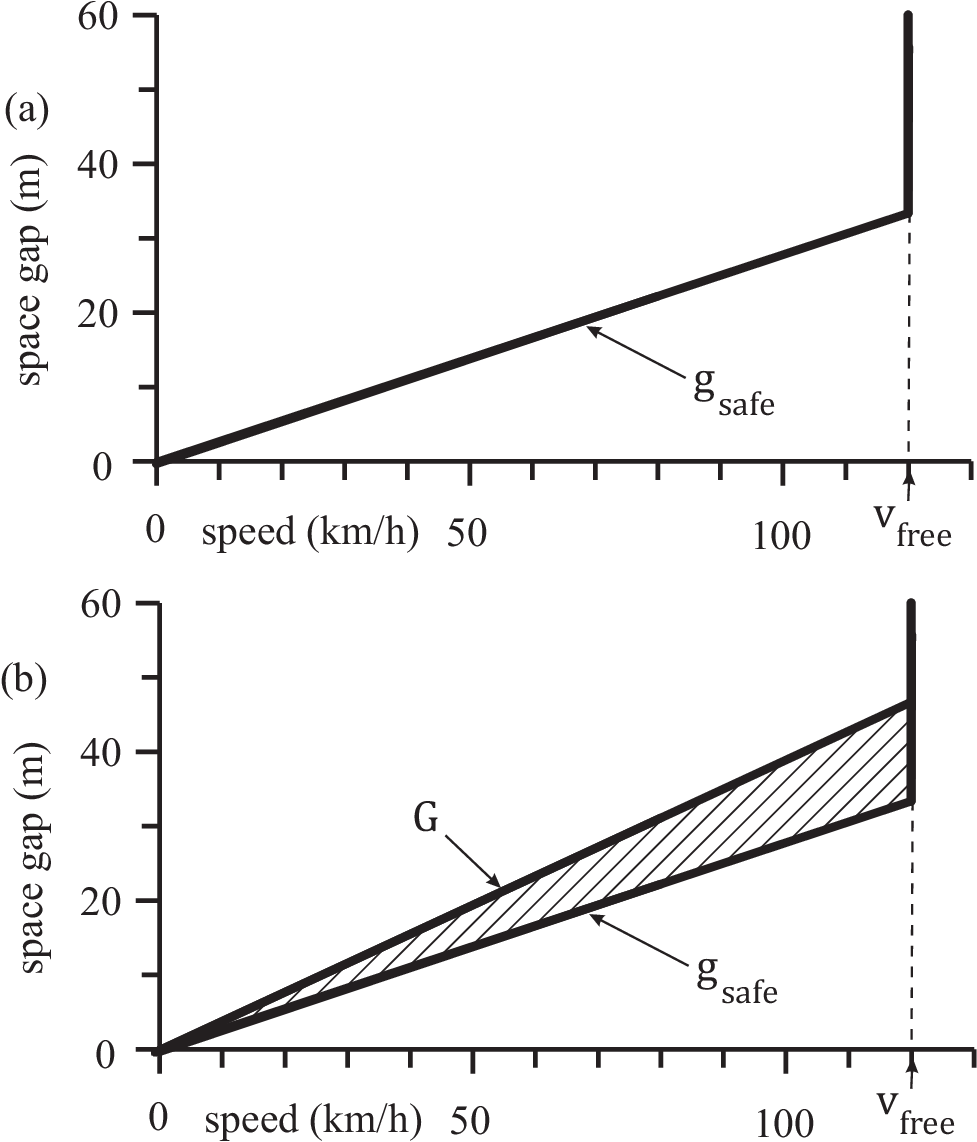}
\end{center}
\caption[]{Steady states of ACC-model (\ref{ACC_Cl})  (a)
and TPACC-model (\ref{g_v_g_min1})--(\ref{G_g_safe_simple}) (b) in  space-gap--speed plane. $g_{\rm safe}=v\tau_{\rm d}$ for (a) and
$g_{\rm safe}=v\tau_{\rm safe}$ for (b); maximum free flow speed  $v_{\rm free}=$ 120 km/h,    vehicle length
 $d=$ 7.5 m,   $\tau_{\rm d}=$ 1 s, $\tau_{\rm safe}=$ 1 s, $\tau_{\rm G}=$ 1.4 s.
}
\label{ACC_TPACC_steady}   
\end{figure}

Here, we show  that vehicle
overacceleration can occur without explicitly introducing it in
a microscopic traffic flow model. 
Moreover, we illustrate the occurrence of  this
overacceleration effect   in a traffic flow model for which  
  there is no indifferent zone of car-following in the model (Fig.~\ref{ACC_TPACC_steady}(a)). 
	
	We show that this occurs in classical Helly's model~\cite{Helly} applied for ACC-vehicles under the choice of model parameters at which
	there is no string instability and, therefore, no overdeceleration effect in any platoons of the ACC-vehicles.
 The classical ACC-model described by Helly's formula reads as follows
(see, e.g.,~\cite{IoannouChien2002A,Levine1966A_Aut,Liang1999A_Aut,Liang2000A_Aut,Swaroop1996A_Aut,Swaroop2001A_Aut,Rajamani2012A_Aut,Davis2004B9,Davis2014C}):
 \begin{equation}
a(g, v, v_{\ell}) = K_{\rm ACC,1}(g-v\tau_{\rm d})+K_{\rm ACC,2}\Delta v,  
 \label{ACC_Cl} 
\end{equation} 
where  $\tau_{\rm d}$ is a desired time headway
of an ACC-vehicle to the 
 preceding 
 vehicle; vehicle acceleration and speed are limited 
by maximum acceleration $a=a_{\rm max}$ and $v=v_{\rm free}$, 
respectively; $K_{\rm ACC,1}$ and $K_{\rm ACC,2}$  are positive dynamic   coefficients of  
   ACC-vehicle adaptation that are chosen   to  satisfy
  condition for string stability found by Liang and Peng~\cite{Liang1999A_Aut} 
	\begin{equation}
K_{\rm ACC,2}>(2-K_{\rm ACC,1}(\tau_{\rm d})^{2})/2\tau_{\rm d}.
 \label{ACC_stability}
 \end{equation}
At chosen $\tau_{\rm d}$, $K_{\rm ACC,1}$, and $K_{\rm ACC,2}$ (see the choice of values of
 $K_{\rm ACC,1}$, and $K_{\rm ACC,2}$
in Appendix~\ref{ACC_safety_S}),    no string instability occurs for any speed in simulations of traffic flow consisting of 
the ACC-vehicles (Fig.~\ref{ACC_safety_acc})~\footnote{It should be emphasized that in traffic of
 human-driving vehicles, each driver exhibits a finite reaction time. For this reason,
overdeceleration can occur leading to traffic flow instability. Contrary to human-driving vehicles,
 a reaction time  for automated vehicles can be considered negligible.
  Consequently,   for a chosen desired time headway
	$\tau_{\rm d}$  in the ACC vehicle model (\ref{ACC_Cl}),
	we can select the dynamic coefficients $K_{\rm ACC,1}$ and $K_{\rm ACC,2}$ in accordance with  (\ref{ACC_stability}) such that
no overdeceleration of the vehicles  and thus no string instability occurs. }.

\subsection{Traffic Breakdown at Bottleneck in Helly's Model for ACC}

 Simulations of traffic flow consisting of 100$\%$ ACC-vehicles  (\ref{ACC_Cl})
 moving on single-lane road with an on-ramp bottleneck   (road and on-ramp models are in Appendix~\ref{Road_S}) show (Fig.~\ref{ACC_safety_acc}(a))
 that due to the merging of on-ramp vehicles, there is a local speed decrease
in free flow at the bottleneck. This free flow is in a metastable state with respect to
an F$\rightarrow$S transition. Indeed,  at time instant $t=T_{\rm ind}$ the on-ramp inflow rate $q_{\rm on}$  has been increased
on value $\Delta q_{\rm on}$   during a short time interval  $\Delta t$.
Through the application of this on-ramp inflow impulse $\Delta q_{\rm on}$, traffic breakdown 
(F$\rightarrow$S transition)    has been induced  at the bottleneck. 
 Synchronized flow occurring at the bottleneck remains after the flow rate
$q_{\rm on}$ decreases to its initial value. Due to the
F$\rightarrow$S transition, a widening synchronized flow pattern (WSP) is realized at the bottleneck
(Fig.~\ref{ACC_safety_acc}(a)). In other words, we have shown that 
free flow  of 100$\%$ ACC-vehicles is indeed  in a metastable state with respect to
an F$\rightarrow$S transition.
In ACC-model   (\ref{ACC_Cl}) used in these simulations there is no vehicle overdeceleration and, respectively, no string instability.

\begin{figure}  
\begin{center}
\includegraphics[width = 8 cm]{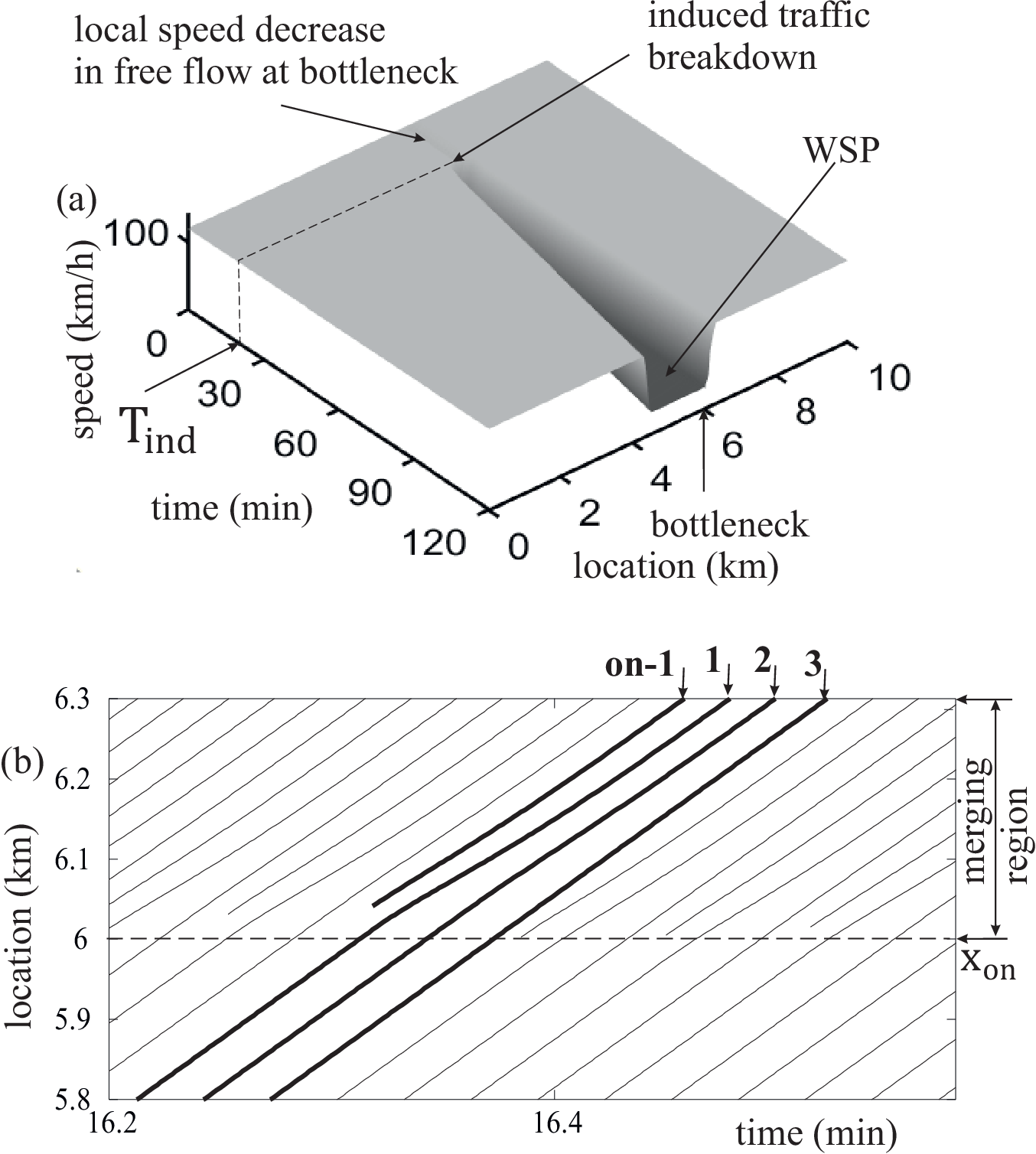}
\end{center}
\caption[]{Overacceleration through safely acceleration in   ACC model (\ref{ACC_Cl}) 
on single-lane road with on-ramp bottleneck at  the
  flow rate in free flow upstream of the bottleneck $q_{\rm in}=$ 2000 vehicles/h
	and on-ramp inflow rate
	$q_{\rm on}=$ 920  vehicles/h. Parameters of model  (\ref{ACC_Cl}): $\tau_{\rm d}=$ 1 s,
	$K_{\rm ACC,1}=  0.3 \ s^{-2}$ (choice of   $K_{\rm ACC,2}$ is explained in Appendix~\ref{ACC_safety_S}),
	  $v_{\rm free}=$ 120 km/h,
 $d=$ 7.5 m,   maximum vehicle acceleration  $a_{\rm max}=$ 2.5 $\rm m/s^{2}$,
road length $L=$ 10 km. Parameters of on-ramp bottleneck (model of the bottleneck is given in Appendix~\ref{Road_S}): location of   on-ramp
$x_{\rm on}=$ 6 km, $\lambda_{\rm b}=$ 0.2 s,  $L_{\rm m}=$ 0.3 km. 
(a)  Speed in space and time: Local speed decrease at bottleneck in free flow   and 
WSP
induced in  free flow   at $T_{\rm ind}=$ 20 min through application of  addition on-ramp inflow impulse $\Delta q_{\rm on}=$
280 vehicles/h of duration   $\Delta t=$ 2 min. 
(b) Some of simulated vehicle trajectories
at the bottleneck at time $t < T_{\rm ind}$.  
WSP is a widening SP, SP is a synchronized flow pattern. Adapted from~\cite{KKl2025A}. 
}
\label{ACC_safety_acc}
\end{figure}

\begin{figure}   
\begin{center}
\includegraphics[width = 8 cm]{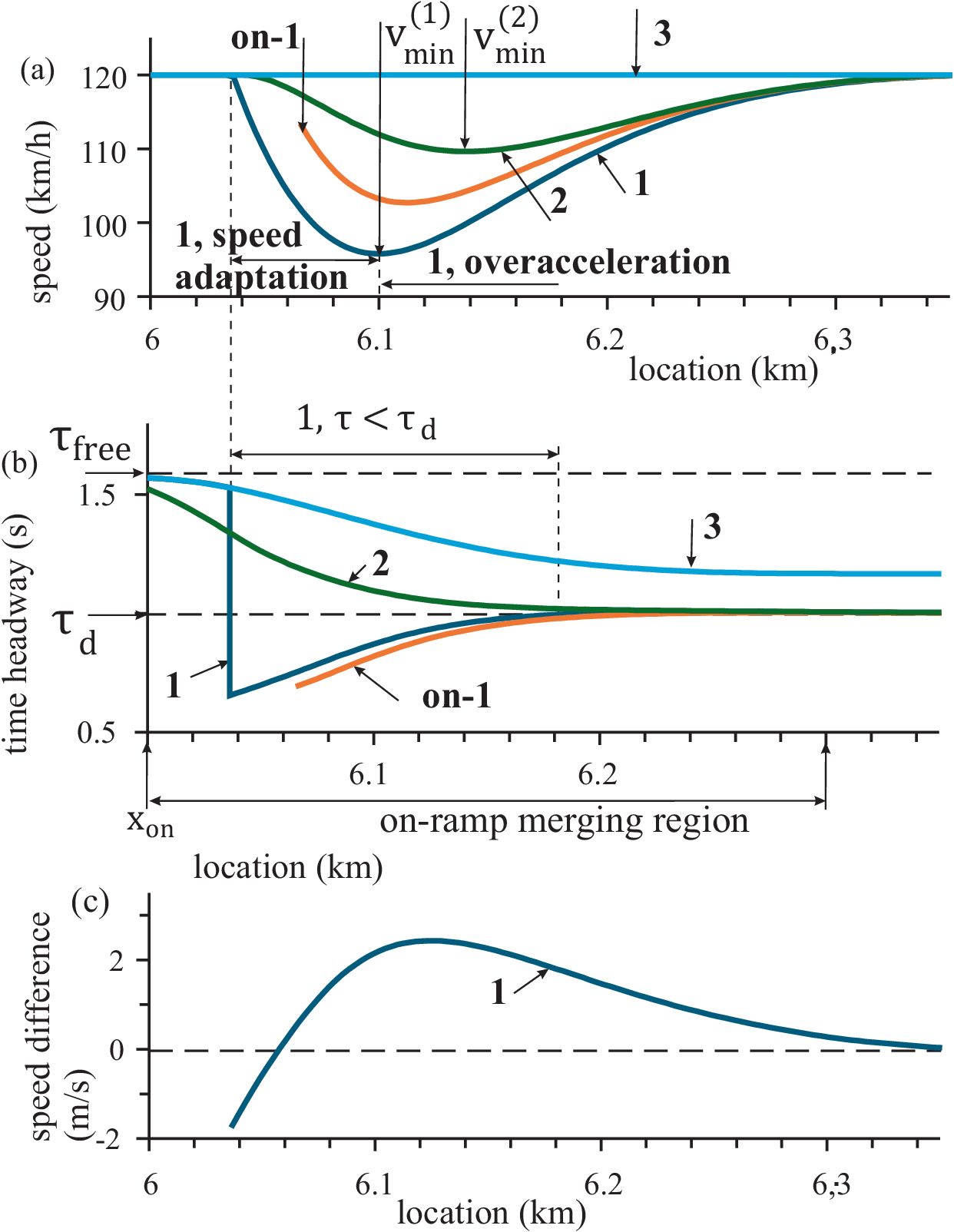}
\end{center}
\caption[]{Continuation of  Fig.~\ref{ACC_safety_acc}(b).
(a), (b) 
Location-functions of speed (a) and time-headway (b)  of some vehicles that numbers are, respectively,
 the same as those in
 Fig.~\ref{ACC_safety_acc}(b).  (c)
Location-function of speed difference $\Delta v= v^{\rm (on-1)}-v^{(1)}$  between speeds of vehicles on-1 and 1.   Adapted from~\cite{KKl2025A}. 
}
\label{ACC_safety_acc_traj}  
\end{figure}

In accordance with the definition of vehicle overacceleration (Sec.~\ref{Cause_OA_sec}), we can expect that
there should be some mechanism of overacceleration causing the nucleation character of 
traffic breakdown 
(F$\rightarrow$S transition).
To study the mechanism of overacceleration, we consider  microscopic speeds along   vehicle trajectories 1--3 in free flow (at $t<T_{\rm ind}$)
(Figs.~\ref{ACC_safety_acc}(b) and~\ref{ACC_safety_acc_traj}).

Behaviors of vehicles 1--3 has sharply changed after on-ramp vehicle on-1 has merged onto the main road (Fig.~\ref{ACC_safety_acc}(b)).
There are  two reasons for this behavior change: (i) The speed of vehicle on-1 just after the merging is considerably lower than the  maximum free flow speed
 $v_{\rm free}$
of vehicles 1--3 far enough upstream of the bottleneck (Fig.~\ref{ACC_safety_acc_traj}(a)); (ii)
  time headway of vehicle 1 to vehicle on-1 has sharply dropped below the desired time headway
$\tau_{\rm d}=$ 1 s (Fig.~\ref{ACC_safety_acc_traj}(b)). Therefore, in accordance with ACC-model  (\ref{ACC_Cl}),
vehicle 1 should decelerate while adapting its speed to the speed of   vehicle on-1 as well as to increase
time headway between vehicles on-1 and 1
($\lq\lq$1, speed adaptation" in Fig.~\ref{ACC_safety_acc_traj}(a)). 

However, shortly later   the speed difference
$\Delta v= v^{\rm (on-1)}-v^{(1)}$  between vehicles on-1 and 1 becomes a positive increasing value (Fig.~\ref{ACC_safety_acc_traj}(c)).
Therefore, vehicle 1 begins also to accelerate (Fig.~\ref{ACC_safety_acc_traj}(a)).
We call this vehicle acceleration as {\it safety acceleration} because  its value should satisfy
some safety conditions at which no collision between  between vehicles on-1 and 1 is possible.

In free-flow conditions at the bottleneck ($t<T_{\rm ind}$ in Fig.~\ref{ACC_safety_acc}(a)),
the safety acceleration of vehicle 1 is on average stronger than speed adaptation at the bottleneck. This is because 
the decrease of the speed of following vehicle 2 due to speed adaptation does not cause  
 upstream propagation of a  speed decrease  
of the following vehicles: Already speed of vehicle 3 following vehicle 2
 remains to be equal to  $v_{\rm free}$ (Fig.~\ref{ACC_safety_acc_traj}(a)).

The safety acceleration of vehicle 1 satisfies the definition of vehicle overacceleration, i.e.,
the safety acceleration can be considered overacceleration
($\lq\lq$1,  overacceleration" in Fig.~\ref{ACC_safety_acc_traj}(a)). Indeed, in accordance with overacceleration definition,
the safety acceleration of vehicle 1 causes the free flow metastability with respect to the 
F$\rightarrow$S transition at the bottleneck (Fig.~\ref{ACC_safety_acc}(a)):
 Contrary to the free flow condition ($t<T_{\rm ind}$ in Fig.~\ref{ACC_safety_acc}(a)),
after the F$\rightarrow$S transition has been induced at the bottleneck
 ($t\geq T_{\rm ind}+\Delta t$ in Fig.~\ref{ACC_safety_acc}(a)),
synchronized flow is self-maintained at the bottleneck.

To explain this, we note that in free flow the mean time-headway between ACC-vehicles $\tau_{\rm free}=$ 1.575 s
(Fig.~\ref{ACC_safety_acc_traj}(b))~\footnote{The mean time-headway between ACC-vehicles
in free flow is found from formulas  $g_{\rm free}=\frac{v_{\rm free}}{q_{\rm in}}-d$
 and $\tau_{\rm free}=g_{\rm free}/v_{\rm free}$.}. Contrary to free flow, we have found that in synchronized flow at the bottleneck
the mean time-headway between ACC-vehicles is considerably shorter $\tau_{\rm syn}\approx \tau_{\rm d}=$ 1 s.
At such short time-headway, synchronized flow is self-maintained at the bottleneck, i.e., no return
S$\rightarrow$F transition can occur at the bottleneck. This means that in synchronized flow
  overacceleration of ACC-vehicles at the bottleneck is on average
 weaker than speed adaptation.

 Thus, due to the sharp drop in the mean time-headway
 between ACC-vehicles (from   1.575 s to about 1 s) occurring 
 when free flow transforms into synchronized flow,     
 a discontinuity in   ACC-vehicle  overacceleration is realized; this is
in accordance with the three-phase traffic theory (Fig.~\ref{Hyp-OA}(b)).
 This explains why 
    safety acceleration of ACC-vehicles at the bottleneck is an example of vehicle overacceleration.
		\begin{itemize}
\item [--]
		Overacceleration     
  explains the nucleation nature of traffic breakdown (F$\rightarrow$S transition) at the bottleneck
 in such a model for which there is  a fixed time-headway in 
steady states of the model, i.e., the model steady states    lie on 
   1D-region in the space-gap--speed plane (Fig.~\ref{ACC_TPACC_steady}(a)).
	\end{itemize}
	
 	 \begin{figure} 
\begin{center}
\includegraphics[width = 8 cm]{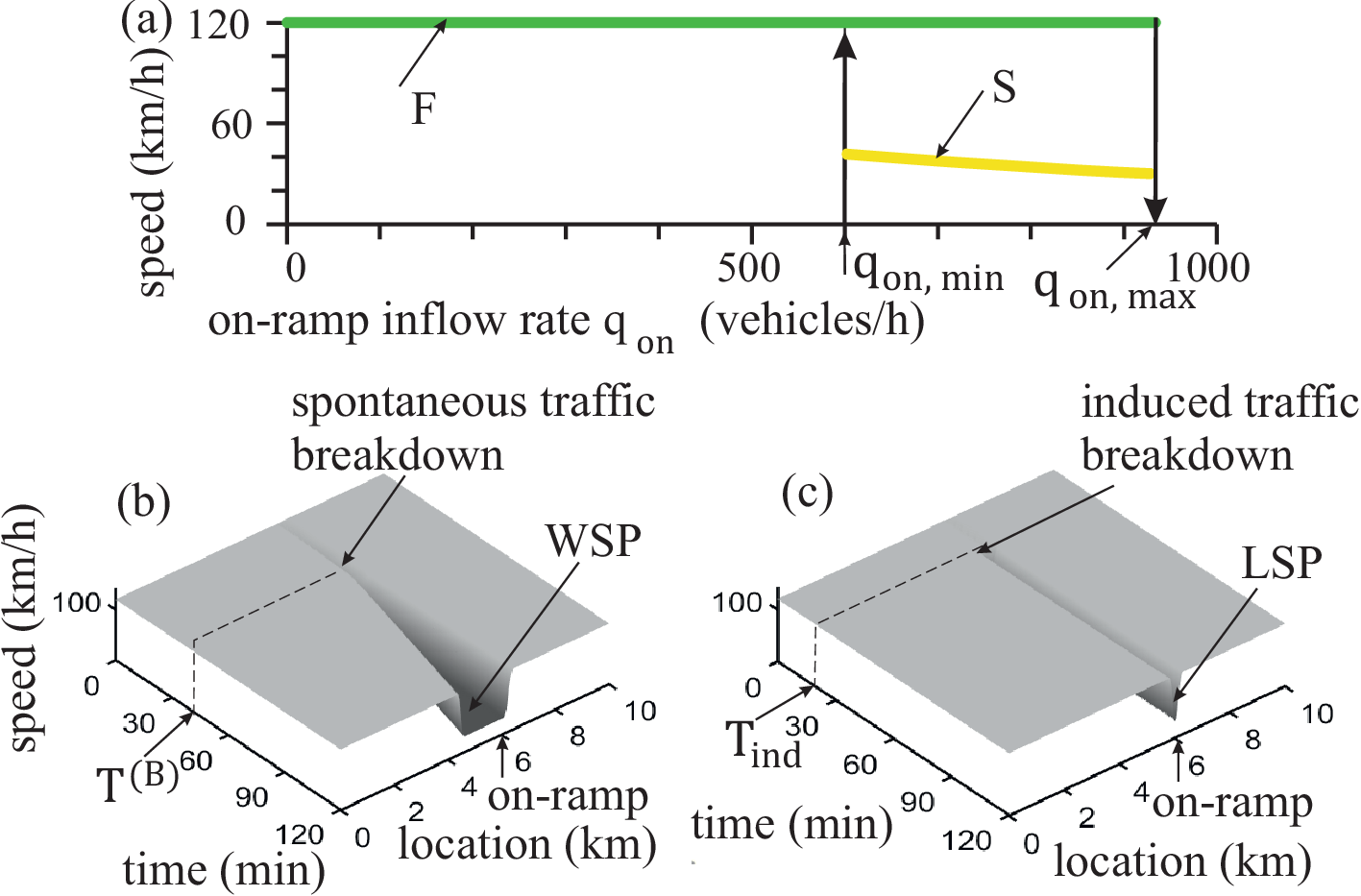}   
\end{center}
\caption[]{Characteristics of highway capacity of
 traffic consisting of   ACC-vehicles moving on single lane-road simulated
 with Helly's model (\ref{ACC_Cl}). 
  (a) Speed--on-ramp-inflow characteristic at $q_{\rm in}=$ 2000  vehicles/h.
	(b), (c) Speed in space and time related to (a)  
	 at $q_{\rm on}=$ 938 vehicles/h (b) that is close to $q_{\rm on}=q_{\rm on, \ max}=$ 936 vehicles/h and at $q_{\rm on}=q_{\rm on, \ min}=$ 781 vehicles/h (c).
	$q_{\rm on, \ max}$ is the maximum on-ramp inflow rate at which  free flow can still exist at the bottleneck;
maximum capacity is equal to $C_{\rm  max}= q_{\rm on, \ max}+q_{\rm in}$.
$q_{\rm on, \ min}$ is the minimum on-ramp inflow rate at which an SP   can still exist
at the bottleneck;
minimum capacity is equal to $C_{\rm  min}= q_{\rm on, \ min}+q_{\rm in}$.
Within the on-ramp inflow range $q_{\rm on, \ min}\leq q_{\rm on}< q_{\rm on, \ max}$, i.e., within the capacity range
$C_{\rm  min}\leq q_{\rm  sum} < C_{\rm  max}$,
 the metastability of
free flow is realized:
 either free flow or the SP can exist at the bottleneck. 
 At $q_{\rm on}<q_{\rm on, \ min}$, i.e.,  at $q_{\rm sum}<C_{\rm   min}$,  no congested patterns can be induced in free flow, whereas
at $q_{\rm on}>q_{\rm on, \ max}$ , i.e.,  at $q_{\rm sum}>C_{\rm   max}$, the SP occurs spontaneously at the bottleneck after a time delay $T^{\rm (B)}$.
In (b), spontaneous traffic breakdown (F$\rightarrow$S transition) occurs after time-delay    $T^{\rm (B)}=$ 43 min.
	In (c), traffic breakdown   has been induced  
  in   free flow   at $T_{\rm ind}=$ 20 min through application of  addition on-ramp inflow impulse $\Delta q_{\rm on}=$
619 vehicles/h of duration   $\Delta t=$ 2 min. 
Other model parameters are the same as those in Fig.~\ref{ACC_safety_acc}. Adapted from~\cite{KKl2025A}.
}
\label{2Z_ACC_short}
\end{figure}

		The above conclusions are also supported by simulations of highway capacity in traffic of ACC-vehicle traffic at  
		  the bottleneck (Fig.~\ref{2Z_ACC_short}(a)). As stated in the three-phase traffic theory (Sec.~\ref{Z_highway_C_Sec})
		there is a  specific highway capacity range (\ref{range_C}).  When the total flow rate $q_{\rm sum}=q_{\rm in}+q_{\rm on}$ exceeds 
		than maximum capacity $C_{\rm max}$,  traffic breakdown   (F$\rightarrow$S transition) occurs  spontaneously
		after a time-delay $T^{\rm (B)}$, resulting in WSP formation at the bottleneck (Fig.~\ref{2Z_ACC_short}(b)).
		 
		Within the capacity range (Fig.~\ref{2Z_ACC_short}(a)), specifically
		when the total flow rate $q_{\rm sum}=q_{\rm in}+q_{\rm on}$ is less than maximum capacity $C_{\rm max}$,
		traffic breakdown   (F$\rightarrow$S transition) can be induced and the resulting
		synchronized flow of ACC-vehicles is self-maintained
		at the bottleneck. 
		When $q_{\rm sum}$
		 decreases, the width of an WSP   at the bottleneck decreases and the WSP shown in Fig.~\ref{2Z_ACC_short}(b) transforms into  a localized synchronized flow pattern (LSP) (Fig.~\ref{2Z_ACC_short}(c)).
			The LSP dissolves due to overacceleration effect when $q_{\rm sum}$ falls below  
			the minimum capacity $C_{\rm min}$.

\section{Overacceleration through Safety Acceleration in  Three-Phase ACC (TPACC) \label{TPACC_St_S}} 

Dynamics rules of motion of an automated vehicle can be chosen  totally different in comparison with
 dynamics rules of motion of
 human-driving vehicles. However, when the dynamic motion of the automated vehicle becomes
 unpredictable for human-driving vehicles, the automated vehicle could be perceived as an obstacle for human-driving vehicles in mixed traffic.
This can cause the decrease in both traffic safety and highway capacity. 

In particular, empirical traffic data show that, contrary to classical ACC-model (\ref{ACC_Cl}), 
rather than
 control a fixed time headway, human-driving vehicles do not control time headway
within   the indifferent zone of 
car-following (\ref{indif_tau})   (Figs.~\ref{IZ_tau_g} and~\ref{IndZone}). To make the motion
of automated vehicles compatible with human-driving vehicles in mixed traffic,  
the author  has introduced a Three-Phase ACC (TPACC) whose basic feature is the existence of the indifferent zone of 
car-following (\ref{indif_tau})  in the rule of 
TPACC-vehicles~\cite{KernerPat1,Kerner2018C}: Rather than a fixed   
time-headway of standard ACC-models, acceleration/deceleration of
 TPACC-vehicle $a$    
does not depend on time-headway to the preceding vehicle
within the time-headway range  (\ref{indif_tau}).
 As shown and explained in~\cite{Kerner2018C,Kerner2021A}, there are at least the following advantages of    TPACC-vehicles in comparison with 
ACC-vehicles described by Helly's formula  (\ref{ACC_Cl}):
\begin{itemize}
\item [--] {\it No string instability}:
Contrary to   ACC-vehicles with a fixed time-headway (\ref{ACC_Cl}), there is no string instability
in platoons of TPACC-vehicles. This is due to the existence of the indifferent zone for car-following (Figs.~\ref{IZ_tau_g} and~\ref{IndZone}).
\item [--] {\it Reduction of the local speed decrease at the  bottleneck}:
The indifferent zone for car-following in TPACC-vehicles causes the reducing of
the local speed decrease at the bottleneck in mixed traffic. 
\item [--] {\it Reduction of the breakdown probability}:
The indifferent zone for car-following in TPACC-vehicles causes the reducing of
the breakdown probability in mixed traffic.
\end{itemize}
The analysis of these advantages  has been already done in~\cite{Kerner2018C,Kerner2021A}. For this reason,  
below we limit by
a consideration of features of overacceleration of TPACC-vehicles.

\subsection{TPACC model \label{TPACC_model_sec}}

In a elementary model of TPACC-vehicle moving in road lane,
 TPACC-vehicle acceleration/deceleration $a$ 
is described by
a system of equations~\cite{Kerner2018C}:  
 \begin{equation}
a =  
K_{\Delta v}\Delta v   \ \textrm{at $g_{\rm safe} \leq g \leq G$}, 
\label{g_v_g_min1}  
\end{equation}
 \begin{equation}
a =  a_{\rm G} \ \textrm{at $ g > G$},
 \label{g_v_g_min2}
\end{equation}
 \begin{equation}  
 a =  a_{\rm safety}(g, v, v_{\ell}) \ \textrm{at $ g < g_{\rm safe}$},
 \label{g_v_g_min3}
\end{equation}
where   Eq.~(\ref{g_v_g_min1}) coincides with Eq.~(\ref{SA_indif}); 
$a_{\rm G}(g, v, v_{\ell})$ is TPACC-vehicle acceleration at large space gaps; 
$a_{\rm safety}(g, v, v_{\ell})$ is a safety TPACC-vehicle 
	acceleration;
 vehicle acceleration and speed are limited 
by maximum acceleration $a=a_{\rm max}$ and $v=v_{\rm free}$, 
respectively; we assume that $G > g_{\rm safe}$ at $v>0$; for simulations,
  we use here  simple speed-functions $G(v)$ and $g_{\rm safe}(v)$:
\begin{equation}
G(v)=v\tau_{\rm G}, \quad g_{\rm safe}(v)=v\tau_{\rm safe}.
\label{G_g_safe_simple}
\end{equation} 
Contrary to the ACC-model (Fig.~\ref{ACC_TPACC_steady}(a)), steady states of TPACC-model  cover  
  a 2D-region in the space-gap--speed plane (Fig.~\ref{ACC_TPACC_steady}(b)).

Following simulations of classical ACC-vehicles in Sec.~\ref{Safety_OA_sec}, we have simulated 		 
traffic breakdown in traffic consisting of 100$\%$ of TPACC-vehicles on single-lane road with the same bottleneck
(Figs.~\ref{Greater_G_induced} and~\ref{Greater_G_traj}). In these simulations,  for functions
 $a_{\rm G}(g, v, v_{\ell})$   and   $a_{\rm safety}(g, v, v_{\ell})$ we have used Helly's model~\cite{Helly} as follows:
  \begin{equation} 
a_{\rm G}(g, v, v_{\ell})=   K_{1}(g-v\tau_{\rm p})+ K_{2}\Delta v,
  \label{a_G-f}  
	\end{equation}  
 \begin{equation}
a_{\rm safety}(g, v, v_{\ell})=   K_{3}(g-g_{\rm safe})+ K_{4}(g, v, \Delta v)\Delta v,
\label{Helly_st2}
\end{equation}
where $K_{1}$, $K_{2}$,  and $K_{3}$  are    positive dynamic coefficients; $\tau_{\rm p}$ is a constant time-headway satisfying condition
$\tau_{\rm safe}<\tau_{\rm p}<\tau_{\rm G}$;
\begin{equation}
K_{4}=\left\{\begin{array}{ll}
K^{(1)}_{4}   \ \textrm{at $\Delta v > 0$}, \\
K^{(2)}_{4} \frac{v\tau_{\rm safe}}{g} \ \textrm{at $ \Delta v \leq 0$}, \\
\end{array} \right.
\label{Helly_st3}
\end{equation}
   $K^{(1)}_{4}$,  and $K^{(2)}_{4}$  are    positive dynamic coefficients; explanations of term
	$\frac{v\tau_{\rm safe}}{g}$ in (\ref{Helly_st3}) are given in Appendix~\ref{ACC_safety_S}; explanations of  
 the choice of dynamic coefficients $K_{2}$, $K_{3}$, $K^{(1)}_{4}$,  and $K^{(2)}_{4}$ in simulations 
are given in Appendix~\ref{TPACC_safety_S}.

\subsection{Traffic Breakdown in Flow of TPACC-Vehicles}
	
	There is no TPACC-vehicle overdeceleration  and, therefore, no string instability in
platoons of TPACC-vehicles described by TPACC-model
  (\ref{g_v_g_min1})--(\ref{Helly_st3}). Nevertheless, as in traffic of ACC-vehicles (Fig.~\ref{ACC_safety_acc}),
	free flow of   TPACC-vehicles is in metastable state with respect of the F$\rightarrow$S transition at the bottleneck
 (Fig.~\ref{Greater_G_induced}(a)).   In accordance with the overacceleration definition (Sec.~\ref{Cause_OA_sec}), 
this means that there should be a mechanism of overacceleration in
TPACC-model (\ref{g_v_g_min1})--(\ref{Helly_st3}). 
 As in free flow of ACC-vehicles (Fig.~\ref{ACC_safety_acc}),
this overacceleration mechanism is caused by safety acceleration of TPACC-vehicles. 
 To see this, we consider some TPACC vehicle trajectories (Fig.~\ref{Greater_G_traj}),
 which are highlighted in bold in Fig.~\ref{Greater_G_induced}(b).

\begin{figure}
\begin{center}
\includegraphics[width = 8 cm]{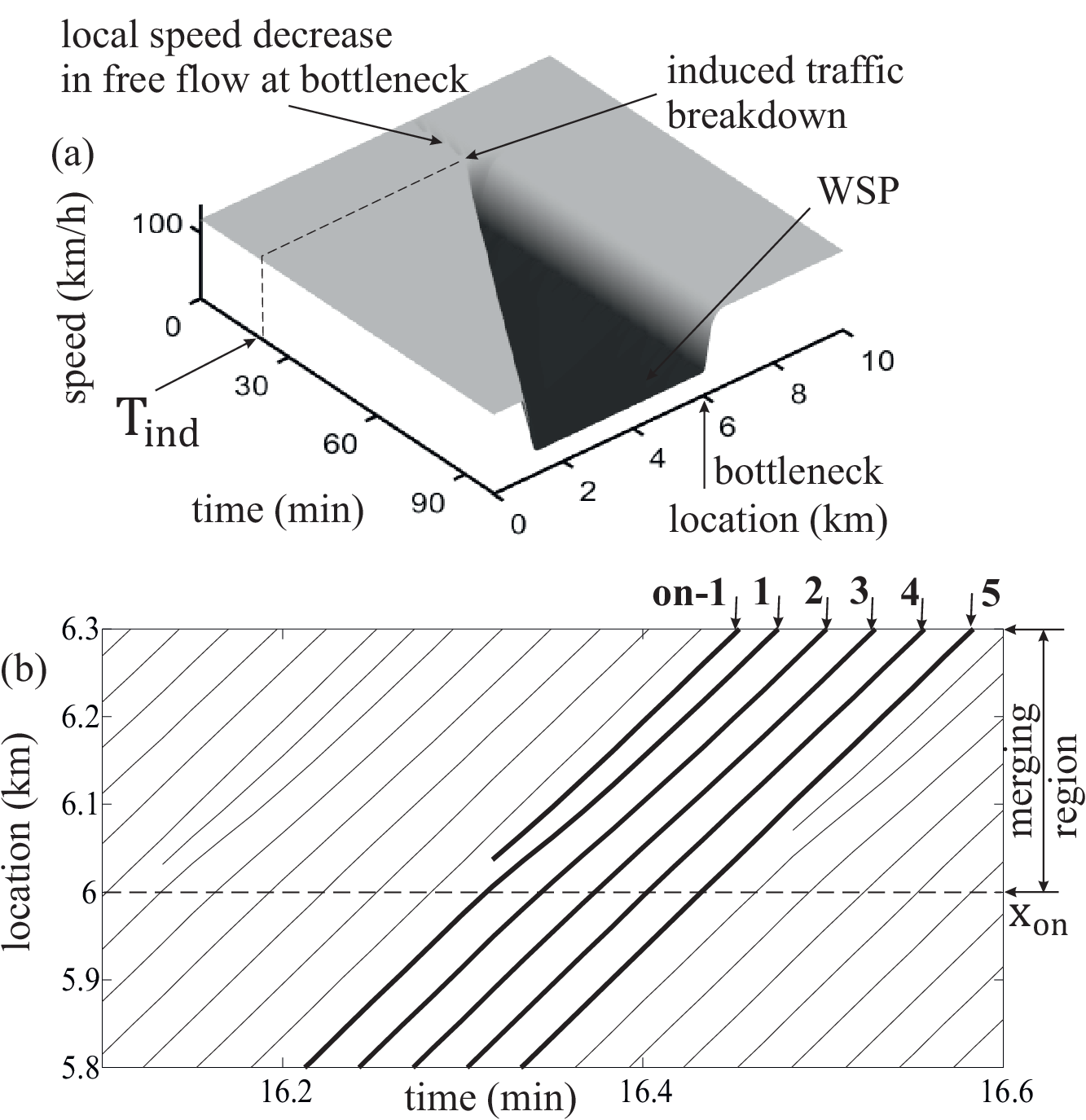}
\end{center}
\caption[]{Induced traffic breakdown in traffic of 100$\%$ of TPACC-vehicles
 on single-lane road with on-ramp bottleneck.
 Simulations of model (\ref{g_v_g_min1})--(\ref{Helly_st3})  at  $\tau_{\rm G}=$ 1.4 s,  $\tau_{\rm p}=$ 1.3 s,
$\tau_{\rm safe}=$ 1 s, $q_{\rm in}=$ 2000 vehicles/h, and $q_{\rm on}=$ 350 vehicles/h. 
Other model parameters are  $K_{1}=   0.3 \ s^{-2}$, $K_{2}= K^{(1)}_{4}= 0.6 \ s^{-1}$, 
$K_{3}= 0.5 \ s^{-2}$, 
 $K^{(2)}_{4}= 1 \ s^{-1}$.
 (a)  Speed in space and time: Local speed decrease at bottleneck in free flow   and  WSP  
induced in  free flow   at $T_{\rm ind}=$ 20 min through application of  addition on-ramp inflow impulse $\Delta q_{\rm on}=$
250 vehicles/h of duration   $\Delta t=$ 1 min.  (b) Some of simulated vehicle trajectories 
at  bottleneck at time $t < T_{\rm ind}$.  Road and on-ramp model parameters are the same as in Fig.~\ref{ACC_safety_acc}
	(see Appendix~\ref{Road_S}). Adapted from~\cite{KKl2025A}.
}
\label{Greater_G_induced}
\end{figure}

	\begin{figure}   
\begin{center}
\includegraphics[width = 8 cm]{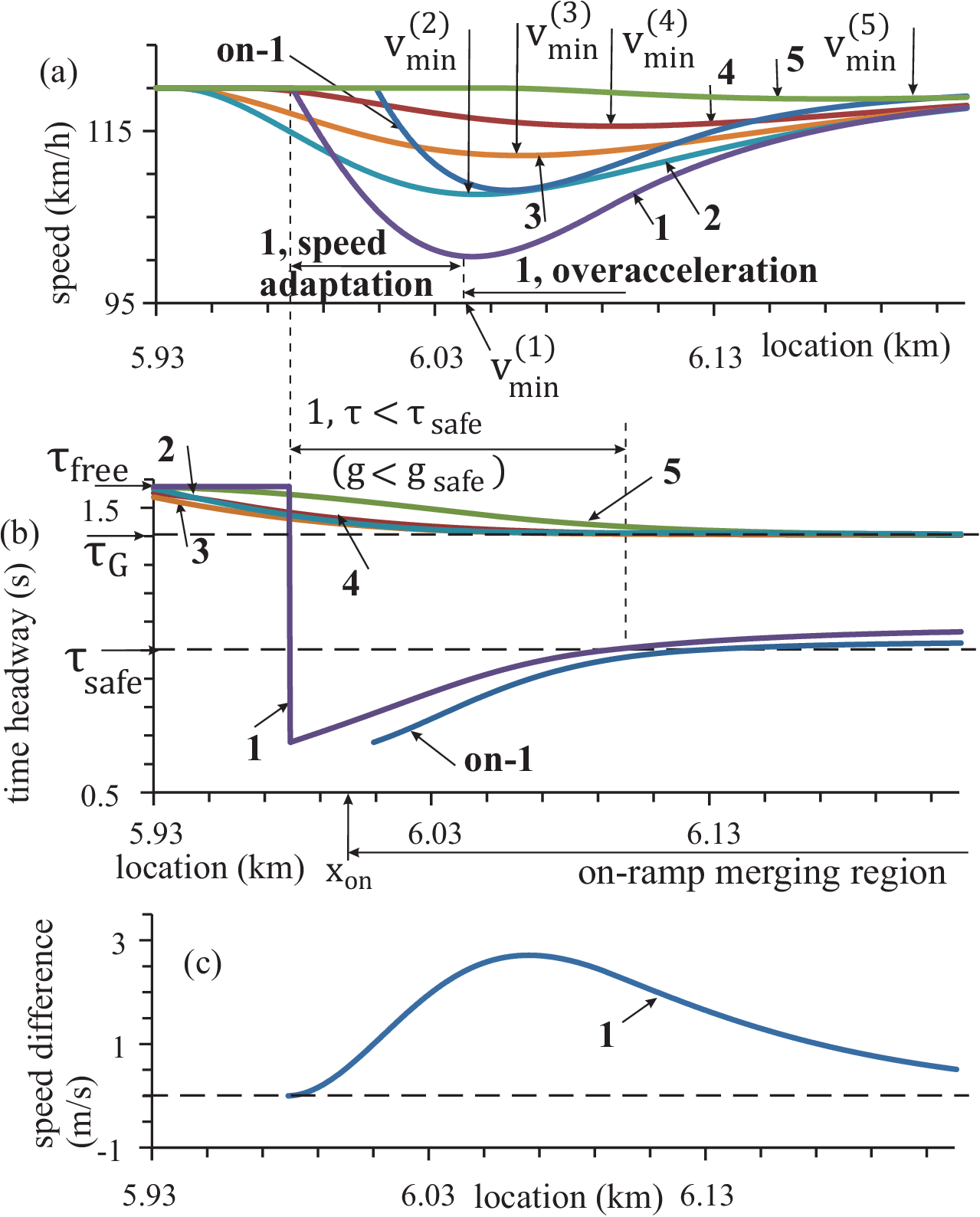}
\end{center}
\caption[]{Continuation of Fig.~\ref{Greater_G_induced}(b).   
(a), (b) 
Location-functions of speed (a) and time-headway (b)  of some vehicles.  (c)
Location-function of speed difference $\Delta v= v^{\rm (on-1)}-v^{(1)}$  between speeds of vehicles on-1 and 1.
 Vehicle numbers
are labeled by the same numbers as those  in Fig.~\ref{Greater_G_induced}(b).     Adapted from~\cite{KKl2025A}.
}
\label{Greater_G_traj}
\end{figure}

 After on-ramp vehicle on-1 has merged from the on-ramp onto the main road,
time-headway of the following vehicle 1 drops below the safety time-headway $\tau_{\rm safe}$ (Fig.~\ref{Greater_G_traj}(b)).
This results in strong deceleration of vehicle 1 ($\lq\lq$1, speed adaptation" in Fig.~\ref{Greater_G_traj}(a)).
Respectively, vehicles 2--5 following vehicle 1 also decelerate.
Due to the growth of the speed difference $\Delta v= v^{\rm (on-1)}-v^{(1)}$  between speeds of vehicles on-1 and 1, the deceleration of
vehicle 1 changes to acceleration.  For the same reason, as explained for classical ACC-vehicles (Fig.~\ref{ACC_safety_acc_traj}(a)),
this safety acceleration of vehicle 1 is overacceleration ($\lq\lq$1, overacceleration" in 
Fig.~\ref{Greater_G_traj}(a)): The safety acceleration of vehicle 1 causes the metastability
of free flow with respect to the F$\rightarrow$S transition (Fig.~\ref{Greater_G_induced}(a)).

\section{Cooperation of Different Overacceleration Mechanisms in Road Lane    \label{Adv_TPACC_model_sec}}  
 
\begin{figure}
\begin{center}
\includegraphics[width = 8 cm]{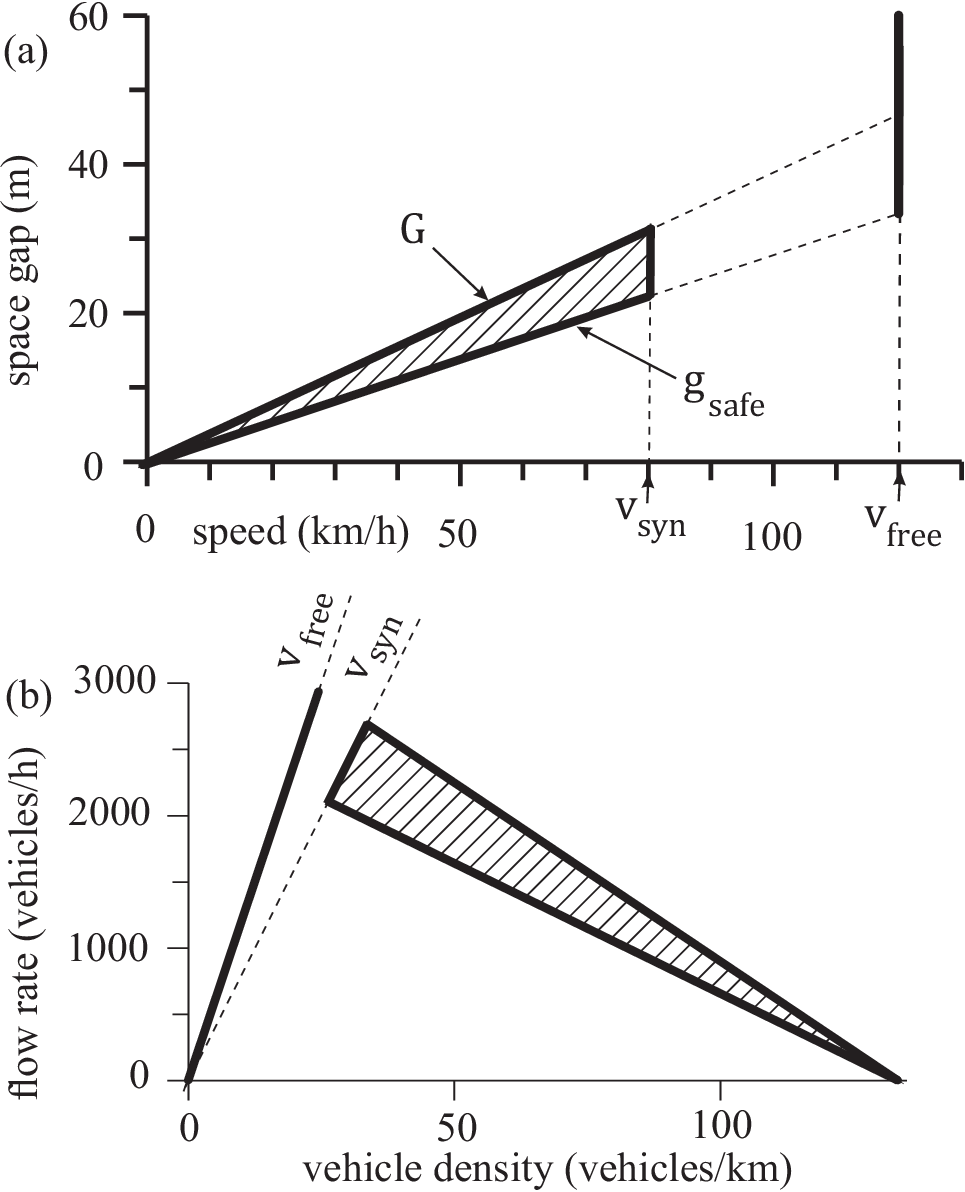}
\end{center}
\caption[]{Steady states of model (\ref{g_v_g_min1_ad})--(\ref{Helly_st2_ad})
 in space-gap--speed   (a)
and flow--density (b) planes.
$v_{\rm free}=$ 120 km/h,  $\tau_{\rm safe}=$ 1 s, $\tau_{\rm G}=$ 1.4 s, $v_{\rm syn}=$ 80 km/h, $d=$ 7.5 m.
}
\label{Nucleation_Theory_1}
\end{figure}

It must be emphasized that overacceleration in a road lane through
 safety acceleration (Sec.~\ref{TPACC_model_sec})  occurs without explicitly introducing
it in the model. 
 In vehicular traffic in road lane, 
there can be other overacceleration mechanisms. One of them is an  overacceleration mechanism caused by a random
vehicle acceleration  occurring   within the indifferent region of car-following  (\ref{indif_g}). In stochastic 
three-phase traffic models~\cite{KKl,KKW,KKl2003A,KKl2004A}, we have simulated this random overacceleration mechanism through model fluctuations. In deterministic microscopic
  models, in which there are no model fluctuations, we  can simulate such overacceleration mechanism while explicitly introducing
it  in the model through 
  a   model
of vehicle  overacceleration   $a_{\rm OA}$  introduced in~\cite{Kerner2023B}, which is described  by equation
\begin{equation}
a_{\rm OA}=\alpha   \Theta (v - v_{\rm syn}),
\label{a_OA}
\end{equation}
where $\alpha$ is a   coefficient of    overacceleration;  $\Theta (z) =0$ at $z<0$ and $\Theta (z) =1$ at $z\geq 0$;
 $v_{\rm syn}$ is a given synchronized flow speed, where $v_{\rm syn}<v_{\rm free}$ (Fig.~\ref{Nucleation_Theory_1}).
The model of vehicle  overacceleration   $a_{\rm OA}$ (\ref{a_OA}) corresponds to the discontinuity of the probability
of vehicle acceleration  (Fig.~\ref{Hyp-OA}(a)). 

It can be assumed that
 within the space-gap range
$g_{\rm safe} \leq g \leq G$  the less the  difference $g-g_{\rm safe}$ between the  space-gap and the safe space-gap, the smaller should be the probability
of   overacceleration and, therefore,   we apply a space-gap dependence of the coefficient of overacceleration $\alpha$~\cite{KKl2025A}:
 \begin{equation}
\alpha =\left\{\begin{array}{ll} 
(\alpha_{0}-\alpha_{1})\Biggl(\frac{g-g_{\rm safe}}{G-g_{\rm safe}}\Biggr)^{k} +  \alpha_{1}   \
  \textrm{at $g_{\rm safe} \leq g \leq G$}, \\  
\alpha_{0}    \ \textrm{at $ g > G$}, \\
0 \  \textrm{at $g < g_{\rm safe}$  }, \\
\end{array} \right.
\label{a_OA_gap}
\end{equation}
where $k$, $\alpha_{0}$, and $\alpha_{1}$  are positive  parameters, $\alpha_{0}>\alpha_{1}$.

In a elementary traffic flow  model that uses vehicle  overacceleration mechanism   $a_{\rm OA}$ (\ref{a_OA}), 
   vehicle acceleration/deceleration $a$   in a road lane
is described by the following  
 system of  equations: 
 \begin{equation}
a = a_{\rm OA} +
K_{\Delta v}\Delta v   \ \textrm{at $g_{\rm safe} \leq g \leq G$}, 
\label{g_v_g_min1_ad}  
\end{equation}
\begin{equation} 
a_{\rm G}(g, v, v_{\ell})= a_{\rm OA} +  K_{1}(g-v\tau_{\rm G})+ K_{2}\Delta v  \ \textrm{at $ g > G$},
  \label{a_G-f_ad}  
	\end{equation} 
	\begin{equation}
a_{\rm safety}(g, v, v_{\ell})=   K_{3}(g-g_{\rm safe})+ K_{4}(g, v, \Delta v)\Delta v   \ \textrm{at $ g < g_{\rm safe}$},
\label{Helly_st2_ad}
\end{equation}
where $a_{\rm OA}$ is given by Eqs.~(\ref{a_OA}), (\ref{a_OA_gap}); 
functions $G(v)$ and $g_{\rm safe}(v)$ are found from   (\ref{G_g_safe_simple});  
coefficient $K_{4}$ is determined from Eq.~(\ref{Helly_st3}).
Steady states of the  TPACC-model (\ref{g_v_g_min1_ad})--(\ref{Helly_st2_ad}) are shown in Fig.~\ref{Nucleation_Theory_1}.

It must be emphasized that model (\ref{g_v_g_min1_ad})--(\ref{Helly_st2_ad}) can also be used for simulations of  
 traffic flow consisting of both human-driving vehicles~\footnote{A generalized microscopic model for human-driving vehicles, which
can simulate moving jam emergence,   will be considered in Sec.~\ref{Gen_Model_Sec}.} and TPACC-vehicles. This underlines the main objective of choosing the TPACC model for automated vehicles, whose movement should be consistent with the movement of human-driven vehicles.

	\begin{figure}
\begin{center}
\includegraphics[width = 8 cm]{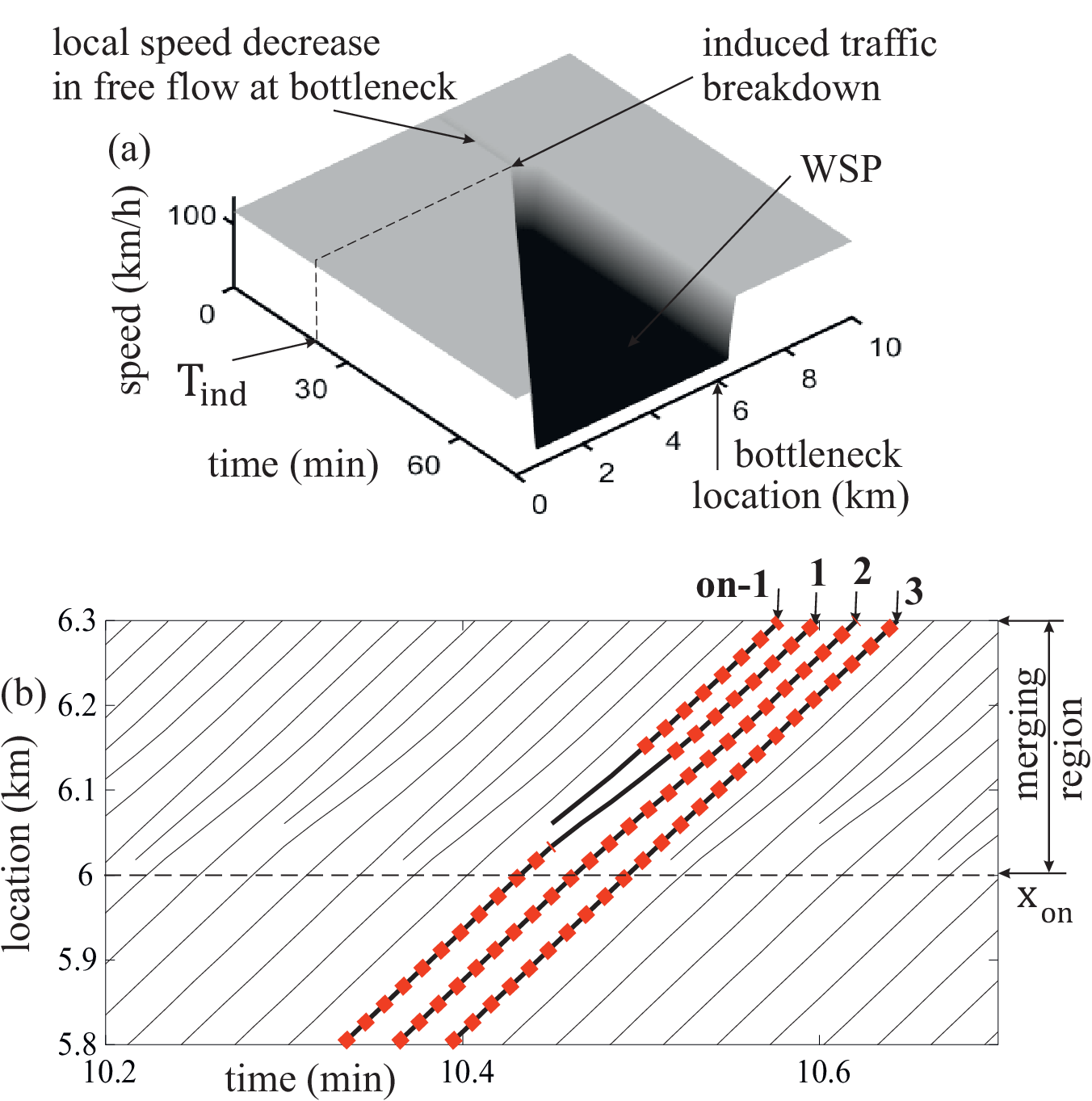}
\end{center}
\caption[]{Simulations of cooperation of different mechanisms of overacceleration on single-lane road.
Induced traffic breakdown on single-lane road with on-ramp bottleneck.
Simulations of model (\ref{a_OA})--(\ref{Helly_st2_ad}), (\ref{G_g_safe_simple})
with model parameters:   $\alpha_{0}=$ 2 $\rm m/s^{2}$,
$\alpha_{1}=$ 0.1 $\rm m/s^{2}$, $k=$ 1, $\tau_{\rm G}=$ 1.4 s, $q_{\rm in}=$ 2000 vehicles/h, $q_{\rm on}=$ 800 vehicles/h; other model parameters are the same as those in
 Fig.~\ref{Greater_G_induced}. 
 (a)  Speed in space and time: Local speed decrease at bottleneck in free flow  and WSP
induced in the free flow   at $T_{\rm ind}=$ 20 min through application of  addition on-ramp inflow impulse $\Delta q_{\rm on}=$
100 vehicles/h of duration   $\Delta t=$ 2 min.  (b) Some of simulated vehicle trajectories within  local speed decrease in free flow
at  bottleneck at time $t < T_{\rm ind}$
  before traffic breakdown has been induced; parts of trajectories within which condition $a_{\rm OA}>0$ is satisfied
  are marked through colored red squares.     Adapted from~\cite{KKl2025A}.
}
\label{Greater_G_induced_alpha}
\end{figure}

				 \begin{figure}  
\begin{center}
\includegraphics[width = 8 cm]{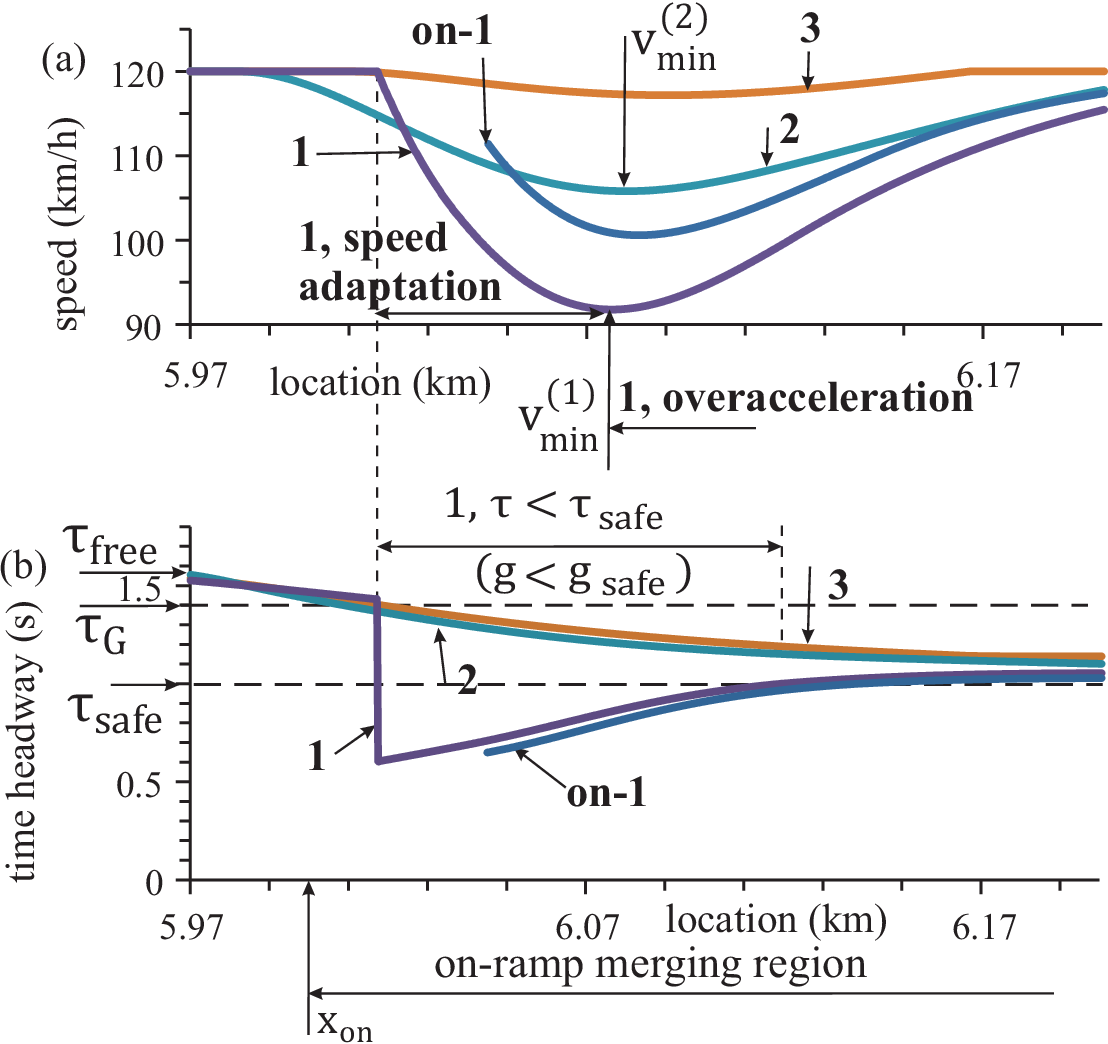}
\end{center}
\caption[]{Continuation of  Fig.~\ref{Greater_G_induced_alpha}(b).
(a), (b) Location-dependencies of
microscopic characteristics for some of the vehicles whose numbers are the same as those in  Fig.~\ref{Greater_G_induced_alpha}(b), respectively: (a)
 vehicle speeds; (b) vehicle time-headway.    Adapted from~\cite{KKl2025A}.
 }
\label{Greater_G_alpha_traj}
\end{figure}

\begin{figure} 
\begin{center}
\includegraphics[width = 8 cm]{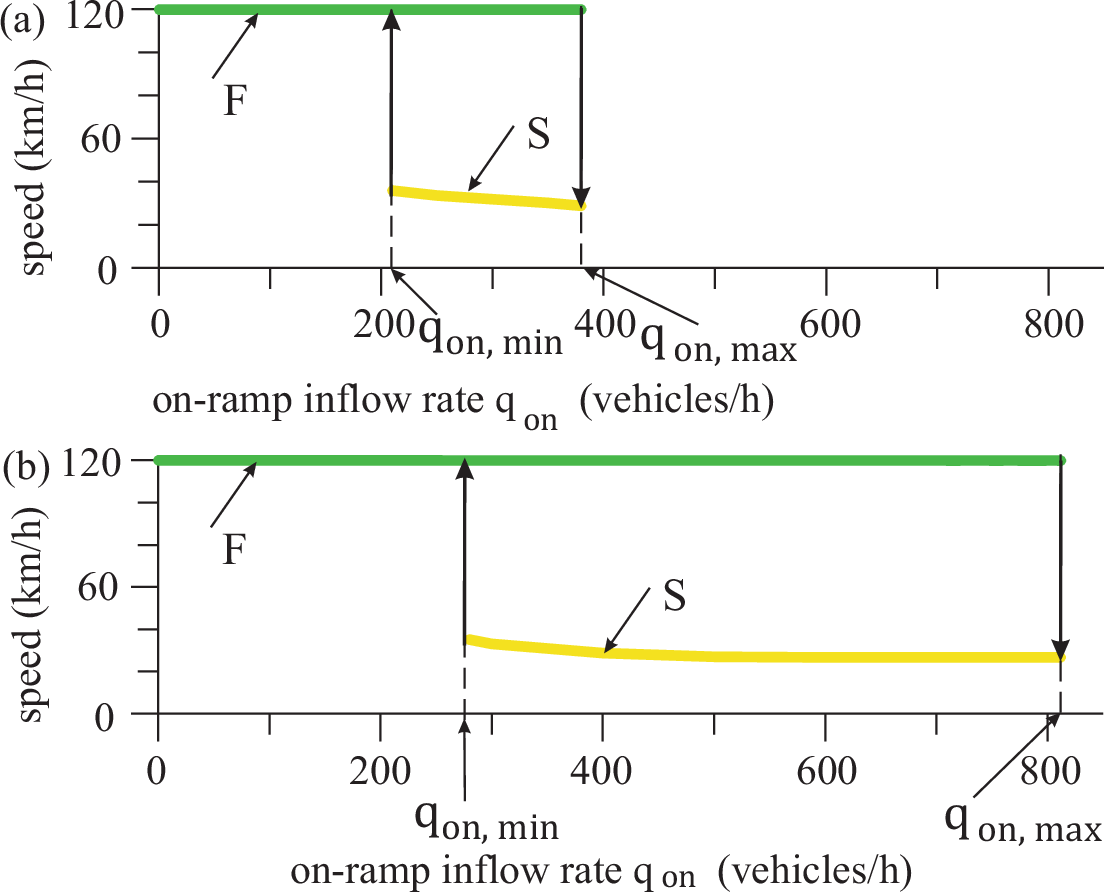}
\end{center}
\caption[]{Effect of cooperation of different mechanisms of overacceleration on maximum highway capacity $C_{\rm max}$
of vehicular traffic on single-lane road with the on-ramp bottleneck   
  calculated at   $q_{\rm in}=$ 2000 vehicles/h for model (\ref{g_v_g_min1})--(\ref{Helly_st3}) (a) and for model
	(\ref{g_v_g_min1_ad})--(\ref{Helly_st2_ad}) (b).
(a) Overacceleration mechanism through safety acceleration only,  model and parameters of Fig.~\ref{Greater_G_induced}.
(b) Cooperation of overacceleration mechanisms through safety acceleration  
and due to overacceleration 
 $a_{\rm OA}$ (\ref{a_OA}),  (\ref{a_OA_gap}); model and parameters of Fig.~\ref{Greater_G_induced_alpha}.
  F -- free flow, S -- synchronized flow.
$q_{\rm on}=q_{\rm on, \ min}$ is a minimum on-ramp flow rate
related to  a minimum highway capacity
$C_{\rm min}=q_{\rm in}+q_{\rm on, \ min}$,
whereas   $q_{\rm on}=q_{\rm on, \ max}$ is a maximum on-ramp flow rate
 related to  a maximum highway capacity
$C_{\rm max}=q_{\rm in}+q_{\rm on, \ max}$. 
Within the on-ramp inflow range $q_{\rm on, \ min}\leq q_{\rm on}< q_{\rm on, \ max}$, the metastability of
free flow is realized, i.e.,
 either free flow or an SP can exist at the bottleneck; 
 at $q_{\rm on}<q_{\rm on, \ min}$ no congested patterns can be induced in free flow, whereas
at $q_{\rm on}>q_{\rm on, \ max}$ the SP occurs spontaneously at the bottleneck after a time delay $T^{\rm (B)}$.
Calculated parameters: $(q_{\rm on, \ min}, \ q_{\rm on, \ max})=$ (217, 372) vehicles/h (a),
  (280, 807) vehicles/h (b).     Adapted from~\cite{KKl2025A}.
}
\label{2Z_Fig_short} 
\end{figure}

Simulations  of the model (\ref{g_v_g_min1_ad})--(\ref{Helly_st2_ad})  presented in Figs.~\ref{Greater_G_induced_alpha}
and~\ref{Greater_G_alpha_traj} exhibit qualitatively the same features as those
found for TPACC-model (\ref{g_v_g_min1})--(\ref{Helly_st3}) 
(see explanations in Sec.~\ref{TPACC_model_sec}). However, there is the following important quantitative difference between these two models: At the same chosen flow rate on the main road $q_{\rm in}$, maximum capacity $C_{\rm max}$ in TPACC-model
(\ref{g_v_g_min1})--(\ref{Helly_st3}) (Fig.~\ref{2Z_Fig_short}(a))
is considerably less than  $C_{\rm max}$ in the model  (\ref{a_OA})--(\ref{Helly_st2_ad}) 
  (Fig.~\ref{2Z_Fig_short}(b)). 
	
	  This result can be explained as follows. In TPACC-model
		(\ref{g_v_g_min1})--(\ref{Helly_st3}), there is only the
		overacceleration mechanism caused by safety acceleration explained in Sec.~\ref{TPACC_model_sec}. Contrarily, 
		in the model (\ref{g_v_g_min1_ad})--(\ref{Helly_st2_ad}), in addition to this 
		overacceleration mechanism, there is the overacceleration mechanism $a_{\rm OA}$ (\ref{a_OA}),  (\ref{a_OA_gap}).
		Therefore,   within the local speed decrease at the bottleneck there is
 a cooperation of the 
  two different mechanisms of overacceleration in the model  (\ref{g_v_g_min1_ad})--(\ref{Helly_st2_ad})~\footnote{A spatiotemporal character 
		of the overacceleration cooperation, whose consideration has been omitted here, can be found in~\cite{KKl2025A}.}:
		\begin{itemize}
		\item [(i)] the overacceleration mechanism $a_{\rm OA}$ (\ref{a_OA}),  (\ref{a_OA_gap}) and
		\item [(ii)] the overacceleration mechanism caused by safety acceleration.
		\end{itemize}
		Contrary to the model (\ref{g_v_g_min1_ad})--(\ref{Helly_st2_ad}), there is no overacceleration cooperation in TPACC-model (\ref{g_v_g_min1})--(\ref{Helly_st3}).
	Due to 	the {\it overacceleration cooperation} in the model (\ref{g_v_g_min1_ad})--(\ref{Helly_st2_ad}),
	overacceleration overcomes
	on average speed adaptation ($\lq\lq$1, speed adaptation" in Fig.~\ref{Greater_G_alpha_traj}) up to much greater on-ramp
	inflow rate $q_{\rm on}=q_{\rm on, \ max}=$ 807 vehicles/h
	(Fig.~\ref{2Z_Fig_short}(b)) than the value 
		$q_{\rm on}=q_{\rm on, \ max}=$ 372 
		vehicles/h (Fig.~\ref{2Z_Fig_short}(a)) reachable  for TPACC-model (\ref{g_v_g_min1})--(\ref{Helly_st3}).
		Thus, the overacceleration cooperation can be very important for the increase in maximum capacity $C_{\rm max}$ of 
		traffic flow~\footnote{For a consideration  of the effect of overacceleration on traffic flow consisting of TPACC-vehicles, we have applied here   simplified  
		model (\ref{g_v_g_min1})--(\ref{Helly_st3}) or model (\ref{g_v_g_min1_ad})--(\ref{Helly_st2_ad}). For real TPACC-applications,
		dynamic behavior of characteristics of   
  TPACC-vehicles, like values $G$ and $g_{\rm safe}$   in Eq.~(\ref{g_v_g_min1}), should depend on current traffic situation, in particular,
	on  the speed difference between vehicles $\Delta v$~\cite{KernerPat1}.    
	However, a consideration of this subject is out of scope of this review that could represent very interesting tasks for future traffic research.}.

\section{Overacceleration  through   Lane-Changing in Traffic of Automated Vehicles  
\label{Lane_OA_sec}}  

The  mechanism for overacceleration through vehicle lane-changing on multi-lane roads, which was
 predicted already in first publications of three-phase traffic theory~\cite{Kerner1999B,Kerner1999A,Kerner1999C}, has been   used
in many simulations of stochastic microscopic three-phase models (see references in~\cite{KernerBook1,KernerBook2,KernerBook4,KernerReviews}). 

In this section, we show that features of vehicle overacceleration  and its effect of traffic flow known for
traffic of human-driving vehicle~\cite{KernerBook1,KernerBook2,KernerBook4,KernerReviews} are qualitatively the same for traffic 
consisting of automated vehicles~\cite{Kerner2023A}.
For motion of automated vehicles in a road lane, we limit here by consideration of the
 classical ACC-model   (\ref{ACC_Cl}) by Helly~\cite{Helly}. Qualitatively the same results have been found, if instead of classical ACC-model
    (\ref{ACC_Cl})
we use  TPACC-model of  Sec.~\ref{TPACC_St_S}.  
Through simulations of automated vehicles moving  on two-lane road with the bottleneck 
we show that there is a cooperation between
the  mechanism for overacceleration through lane-changing with the mechanism of overacceleration through safety acceleration
studied for single-lane road in Sec.~\ref{Safety_OA_sec}.
 As in Sec.~\ref{Safety_OA_sec}, we choose parameters of automated vehicles at which no
vehicle overdeceleration and, consequently, no string instability can occur.
 
   \subsection{Nucleation   of Traffic Breakdown on Two-Lane Road}

 \begin{figure}
\begin{center}
\includegraphics[width = 8 cm]{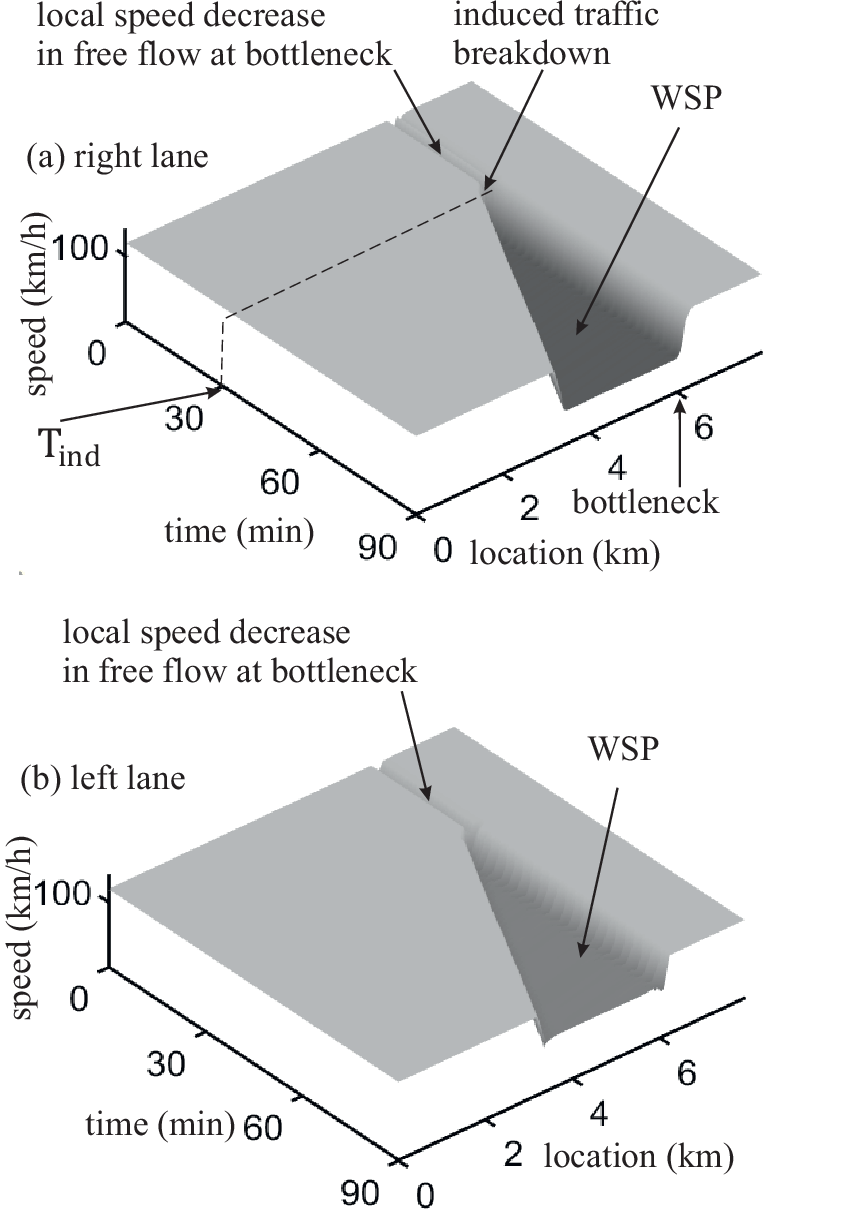}
\end{center}
\caption[]{Simulations of the metastability of free flow with respect to the F$\rightarrow$S transition  in
automated-driving vehicular traffic moving on two-lane road with the bottleneck 
in model (\ref{ACC_Cl}), (\ref{RL})--(\ref{g_prec_ACC}) at $q_{\rm in}=$ 2571 (vehicles/h)/lane, 
$q_{\rm on}=$ 720  vehicles/h. (a, b)
Speed in space and time in  the right lane (a) and left lane (b).
At time instant $T_{\rm ind}$, through application of on-ramp inflow impulse traffic breakdown (F$\rightarrow$S transition)
has been induced at the bottleneck.
 Parameters of on-ramp inflow-rate impulse inducing F$\rightarrow$S transition   at bottleneck: $T_{\rm ind}=$ 30 min,
$\Delta q_{\rm on}=$ 180 vehicles/h, $\Delta t=$ 2 min.
Parameters of automated vehicles: 
$\tau_{\rm d}=$ 1 s,
$K_{\rm ACC,1}= 0.3 \ s^{-2}$, $K_{\rm ACC,2}= 0.9 \ s^{-1}$, $v_{\rm free}=$ 120 km/h, $d=$ 7.5 m.  Road length $L=$ 8 km. 
Lane-changing parameters:
$\delta_{1}=$ 1 m/s, $\delta_{2}=$ 5 m/s, $\tau_{1}=$ 0.6 s, $\tau_{2}=$ 0.2 s. 
  Other model parameters are the same as those in Fig.~\ref{ACC_safety_acc}.    Adapted from~\cite{Kerner2023A}.
}
\label{Induced_F_S_Bottl}
\end{figure} 

  \begin{figure}
\begin{center}
\includegraphics[width = 8 cm]{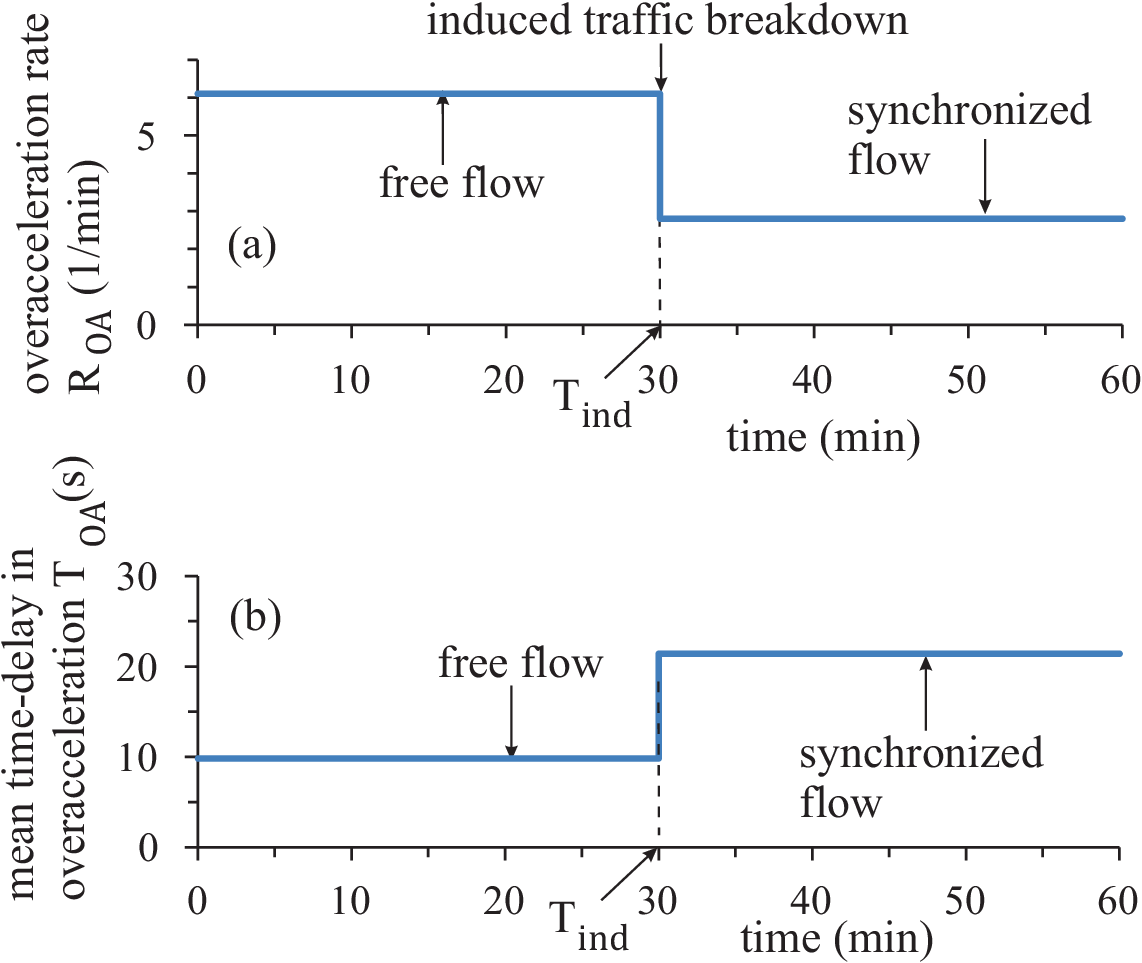}
\end{center}
\caption[]{Continuation of Fig.~\ref{Induced_F_S_Bottl}. Induced traffic breakdown (induced F$\rightarrow$S transition):
(a, b)   Time-dependencies of the averaged overacceleration rate $R_{\rm OA}$ (a)  
and  the mean time-delay in  overacceleration   $T_{\rm OA}$ (b).    Adapted from~\cite{Kerner2023A}.
}
\label{Induced_F_S_Bottl_tr2}
\end{figure}

	\begin{figure}
\begin{center}
\includegraphics[width = 8 cm]{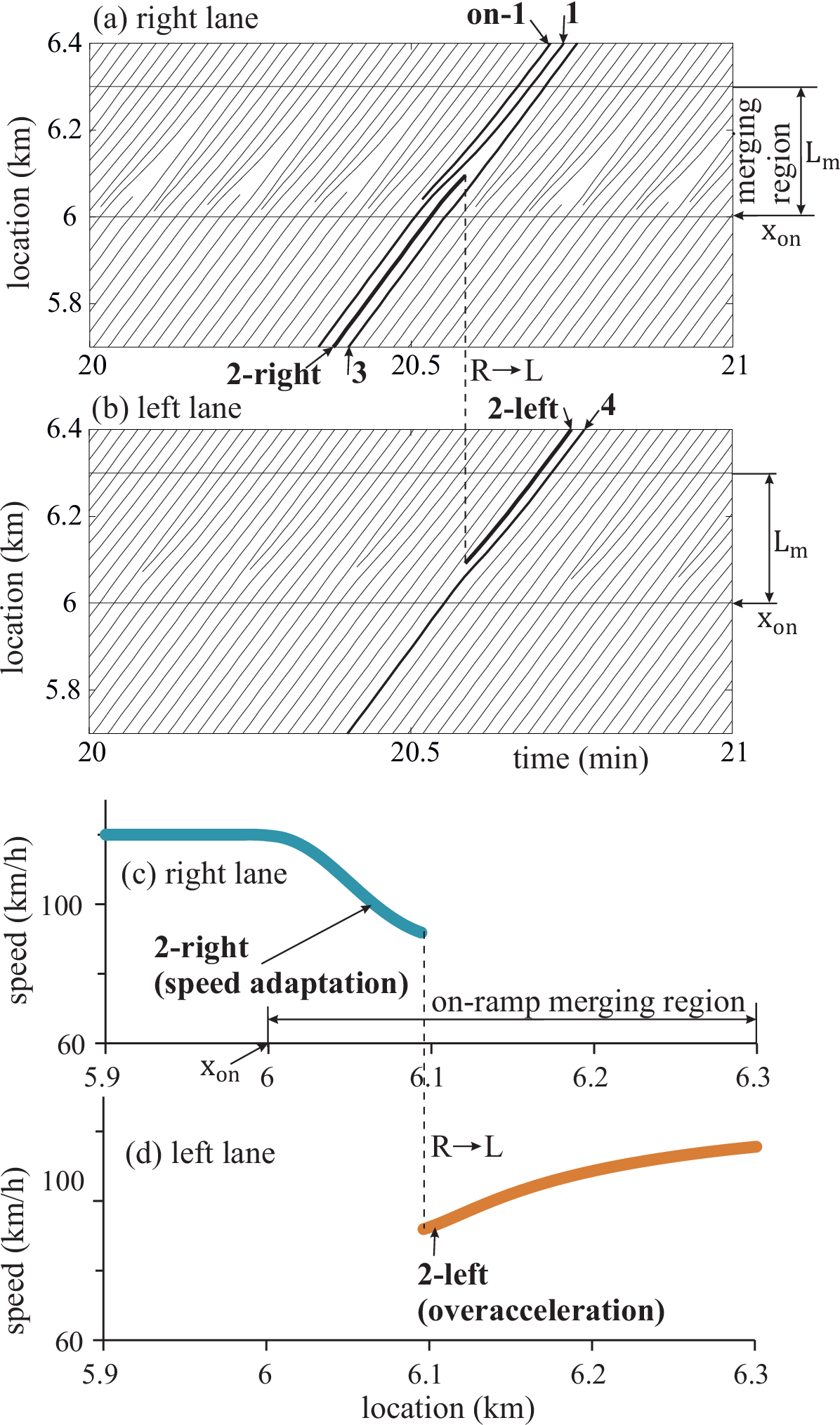}
\end{center}
\caption[]{Continuation of Fig.~\ref{Induced_F_S_Bottl}.  (a, b) Simulated vehicle trajectories within  local speed decrease in free flow
at  bottleneck in  the right lane (a) and left lane (b) at time $t < T_{\rm ind}$.
 (c, d)
Location-functions of speed  of vehicle 2 labeled by $\lq\lq$2-right" in the right lane (c) and by
$\lq\lq$2-left" in left lane 
 (d)      in (a, b).
R$\rightarrow$L lane-changing of vehicle 2  is marked by   dashed vertical lines R$\rightarrow$L.    Adapted from~\cite{Kerner2023A}.
}
\label{Free_Flow_Bottl_tr}
\end{figure}

	\begin{figure} 
\begin{center}
\includegraphics[width = 8 cm]{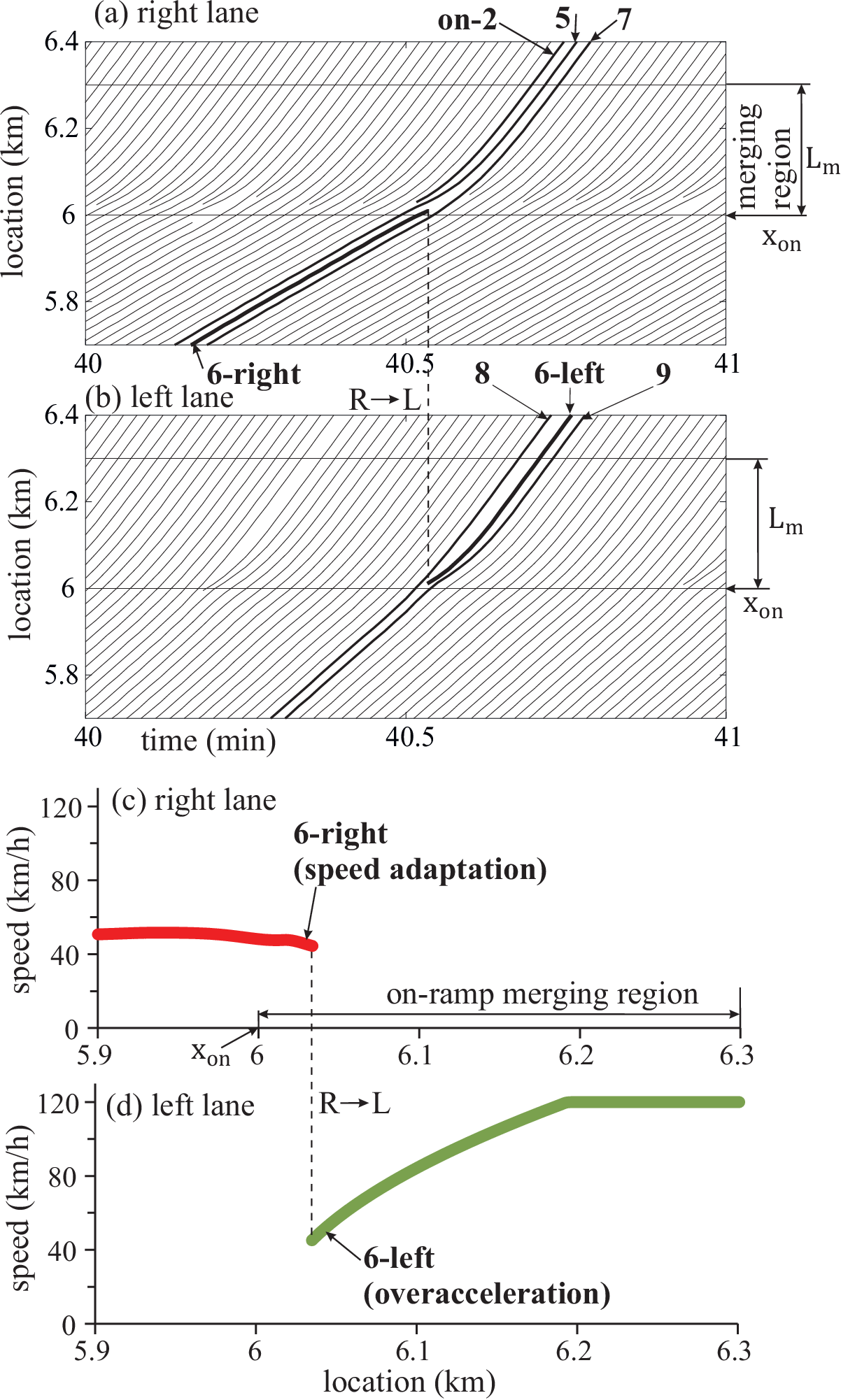}
\end{center}
\caption[]{Continuation of Fig.~\ref{Induced_F_S_Bottl}.    (a, b) Simulated vehicle trajectories in synchronized flow
at  bottleneck in the right lane (a) and left lane (b) at time $t > T_{\rm ind} + \Delta t$.
 (c, d)
Location-functions of speed  of vehicle 6 labeled by $\lq\lq$6-right" in the right lane (c) and by
$\lq\lq$6-left" in left lane 
 (d)      in (a, b).
R$\rightarrow$L lane-changing of vehicle 6  is marked by   dashed vertical lines R$\rightarrow$L.   Adapted from~\cite{Kerner2023A}.
}
\label{Induced_F_S_Bottl_tr}
\end{figure}

 For lane-changing 
of automated vehicles moving  on two-lane road with the bottleneck we use 
 well-known   incentive lane changing rules from the right to left lane R$\rightarrow$L
(\ref{RL}) and from the left to right lane L$\rightarrow$R
(\ref{LR}) as well as safety conditions (\ref{g_prec_ACC}) well-known for human-driving vehicles (see, e.g.,~\cite{Nagel1998}) 
\begin{eqnarray}
\label{RL}
R \rightarrow L: v^{+}(t) \geq v_{\ell}(t)+\delta_{1} \ {\rm and} \ v(t) \geq v_{\ell}(t), \\
L \rightarrow R: v^{+}(t) \geq v_{\ell}(t)+\delta_{2} \ {\rm or} \  v^{+}(t) \geq v(t)+\delta_{2},
\label{LR} \\
g^{+}(t)   \geq\ v(t) \tau_{2}, \quad g^-(t) \geq\ v^{-}(t) \tau_{1}.
\label{g_prec_ACC}
\end{eqnarray} 
The automated vehicle changes to the faster target lane 
with the objective to pass a slower automated vehicle in the current lane     if   time headway to  preceding and following vehicles in the target  lane   are not shorter than some given safety time headway $\tau_{1}$
and $\tau_{2}$. In   (\ref{RL})--(\ref{g_prec_ACC}),
superscripts $+$  and $-$ denote, respectively, the preceding and the following vehicles in the target lane;
 $\tau_{1}$, $\tau_{2}$,   $\delta_{1}$, $\delta_{2}$  are positive constants~\footnote{It should be noted that  
in (\ref{RL}), (\ref{LR}) the value $v^{+}$ at $g^{+} > L_{\rm a}$ and the value $v_{\ell}$ at $g > L_{\rm a}$ are replaced by $\infty$, where $L_{\rm a}$ is a look-ahead distance; in simulations, we have used
$L_{\rm a}=$ 80 m. However,   due to large flow rates used in simulations     
both condition $g^{+} > L_{\rm a}$ and condition   $g > L_{\rm a}$ are not satisfied. }.

Simulations of   the model
(\ref{ACC_Cl}), (\ref{RL})--(\ref{g_prec_ACC}) show that
 there is the free flow metastability with respect to the F$\rightarrow$S transition at the bottleneck
on {\it both} lanes of the road (Fig.~\ref{Induced_F_S_Bottl}). In accordance with the definition of vehicle overacceleration (Sec.~\ref{Definition_OA_sec}), 
this means that there are some
vehicle acceleration behaviors that can be considered vehicle overacceleration.

\subsection{Discontinuous Rate of Overacceleration through Lane-Changing} 

   To understand the role of
 lane-changing rules in the occurrence
of vehicle overacceleration, we make a microscopic analysis of the effect of lane-changing on vehicle overacceleration
(Figs.~\ref{Induced_F_S_Bottl_tr2}--\ref{Over_Following_tr_S}). Before the F$\rightarrow$S transition
has been induced at the bottleneck ($0\leq t < T_{\rm ind}$), 
there is R$\rightarrow$L lane-changing with the rate $R_{\rm RL}\approx$ 6.1 $\rm min^{-1}$ (Fig.~\ref{Induced_F_S_Bottl_tr2}(a)).
When due to the F$\rightarrow$S transition free flow transforms into synchronized flow (Fig.~\ref{Induced_F_S_Bottl_tr}(a)),
R$\rightarrow$L lane-changing rate $R_{\rm RL}$ drops sharply to
  $R_{\rm RL}\approx$ 2.8 $\rm min^{-1}$ (Fig.~\ref{Induced_F_S_Bottl_tr2} (a)).
	  
		R$\rightarrow$L lane-changing of a vehicle that has initially decelerated in the right lane
(vehicle 2-right in Figs.~\ref{Free_Flow_Bottl_tr} (a, c) and vehicle
 6-right in Figs.~\ref{Induced_F_S_Bottl_tr} (a, c) have decelerated before R$\rightarrow$L lane-changing) leads to
the acceleration of the vehicle in the left lane.   The vehicle acceleration under consideration is solely determined
		by    R$\rightarrow$L lane-changing of the vehicle. Therefore, 
		the rate of   the vehicle acceleration  denoted by $R_{\rm OA}$, which is  caused by R$\rightarrow$L lane-changing, is given by
		formula
		\begin{equation}
	R_{\rm OA}=R_{\rm RL}. 
	\label{R_OA_R_F}
	\end{equation}
		  When free flow transforms into synchronized flow,	vehicle acceleration caused by R$\rightarrow$L lane-changing exhibits the discontinuous character: In 
	  accordance with (\ref{R_OA_R_F}), there is the
	discontinuity in the rate
	of   vehicle acceleration   $R_{\rm OA}$. Therefore, this vehicle acceleration can be considered
	overacceleration through R$\rightarrow$L lane-changing and, respectively, $R_{\rm OA}$ can be considered the
	discontinuous rate of overacceleration.  
	The mean time delay in overacceleration denoted by 
	$T_{\rm OA}$   is equal to $1/R_{\rm OA}$.  In free flow
	$T_{\rm OA}\approx$ 9.84 s, whereas in synchronized flow $T_{\rm OA}\approx$ 21.4 s (Fig.~\ref{Induced_F_S_Bottl_tr2}(b)).

	As in traffic of human-driving vehicles, there is a
	spatiotemporal competition between overacceleration and speed adaptation
	in traffic consisting of automated vehicles. Indeed,
	there is a tendency to free flow through overacceleration. Simultaneously, there is  the opposite
	tendency to synchronized flow through speed adaptation. This competition between overacceleration  
	and speed adaptation   occurring in space and time is considered in Secs.~\ref{Com_SF_Sec} and~\ref{Com_FS_Sec}.
	
  \subsection{Competition of Overacceleration with Speed Adaptation in Free Flow   \label{Com_SF_Sec}}

\begin{figure} 
\includegraphics[width = 8 cm]{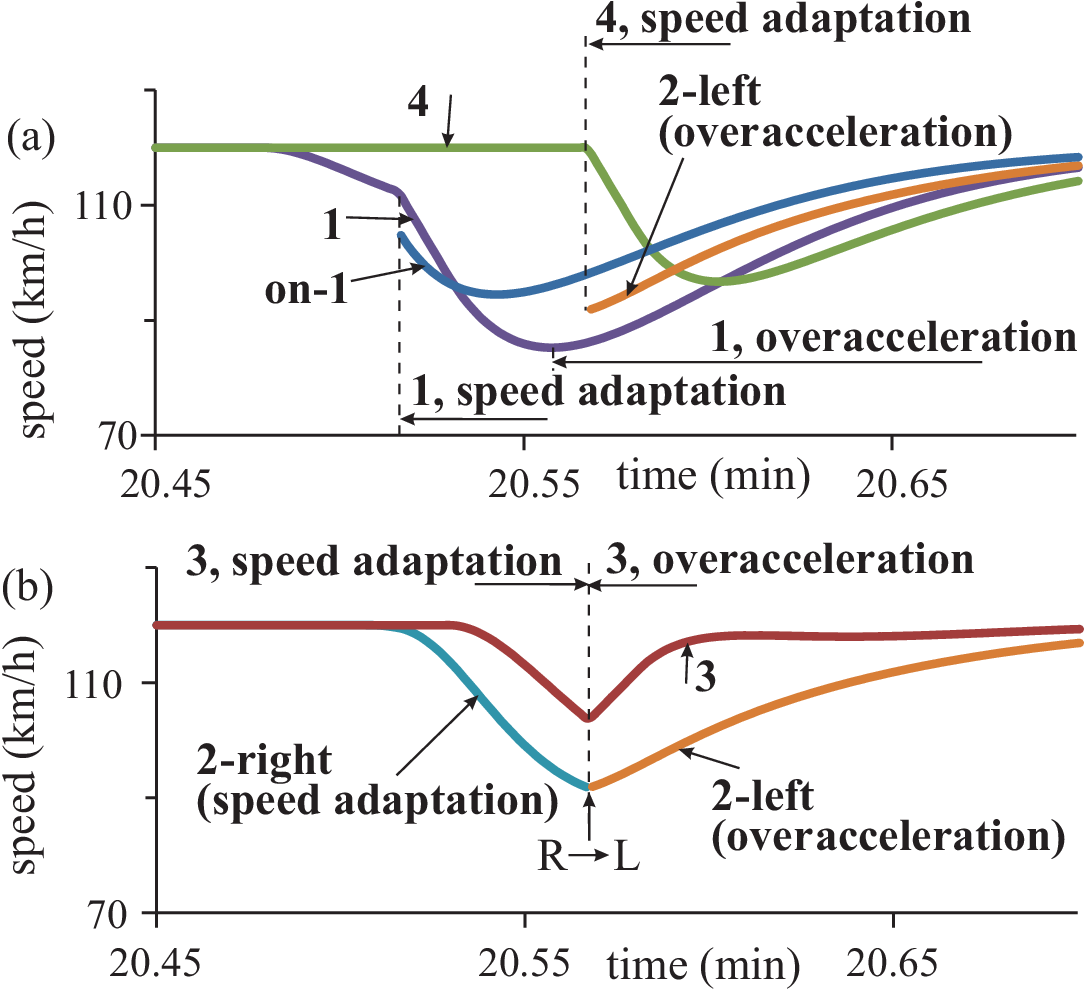}
\caption[]{Simulations of spatiotemporal competition between overacceleration and speed adaptation in free flow:
Tendency to free flow through overacceleration.
Time-functions of speed for vehicle trajectories presented in Figs.~\ref{Free_Flow_Bottl_tr} (a, b)  labeled by the same numbers, 
respectively. Adapted from~\cite{Kerner2023A}. 
}
\label{Over_Following_tr_F}  
\end{figure}

 When on average overacceleration overcomes speed adaptation, free flow is self-maintained at the bottleneck.
This is realized at time $t<T_{\rm ind}$ before the F$\rightarrow$S transition has been induced at the bottleneck
(Fig.~\ref{Induced_F_S_Bottl}). To explain this phenomenon, we consider some vehicle trajectories in 
Fig.~\ref{Over_Following_tr_F}.

As on single-lane road with the bottleneck (Sec.~\ref{Safety_OA_sec}), there is mechanism of overacceleration caused by
safety acceleration on two-lane road with the bottleneck: An example is vehicle 1 in Fig.~\ref{ACC_safety_acc_traj} and vehicle 1 in
Fig.~\ref{Over_Following_tr_F}(a).  In both cases,
after vehicle on-1 has merged from the on-ramp, 
 vehicle 1 decelerates   (1, speed adaptation
in Fig.~\ref{Over_Following_tr_F}(a)). Shortly later vehicle 1  accelerates
 while following   vehicle on-1. 
As  explained in Sec.~\ref{Safety_OA_sec}, in this case the safety acceleration of vehicle 1 should be considered overacceleration
 (1, overacceleration
in Fig.~\ref{Over_Following_tr_F}(a)). 

However, in comparison with single-lane road (Sec.~\ref{Safety_OA_sec}), there is a cooperation of the mechanism of overacceleration through
safety acceleration with a mechanism of overacceleration caused by lane-changing that is as follows.
Through R$\rightarrow$L lane-changing of vehicle 2 (2-left (overacceleration)
in Fig.~\ref{Over_Following_tr_F}(b)) the following vehicle 3
in the right lane  accelerates
($\lq\lq$3, overacceleration" in Fig.~\ref{Over_Following_tr_F}(b)). This overacceleration is 
 trying to maintain
the free-flow state at the bottleneck. 
Due to a high rate of R$\rightarrow$L lane-changing in free flow, respectively, the high rate of
overacceleration $R_{\rm OA}$ (\ref{R_OA_R_F}) (Fig.~\ref{Induced_F_S_Bottl_tr2}), 
overacceleration overcomes on average speed adaptation.
This results in the self-maintaining of free flow at the bottleneck.

\subsection{Competition of Overacceleration with Speed Adaptation in Synchronized Flow  \label{Com_FS_Sec}}

\begin{figure} 
\begin{center}
\includegraphics[width = 8 cm]{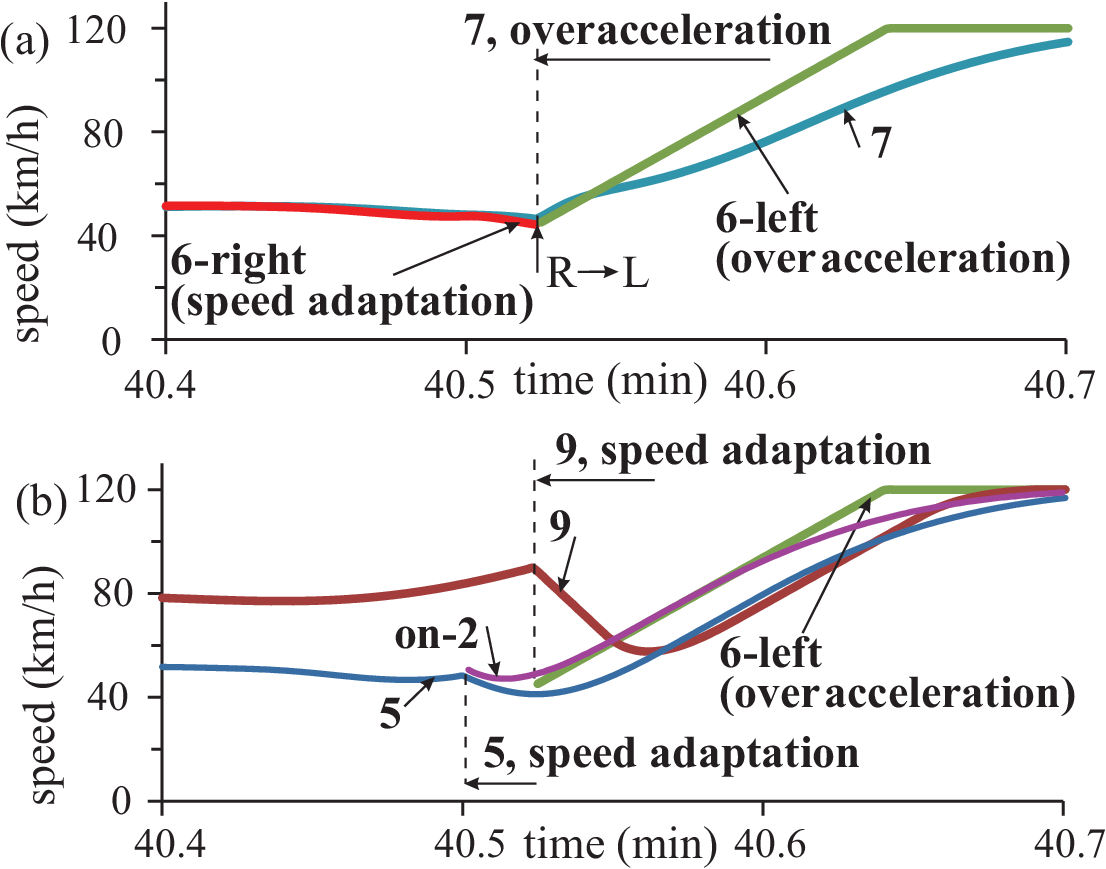}
\end{center}
\caption[]{Simulations of spatiotemporal competition between overacceleration and speed adaptation in synchronized flow:
Tendency to synchronized flow through speed adaptation.
Time-functions of speed for vehicle trajectories presented in Figs.~\ref{Induced_F_S_Bottl_tr} (a, b) labeled by the same numbers, 
respectively.   Adapted from~\cite{Kerner2023A}.
}
\label{Over_Following_tr_S}  
\end{figure}

 At $t>T_{\rm ind}+\Delta t$, i.e., after the induced F$\rightarrow$ transition has occurred,   synchronized
flow is   at the bottleneck. In synchronized flow,  overacceleration tries to transform synchronized flow to
a free-flow state: For example, due to R$\rightarrow$L lane-changing of vehicle 6 
(6-left (overacceleration)
in Fig.~\ref{Over_Following_tr_S}(a)) the following vehicle 7  
  accelerates
(7, overacceleration in Fig.~\ref{Over_Following_tr_S}(b)).
Contrary to overacceleration,  speed adaptation tries to maintain the synchronized
flow state (Fig.~\ref{Over_Following_tr_S}(a)). For example,   vehicle on-2 merging from the on-ramp
 forces the following
vehicle 5   to decelerate (Fig.~\ref{Over_Following_tr_S}(b)).

However, due to discontinous overacceleration, i.e., a considerably lower rate of
overacceleration $R_{\rm OA}$ (\ref{R_OA_R_F}) in synchronized flow than $R_{\rm OA}$
 is in free flow (Fig.~\ref{Induced_F_S_Bottl_tr2}), 
speed adaptation overcomes on average overacceleration.
This results in the self-maintaining of synchronized flow at the 
bottleneck~\footnote{It should be noted that there is a dual role of lane-changing that is as follows. In
free flow, R$\rightarrow$L lane-changing of vehicle 2 leads to overacceleration
(2-left (overacceleration) in Fig.~\ref{Over_Following_tr_F}(a)).
  Contrarily, the same lane-changing of vehicle 2 causes
speed adaptation in the left lane. Indeed, the following vehicle
4 in the left lane  must
decelerate (4, speed adaptation in Fig.~\ref{Over_Following_tr_F}(a)), while adapting
its speed to the speed of slower vehicle 2 that has just changed
from the right lane to the left lane.
Speed adaptation caused by a dual role of lane-changing
occurs also in synchronized flow. An example is R$\rightarrow$L
lane-changing of vehicle 6 (6-left (overacceleration) in
Fig.~\ref{Over_Following_tr_S}(b)): This vehicle forces  the following vehicle 9 in the
left lane  to decelerate (9,
speed adaptation in Fig.~\ref{Over_Following_tr_S}(b)).}.
 Thus, as in traffic of automated vehicles moving on single-lane road  with the bottleneck  (Sec.~\ref{Safety_OA_sec}), 
the cause of the free-flow metastability with respect
to the F$\rightarrow$S transition on two-lane road with the bottleneck (Fig.~\ref{Induced_F_S_Bottl}) is a spatiotemporal
competition between discontinuous overacceleration  and speed adaptation.

\section{Simulations of Empirical Induced Traffic Breakdown  Through   Helly's Model for Automated Vehicles \label{AV_MSP_S}}

Here we show that classical Helly's model (\ref{ACC_Cl}) used   for simulations of
 traffic flow of automated vehicles in Secs.~\ref{Safety_OA_sec} and~\ref{Lane_OA_sec} can simulate 
empirical results about the nucleation nature of traffic breakdown (F$\rightarrow$ transition)
in traffic consisting of human-driving vehicles as presented in Fig.~\ref{20041998_MSP}(a).
This is   because under string stability conditions classical Helly's model
 can  simulate traffic breakdown 
through competition of  discontinuous overacceleration  
with speed adaptation discussed  in Sec.~\ref{Com_SF_Sec}.

 \begin{figure} 
\begin{center}
\includegraphics[width = 7.5 cm]{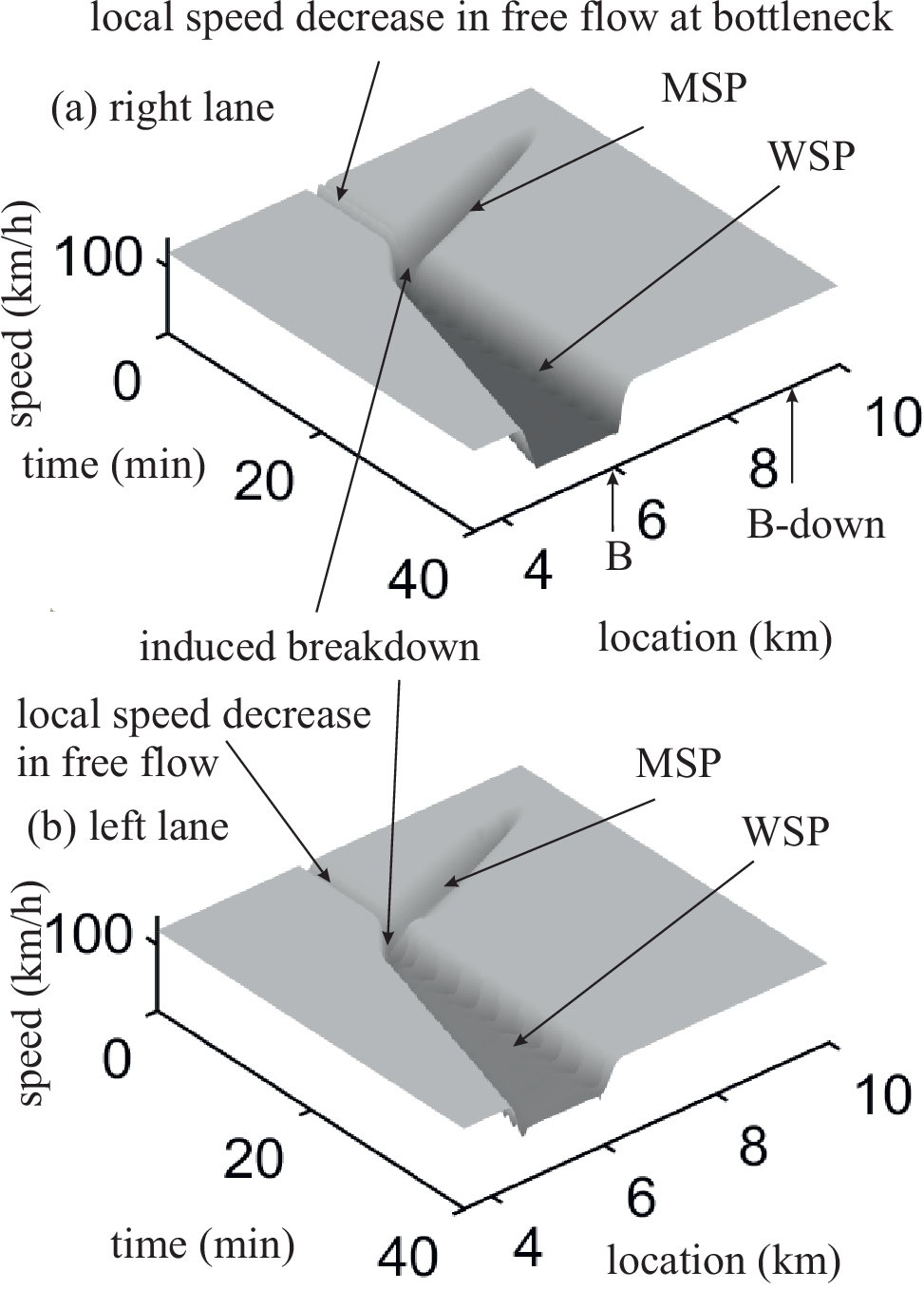}
\end{center}
\caption[]{Simulations of induced F$\rightarrow$S transition  through upstream propagation of MSP 
to upstream bottleneck with classical Helly's model (\ref{ACC_Cl}), (\ref{RL})--(\ref{g_prec_ACC}) for automated vehicles moving
 on two-lane road with two identical  bottlenecks: Speed in space and time in the right
lane  (a) and  left lane
 (b). $q_{\rm in}=$ 2571
(vehicles/h)/lane.   Parameters of upstream bottleneck (B) are $x_{\rm on}=$ 
  6 km, $L_{\rm m}=$ 0.3 km, $q_{\rm on}=$ 720 vehicles/h; 
parameters of downstream bottleneck (B-down) are $x^{\rm (down)}_{\rm on}=$ 9 km,
$L^{\rm (down)}_{\rm m}=$ 0.3 km, $q^{\rm (down)}_{\rm on}=$ 0;
road length $L=$ 10 km.   
The MSP   has been induced  
  in   free flow  at the downstream bottleneck (B-down) at  $t=T^{\rm (down)}_{\rm ind}=$ 5 min through application of
	the on-ramp inflow impulse $\Delta q^{\rm (down)}_{\rm on}=$ 900 vehicles/h of duration $\Delta t^{\rm (down)}=$ 1 min.
The MSP  induced
at the downstream bottleneck (B-down)     propagates upstream; reaching
the upstream on-ramp bottleneck (B) the MSP induces the
F$\rightarrow$S transition at the bottleneck. The two bottlenecks B-down and B correspond to the same on-ramp bottleneck used above
   in Figs.~\ref{ACC_safety_acc}--\ref{Over_Following_tr_S}. 
Other model parameters are the same as those in Fig.~\ref{Induced_F_S_Bottl}.
Adapted from~\cite{Kerner2023A}.
}
\label{MSP_Breakdown} 
\end{figure}

\begin{figure} 
\begin{center}
\includegraphics[width = 8 cm]{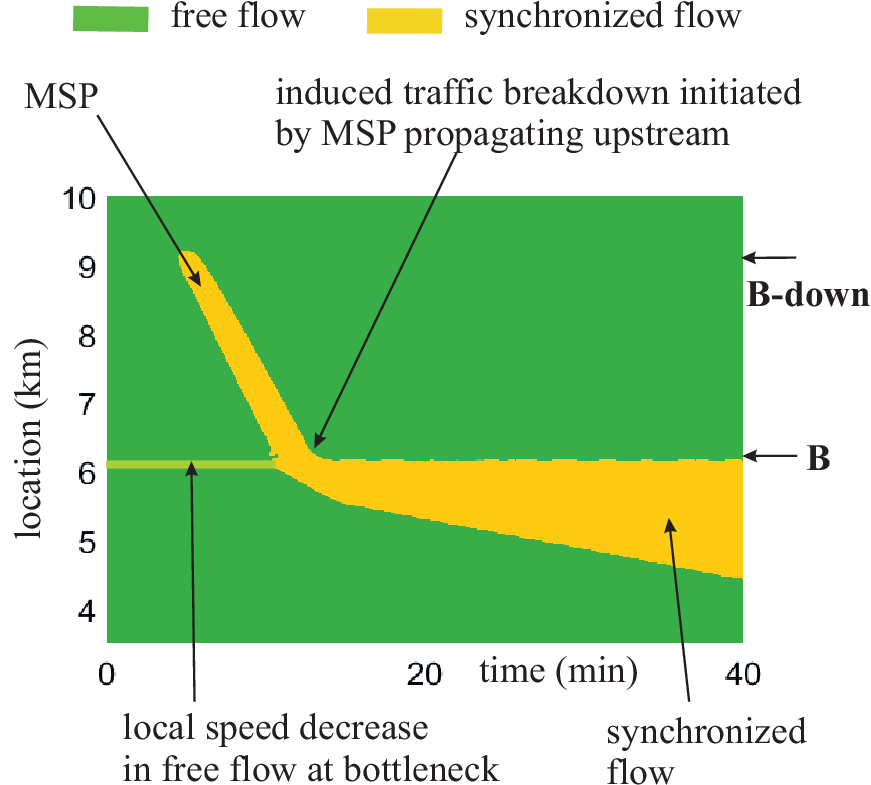}
\end{center}
\caption[]{Continuation of Fig.~\ref{MSP_Breakdown}.
Simulations of automated-driving vehicular traffic that reproduce empirical
   breakdown  nucleation features measured  in  real human-driving   traffic (Fig.~\ref{20041998_MSP}(a)). 
Speed data of Fig.~\ref{MSP_Breakdown} averaged across  two-lane road   are
presented  in space and time in free flow (green) and synchronized flow (yellow). Adapted from~\cite{Kerner2023A}.
}
\label{MSP_color}
\end{figure}

We have used the same model for automated vehicles moving on two-lane road
(\ref{ACC_Cl}), (\ref{RL})--(\ref{g_prec_ACC}) and the same model parameters as used in simulations
presented in Figs.~\ref{Induced_F_S_Bottl}--\ref{Over_Following_tr_S}. However, rather than a single on-ramp bottleneck (Fig.~\ref{Induced_F_S_Bottl}), there are now two identical upstream and downstream on-ramp bottlenecks labeled by B and B-down, respectively (Figs.~\ref{MSP_Breakdown} and~\ref{MSP_color}). The downstream bottleneck B-down is used to induce
an MSP at this bottleneck (see explanation in caption to Fig.~\ref{MSP_Breakdown}). We see that 
the MSP does induce the F$\rightarrow$S transition at the upstream bottleneck as observed in empirical data presented in
Fig.~\ref{20041998_MSP}(a).   This   is   because when   no string instability can occur 
 in   Helly's model (\ref{ACC_Cl}), then 
traffic breakdown
is the F$\rightarrow$S transition occurring due to a competition between discontinuous overacceleration and speed adaptation.
Due to discontinuous overacceleration, the F$\rightarrow$S transition  exhibits the nucleation nature 
  as observed in empirical observations.

\section{Traffic Breakdown in Flow of Human-Driving Vehicles:  Vehicle Overacceleration,
 Not  Vehicle Overdeceleration \label{OA_Not_OD_Sec}}

In Secs.~\ref{Safety_OA_sec}--\ref{AV_MSP_S}, we have shown that when no vehicle overdeceleration and, therefore, no traffic instability is realized in free flow at the bottleneck, then 
traffic breakdown is the F$\rightarrow$S transition that exhibits the nucleation nature. This theoretical result is in accordance with empirical traffic data. We have also shown that this free flow metastability with respect to the F$\rightarrow$S transition is
due to a competition between speed adaptation and discontinuous   overacceleration, not due to   traffic instability in free flow.

However, contrary to traffic consisting of automated vehicles considered above,
 in traffic flow consisting of human-driving vehicles there are driver's delays. The driver's delays
can cause vehicle overdeceleration and, therefore,   traffic instability. Indeed, empirical data show~\cite{KernerBook1} that in synchronized flow
traffic instability (called S$\rightarrow$J instability) occurs often that
 leads to S$\rightarrow$J transitions, i.e., to  the emergence of moving traffic jams.   
In this section, we investigate whether overdeceleration affects on traffic breakdown
(F$\rightarrow$S transition) in free flow at the bottleneck, or not.
Another objective of this section  is as follows: We would like to study
whether and under which conditions     jam absorption driving  as well as other   approaches, which attempt
 to suppress traffic instability, could be useful for   traffic breakdown  control.

\subsection{Microscopic Model of Human-Driving Vehicles  \label{Gen_Model_Sec}}

  When we have considered traffic consisting of human-driving vehicles
with the   model 
(\ref{g_v_g_min1_ad})--(\ref{Helly_st2_ad}), we have ignored the empirical fact that 
 in {\it synchronized flow}  the driver reaction time can lead  to overdeceleration and, as a result, to
 classical traffic flow instability resulting in moving jam emergence (S$\rightarrow$J transition). 
 To simulate
moving jam emergence in  deterministic model 
  (\ref{g_v_g_min1_ad})--(\ref{Helly_st2_ad}), 
as already made in some stochastic three-phase traffic flow models (e.g.,~\cite{KKl,KKW,KKl2003A,KKl2004A,KKl2009A,KKHS2013_Int1}), 
we assume that dynamic model parameters (at least some of the model parameters)
     can change at a synchronized flow speed that is less than some characteristic speed
denoted by 
 $v_{\rm pinch}$: 
\begin{equation}
\mathcal{P}=\left\{
\begin{array}{ll}
\mathcal{P} \ {\rm at} \ v \geq v_{\rm pinch}
 \\
\mathcal{P}_{\rm pinch} \ {\rm at} \ v < v_{\rm pinch}, \\
\end{array}\right.
  \label{pinch-formula} 
  \end{equation}
	where $\mathcal{P}$ denotes one of the 
 model parameters $\tau_{\rm G}$, $\tau_{\rm safe}$, $K_{1}$, $K_{2}$, $K_{3}$, $K_{4}$, 
   etc. used in  model (\ref{g_v_g_min1_ad})--(\ref{Helly_st2_ad}), whereas 
the subscript $pinch$ in  $\mathcal{P}_{\rm pinch}$ is used to
distinguish values of the same model parameters    for
  low speeds $v< v_{\rm pinch}$. We assume also that in  (\ref{a_OA}) and (\ref{pinch-formula}) condition
	\begin{equation}
 v_{\rm pinch}<v_{\rm syn}
  \label{pinch-syn-formula}
  \end{equation}  
 is satisfied.
In such a generalized model (\ref{g_v_g_min1_ad})--(\ref{Helly_st2_ad}), (\ref{pinch-formula}), rather than functions
(\ref{G_g_safe_simple}),  
  we use   formulations for the speed-functions $g_{\rm safe}(v)$ and $G(v)$
	presented in (\ref{g_safe-min}) and (\ref{G-min}), 
	in which at $v\rightarrow 0$ the space gap tends
	to some minimum space-gap $g_{\rm min}$ between vehicles~\footnote{The use of the minimum space-gap $g_{\rm min}$ is
	   well-known for many standard deterministic traffic flow models (see, e.g.,~\cite{Treiber-Kesting}). } (see Appendix~\ref{G-g-safe_Sec}).

		\subsection{General Congested Pattern (GP)  \label{GP_Common_in_theories}}

\begin{figure}
\begin{center}
\includegraphics[width = 8 cm]{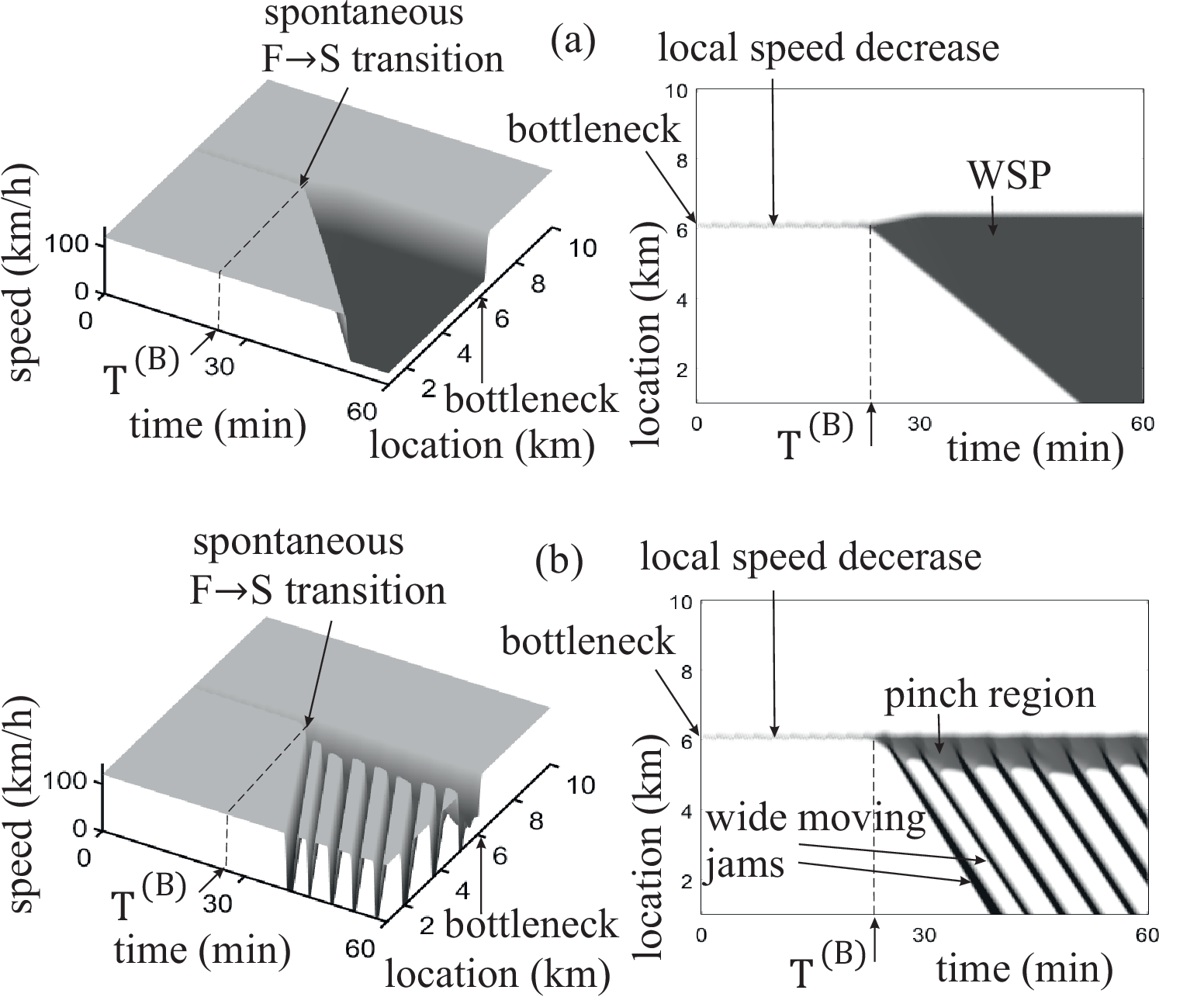}
\end{center}
\caption[]{Vehicle overacceleration versus vehicle overdeceleration: What is the cause of   traffic breakdown 
at a bottleneck?   Left subplots -- simulations of speed in space and time on single-lane road with   on-ramp bottleneck    located 
at  $x_{\rm on}=$ 6 km;
right subplots -- the same
vehicle speed data as those left, respectively, presented by regions with variable shades of gray [shades of gray
vary from white to black when the speed decreases from 120 km/h
(white) to 0 km/h (black)]. In (a),   simulations of model 	 (\ref{g_v_g_min1_ad})--(\ref{Helly_st2_ad})
 at $q_{\rm in}=$ 2250 vehicles/h, $q_{\rm on}=$ 568.5 vehicles/h; other   
model parameters are the same  as those in
 Fig.~\ref{Greater_G_induced_alpha} at which no vehicle overdeceleration is realized. In (b),
simulations of model 	(\ref{g_v_g_min1_ad})--(\ref{Helly_st2_ad}), (\ref{pinch-formula}), (\ref{g_safe-min}), and (\ref{G-min}), in which
vehicle overdeceleration is realized at  $v<v_{\rm pinch}$  with
 model parameters $g_{\rm min}=$ 5 m, $v_{\rm pinch}=$ 36 km/h, at $v<v_{\rm pinch}$ parameters
$K_{3, \rm pinch}= 0.1 \ s^{-2}$ and 
 $K^{(2)}_{4, \rm pinch}= 0.8 \ s^{-1}$; other model parameters   are the same as those in (a).  
	 $\lq\lq$Pinch region" labels a road area within which the pinch effect is realized. Adapted from~\cite{KKl2025A}.
}
\label{not-driver-overreaction_short}
\end{figure}

If in TPACC model
(\ref{g_v_g_min1_ad})--(\ref{Helly_st2_ad}) we choose the flow rate $q_{\rm sum}=q_{\rm in}+q_{\rm on}>C_{\rm max}=q_{\rm in}+q_{\rm on, \ max}$
 (see explanations in caption of Fig.~\ref{2Z_Fig_short}), then  after a time delay $T^{\rm (B)}$
spontaneous traffic breakdown (F$\rightarrow$S transition)
 is realized at the bottleneck (Fig.~\ref{not-driver-overreaction_short}(a)). 

Contrary to   TPACC model
(\ref{g_v_g_min1_ad})--(\ref{Helly_st2_ad}), in  generalized model
(\ref{g_v_g_min1_ad})--(\ref{Helly_st2_ad}), (\ref{pinch-formula}) , (\ref{g_safe-min}), and (\ref{G-min})  at speeds $v<v_{\rm pinch}$ vehicle overdeceleration and, consequently, traffic instability do occur in synchronized flow (S$\rightarrow$J instability)
that development leads to moving jam emergence in synchronized flow
(S$\rightarrow$J transition)~\cite{KernerBook1,KernerBook2,KernerBook3,KernerBook4}).

Therefore, in  generalized model
(\ref{g_v_g_min1_ad})--(\ref{Helly_st2_ad}), (\ref{pinch-formula}) , (\ref{g_safe-min}), and (\ref{G-min}),
we find that after traffic breakdown (F$\rightarrow$S transition) has occurred wide moving jams appear almost immediately in synchronized flow,
i.e., the development of the S$\rightarrow$J instability leads very quickly   to the S$\rightarrow$J transitions 
(Fig.~\ref{not-driver-overreaction_short}(b)). This congested pattern,
which consists of all three-phases F, S, and J of three-phase traffic theory, is called in  a {\it general congested
 pattern (GP)}.
 Here, a question arises: 
\begin{itemize}
\item [--]
Does vehicle overdeceleration (overbraking) and resulting
traffic instability affect    the F$\rightarrow$S transition (traffic breakdown) leading to  GP formation
 at the bottleneck? 
\end{itemize}

\subsection{Does Vehicle Overdeceleration Affect
  Traffic Breakdown leading to GP Formation at Bottleneck?  \label{OA_TrafBreak_Feat_S}}

We have found that  vehicle overdeceleration does not affect the characteristics of
  traffic breakdown  (F$\rightarrow$S transition) at the bottleneck:  
 Neither
the microscopic spatiotemporal evolution of a local speed decrease
in  free flow at the bottleneck during the time interval $t<T^{\rm (B)}$ nor the value of the
time delay $T^{\rm (B)}$ in the traffic breakdown  changes when, instead of the
model (\ref{g_v_g_min1_ad})--(\ref{Helly_st2_ad}), in which no vehicle overdeceleration occurs (Fig.~\ref{not-driver-overreaction_short}(a)),
the model (\ref{g_v_g_min1_ad})--(\ref{Helly_st2_ad}), (\ref{pinch-formula}), (\ref{g_safe-min}) and (\ref{G-min}), in which  vehicle overdeceleration occurs, is used for the simulations (Fig.~\ref{not-driver-overreaction_short}(b)).

\begin{figure}
\begin{center}
\includegraphics[width = 8 cm]{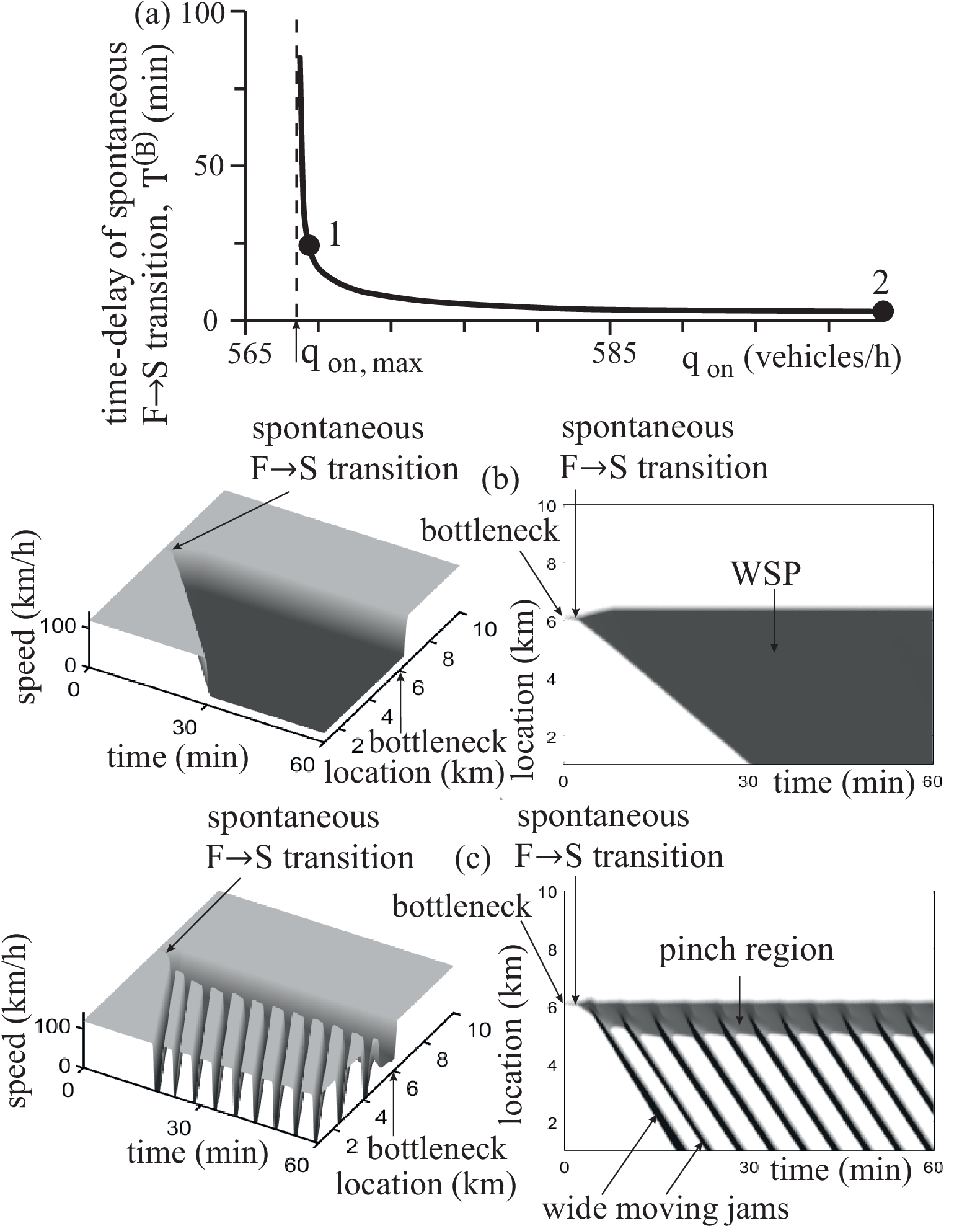}
\end{center}
\caption[]{Explanation of a possible  confusion by 
  simulations of traffic congestion.
Simulations of the same models as those in Figs.~\ref{not-driver-overreaction_short}(a) and~\ref{not-driver-overreaction_short}(b) 
 made on the same single-lane road with the same on-ramp bottleneck
as well as the same flow rate 
  $q_{\rm in}=$ 2250 vehicles/h.   
(a) Time-delay of spontaneous traffic breakdown (F$\rightarrow$S transition) $T^{\rm (B)}$
as function of on-ramp inflow rate $q_{\rm on}$;
	function $T^{\rm (B)}(q_{\rm on})$
		 does not depend on whether   model (\ref{g_v_g_min1_ad})--(\ref{Helly_st2_ad}) used in   Fig.~\ref{not-driver-overreaction_short}(a)
		or   model (\ref{g_v_g_min1_ad})--(\ref{Helly_st2_ad}), (\ref{pinch-formula}), (\ref{g_safe-min}), and (\ref{G-min}) used in   Fig.~\ref{not-driver-overreaction_short}(b) has been applied; $q_{\rm on, \ max}=$ 568 vehicles/h.
       (b), (c) Simulations of speed in space and time (left subplots) at  
			$q_{\rm on}=$ 600 vehicles/h with the model used in   Fig.~\ref{not-driver-overreaction_short}(a) for (b)
			and with the model used in   Fig.~\ref{not-driver-overreaction_short}(b) for (c);   
right subplots -- the same
vehicle speed data as those left, respectively, presented by regions with variable shades of gray [shades of gray
vary from white to black when the speed decreases from 120 km/h
(white) to 0 km/h (black)]. In (a), point 1 is related to $q_{\rm on}=$ 568.5 vehicles/h used in
Figs.~\ref{not-driver-overreaction_short}(a) and~\ref{not-driver-overreaction_short}(b) for which $T^{\rm (B)}\approx$ 24 min, 
whereas point 2 is related to  (b) and (c) for which $T^{\rm (B)}\approx$ 3 min. Adapted from~\cite{KKl2025A}.
}
\label{overreaction-OA}
\end{figure}

  This conclusion remains  valid, when on-ramp inflow rate
   $q_{\rm on}$ increases. To show this, we consider the on-ramp inflow rate dependence of the time-delay of traffic breakdown
  $T^{\rm (B)}$. Simulations show that at a given value $q_{\rm in}$,
the dependence $T^{\rm (B)}(q_{\rm on})$ is a strong falling function  (Fig.~\ref{overreaction-OA}(a)).
Point 1 in Fig.~\ref{overreaction-OA}(a) is
 related to simulations in Figs.~\ref{not-driver-overreaction_short} (a) and (b). 
Point 2 with a larger value $q_{\rm on}$  in Fig.~\ref{overreaction-OA}(a) is
 related to simulations presented in Figs.~\ref{overreaction-OA}(b) and (c).
We find that independent of value $q_{\rm on}$ qualitative characteristics of
traffic breakdown (F$\rightarrow$S transition) and   time delay    $T^{\rm (B)}$ of the breakdown
do  not depend on whether   there is vehicle overdeceleration   in traffic model
 (Figs.~\ref{not-driver-overreaction_short}(b)
and~\ref{overreaction-OA}(c) are related to simulations of model (\ref{g_v_g_min1_ad})--(\ref{Helly_st2_ad}), (\ref{pinch-formula}), (\ref{g_safe-min}), and (\ref{G-min})), or, in contrary, there is
no vehicle overdeceleration    in traffic model  (Figs.~\ref{not-driver-overreaction_short}(a)
and~\ref{overreaction-OA}(b) are related to simulations of model (\ref{g_v_g_min1_ad})--(\ref{Helly_st2_ad})).
  \begin{itemize}
	\item[--]
 Vehicle overdeceleration 
 does not influence on the microscopic features of 
the development of the F$\rightarrow$S transition (traffic breakdown).
\end{itemize}

Thus,  
the microscopic features of traffic breakdown do not depend on whether
vehicle  overdeceleration  is incorporated in the three-phase traffic flow model or not:
The physics of the F$\rightarrow$S transition is solely determined by a competition between discontinuous overacceleration and speed adaption.
However, vehicle overdeceleration, which is realized
 in   synchronized flow resulting from the breakdown,  causes  
the emergence of wide  moving jams
   (S$\rightarrow$J transitions) (Fig.~\ref{not-driver-overreaction_short}(b)): 
  \begin{itemize}
	\item [--]
 Wide moving jams result from 
a sequence of the F$\rightarrow$S$\rightarrow$J transitions.
\end{itemize}

	\subsection{A Common Feature of Standard and Three-Phase Traffic Theories   
		\label{OD_Cl-3Phase_Sec}}
		
		It must be emphasized that there is one important  common feature of standard traffic theory (see, e.g.,~\cite{GM_Com1,GM_Com2,GM_Com3,KS,KS1,KS2,KS4,Newell,Gipps1981,Gipps1986,Wiedemann,Payne_1,Payne_2,Aw-Raschle,Nagel_S,Bando_1,Bando,Bando_2,Bando_3,Nagatani_1,Nagatani_2,Treiber,Krauss,Kra_PhD,Jiang2001,Barlovic,Chen2012A,Chen2012B,Chen2014,KK1993,KK1994,KKK1997,Helbing1999A,Helbing2001A,Helbing2002A}
	and reviews~\cite{Gartner2,ElefteriadouBook2014_Int1,Sch,Brockfeld2003,Bellomo,Ferrara2018A,Leu,Mahnke,MahnkeKLub2009A,Schadschneider2011,Saifuzzaman2015A,Pa1983,Newell1963,Nagel2003A,Nagatani_R,Gazis,Barcelo2010,Treiber-Kesting,DaihengNi,MakridisZhang,Kessels,Schadschneider,Chowdhury,Helbing,Mannering1998,Brackstone1999,Shvetsov2003,Maerivoet2005})
	and   three-phase traffic theory~\cite{KernerBook1,KernerBook2,KernerBook3,KernerBook4,Kerner2018B}: 
 	This common feature  is the existence of   vehicle overdeceleration   in both contradictory theories. 
	In turn, the vehicle overdeceleration can
  cause   traffic  instability leading to moving jam emergence: In  
	the  standard traffic models and the models in the framework of three-phase traffic theory,
	 moving jams in vehicular traffic
appear due to traffic instability caused by vehicle overdeceleration.

 Due to this common feature, 
there is qualitative the same dynamics of moving traffic jams in simulations
of the standard traffic models and the models in the framework of 
three-phase traffic theory: When moving jams emerge due to traffic instability, 
there can be a non-linear interaction of the jams leading to the dissolving of some of the jams.
 This effect, which is often observed in real traffic, is called
pinch effect (Figs.~\ref{not-driver-overreaction_short}(b) and~\ref{overreaction-OA}(c)):
The spatiotemporal non-linear moving jam dynamics, which is well-known
		 since middle of 1990s~\cite{KK1994,KK1995A,Helbing1999A,Helbing2001A,Helbing2002A}, is 
		qualitatively the same  in  
		the standard traffic models and the models in the framework of the three-phase traffic theory
 \begin{itemize}
\item [--] Spatiotemporal distribution of moving traffic jams
within the GP can be qualitatively the same in both  contradictory theories.
 \end{itemize}

 \subsection{Controversial Views on the Origin of Traffic Breakdown  
		\label{Controversial_Sec}}

	However, the origin of traffic breakdown     (Fig.~\ref{overreaction-OA}(c)) is understood 
	qualitatively different in
		standard traffic flow 
		theory (see, e.g., for review~\cite{Gartner2,ElefteriadouBook2014_Int1,Sch,Brockfeld2003,Bellomo,Ferrara2018A,Leu,Mahnke,MahnkeKLub2009A,Schadschneider2011,Saifuzzaman2015A,Pa1983,Newell1963,Nagel2003A,Nagatani_R,Gazis,Barcelo2010,Treiber-Kesting,DaihengNi,MakridisZhang,Kessels,Schadschneider,Chowdhury,Helbing,Mannering1998,Brackstone1999,Shvetsov2003,Maerivoet2005})
		and   three-phase traffic theory~\cite{KernerBook1,KernerBook2,KernerBook3,KernerBook4}:
		\begin{itemize}
		\item[--]  
		Standard traffic flow theory assumes that the origin of traffic breakdown in the free flow
at the bottleneck lies in the traffic instability caused by  vehicle overdeceleration (for a review,
see, e.g.,~\cite{Treiber-Kesting}). The development of traffic instability, in turn, leads to the formation of moving traffic jams  within the GP.
\item[--] 
In contrast, three-phase traffic theory assumes that the origin of traffic breakdown at the bottleneck lies in the F$\rightarrow$S transition, which is caused by the interplay between speed adaptation and discontinuous overacceleration, rather than by traffic instability.
 The moving jams arise later
		in synchronized flow resulting from traffic breakdown, i.e., due to the sequence of F$\rightarrow$S$\rightarrow$J transitions~\cite{KernerBook1,KernerBook2,KernerBook3,KernerBook4}.
 \end{itemize}

In   three-phase traffic theory~\cite{KernerBook1,KernerBook2,KernerBook3,KernerBook4}, the explanation of the GP
	shown in Fig.~\ref{overreaction-OA}(c) is   as follows: There is almost no time-delay between
	the time instant of traffic breakdown $t=T^{\rm (B)}$ and the beginning of moving jam emergence. This means that
	the time-delay
	between the F$\rightarrow$S transition and S$\rightarrow$J transition within
	the sequence of the F$\rightarrow$S$\rightarrow$J transitions
	presented in Fig.~\ref{overreaction-OA}(c) is not possible to distinguish: 
	At the chosen model parameters of   three-phase traffic model
	(\ref{g_v_g_min1_ad})--(\ref{Helly_st2_ad}), (\ref{pinch-formula}), (\ref{g_safe-min}), and (\ref{G-min}),
moving jams occurs due to overdeceleration already in the emergent synchronized flow. 
However, as shown and explained in Sec.~\ref{OA_TrafBreak_Feat_S},  
the emergence of this synchronized flow caused by the F$\rightarrow$S  transition does not depend on whether there is
vehicle  overdeceleration or not.

\subsection{Resolution of
Controversial Views  Regarding  the Origin  of Traffic Breakdown    \label{Resolution_Sec}}

The results of the consideration of 
  traffic breakdown leading to the emergence of moving jams made in Secs.~\ref{GP_Common_in_theories}--\ref{Controversial_Sec} are
{\it theoretical results} derived from simulations of traffic flow models.	
The resolution of
controversial theoretical views on the origin of GP emergence   (Sec.~\ref{Controversial_Sec})  can  be made {\it only}
	from analysis of {\it empirical spatiotemporal traffic data} measured in real traffic.
 
Studies of empirical spatiotemporal data collected on highways  
presented in the books~\cite{KernerBook1,KernerBook2,KernerBook3,KernerBook4} has shown that traffic breakdown is the F$\rightarrow$S transition leading to the emergence of synchronized flow.  Moving jams can  emerge within this synchronized flow, i.e.,
due to a sequence of the F$\rightarrow$S$\rightarrow$J transitions:
 The empirical moving jams do not emerge spontaneously in {\it empirical free flow} at the bottleneck.
\begin{itemize}
\item [--]
The empirical origin of the traffic breakdown---whose subsequent evolution leads 
to the emergence of moving jams---is the F$\rightarrow$S transition,
and not traffic instability due to
vehicle overdeceleration.
\end{itemize}

 Moreover, there are a diverse variety of  empirical spatiotemporal
 traffic data measured in real traffic in which
 no moving jam emergence is observed in synchronized flow resulting from
the   F$\rightarrow$S transition.
 One of the examples of these data is presented in Fig.~\ref{20041998_MSP}(a).
None of the standard traffic theories  (see, e.g., books and reviews~\cite{May,Manual2010,Gartner2,ElefteriadouBook2014_Int1,Da,Sch,Brockfeld2003,Bellomo,Ferrara2018A,Leu,Mahnke,MahnkeKLub2009A,Wh2,Schadschneider2011,Saifuzzaman2015A,Pa1983,Newell1963,Prigogine1971,New,Nagel2003A,Nagatani_R,Ashton,Drew,Gerlough,Gazis,Barcelo2010,Treiber-Kesting,DaihengNi,MakridisZhang,Kessels,Schadschneider,Chowdhury,Helbing,Mannering1998,Brackstone1999,Shvetsov2003,Maerivoet2005,Rakha2009,Piccoli2009,Roess2014,Hegyi2017,Seo2017,MATSim_Nagel2016})  can explain the empirical nucleation nature of
the F$\rightarrow$S transition shown in   
 Fig.~\ref{20041998_MSP}(a).
\begin{itemize}
\item [--]
 The common empirical feature of vehicular traffic is the empirical nucleation nature of
the F$\rightarrow$S transition at the bottleneck, which cannot be explained by  the standard traffic theories. 
\end{itemize}

Thus, the empirical   nucleation nature the F$\rightarrow$S transition
cannot be explained by traffic instability caused by vehicle overdeceleration. For this reason,
approaches  for preventing traffic instability such as  jam absorption driving  (see for review~\cite{ZhengbingHe}) 
   are not capable of preventing real traffic breakdown in free flow.
	However, in Sec.~\ref{S-C-F_sec}, we introduce a concept for how a
	combination of discontinuous overacceleration and jam-absorption driving might restore free traffic flow at the bottleneck.

	Contrary to traffic instability,
	as we have shown in this review,
	the empirical nucleation nature of traffic breakdown (F$\rightarrow$S transition) 
  is explained by a competition of discontinuous overacceleration with speed adaptation in free flow: 
	\begin{itemize}
\item [--]
  Vehicle overacceleration is  the fundamental microscopic mechanism for traffic breakdown control.
\end{itemize}

 \section{Discussion of Results of Mathematical Modeling of Overacceleration   \label{Dis_S}}

	\subsection{Discontinuous Overacceleration as an Inherent Property of Single-Leader Deterministic Car-Following Models \label{Dis_OA_Feature_S}}  
	
First, we emphasize that Secs.~\ref{Safety_OA_sec} and~\ref{TPACC_St_S} of this review deliberately focus on the phenomenon of discontinuous overacceleration within single-leader deterministic car-following models. By narrowing the scope to these models, we  demonstrate that:
(i) Discontinuous overacceleration inherently occurs in single-leader deterministic car-following models without being explicitly or artificially introduced.
(ii) Overacceleration mechanism driven by safety acceleration  becomes   evident when vehicle overdeceleration is not incorporated into the model under free-flow conditions.
This approach allows us to show that these overacceleration mechanisms are inherent mathematical properties resulting directly from single-leader deterministic car-following formulations, applicable to both human-driven and automated vehicles. 
	
To explain these features of discontinuous overacceleration, we consider   simulation results from Helly's model  (\ref{ACC_Cl})
 (Sec.~\ref{Safety_OA_sec}). Vehicle braking, caused by a slower-moving vehicle merging at the on-ramp bottleneck, leads to a wave of speed decrease at the bottleneck (see the speed decrease for vehicles 1 and 2 for Helly's model in Fig.~\ref{ACC_safety_acc_traj}). 
The subsequent overacceleration of vehicles downstream of the bottleneck (labeled for vehicle 1 as ``1, overacceleration'' in Fig.~\ref{ACC_safety_acc_traj}(a))  
 prevents the upstream propagation of the speed wave.   In this case, due to this self-organized process occurring in space and time, only a local speed decrease remains at the bottleneck without further upstream speed-wave propagation. This occurs at $t < T_{\rm ind}$ in Fig.~\ref{ACC_safety_acc}, i.e., before the traffic breakdown (F$\rightarrow$S transition) is induced.

  Discontinuous overacceleration results from   behavioral assumptions
	in the single-leader car-following incorporated in Helly's model without explicitly introducing of the overacceleration in the model.
	The  behavioral assumptions of the classical Helly's model used for ACC (\ref{ACC_Cl}) are as follows:
	A   vehicle control system seeks to minimize both the relative speed difference and the distance headway error simultaneously.
 As we have proven,
 these  behavioral assumptions   leads to two different model solutions at the same initial flow rates at the bottleneck: 
A solution for free flow and another solution for synchronized flow at the bottleneck (Fig.~\ref{2Z_ACC_short}(a)).   When free flow transforms into synchronized flow, 
  the flow rate upstream of the bottleneck decreases abruptly. This flow rate discontinuity 
  means that the mean time-delay in vehicle safety acceleration (overacceleration) downstream of the bottleneck increases discontinuously when free flow transforms into synchronized flow.
 This is consistent
	with the hypothesis about discontinuous overacceleration   (Fig~\ref{Hyp-OA}(b)) discussed in Sec.~\ref{Cause_OA_sec}.
	This is also consistent
	with  the result of mathematical modeling of Helly's model in which there is the drop in the mean time-headway between the vehicles from
	1.575 s to  1 s occurring when free flow transforms into synchronized flow.

	Qualitatively the same conclusion follows from the analysis of the single-leader car-following traffic flow model of TPACC
	(Sec.~\ref{TPACC_St_S}): Vehicle braking, caused by a slower-moving vehicle merging at the on-ramp bottleneck, leads to a wave of speed 
	decrease at the bottleneck (see the speed decrease for 
	vehicles 1--4 for the TPACC model in Fig.~\ref{Greater_G_traj}).  The subsequent overacceleration of vehicles downstream of the bottleneck (labeled for vehicle 1 as ``1, overacceleration'' in Fig.~\ref{Greater_G_traj}(a))   prevents the upstream propagation of the speed wave. As a result,   only a local speed decrease remains at the bottleneck without further upstream speed-wave propagation. This occurs at $t < T_{\rm ind}$ in Fig.~\ref{Greater_G_induced}, i.e., before the traffic breakdown (F$\rightarrow$S transition) is induced.
	\begin{itemize}
	\item[--] In the single-leader deterministic car-following models considered in Secs.~\ref{Safety_OA_sec} and~\ref{TPACC_St_S}, discontinuous overacceleration follows a local decrease in speed at the bottleneck and the resulting emergence of a vehicle deceleration wave---both temporally and causally.
\end{itemize}
 When the speed decrease within the wave exceeds a critical value (nucleus), overacceleration cannot stop the upstream propagation of the speed wave:
 The F$\rightarrow$S transition occurs at the bottleneck, as shown in Figs.~\ref{ACC_safety_acc_traj} and~\ref{Greater_G_induced} at $t > T_{\rm ind}+\Delta t$.
	
	Thus,  discontinuous overacceleration in the Helly's model for ACC and in  the TPACC model results from
	behavioral assumptions
	in the single-leader car-following incorporated in the   models.
	 \begin{itemize}
	\item[--]
	Behavioral assumptions regarding  overacceleration are  the  natural  consequence of   assumptions incorporated
	in single-leader deterministic car-following traffic flow models considered in Secs.~\ref{Safety_OA_sec}
and~\ref{TPACC_St_S}.
	\end{itemize}
	
	It must be emphasized that discontinuous overacceleration  was explicitly introduced into {\it neither} 
	 the Helly's model for ACC {\it nor}   the TPACC model: 
	Discontinuous overacceleration arises   automatically from the dynamic rules of vehicle motion introduced in these 
	single-leader deterministic car-following traffic flow models. For this reason,
  discontinuous overacceleration can be regarded as an inherent  fundamental  property  of
 these single-leader deterministic car-following traffic flow models.  
 \begin{itemize}
\item[--] Discontinuous overacceleration   occurs   in the single-leader deterministic car-following traffic flow models   without explicitly introducing
it in the model.	
\item[--] Discontinuous overacceleration
becomes  evident  when vehicle overdeceleration is not incorporated into the model
 \textit{under free-flow conditions}. This aligns with real-world traffic observations (see Sec.~\ref{OA-OD_Real-World_S}).
\end{itemize} 
	  
	Almost any single-leader deterministic car-following traffic flow model reflects 
	two core features of real-world traffic: (i) the follower vehicle minimizes the relative speed difference to the preceding (leading) vehicle,
	and (ii) the vehicle decelerates when the time headway falls below a safe time headway, whereas it accelerates when the time headway exceeds a certain maximum acceptable time headway.
	
	In Helly's model for ACC (\ref{ACC_Cl}), the safe and maximum time headways are both equal to the desired time headway $\tau_{\rm d}$. In contrast, in the TPACC model
	(\ref{g_v_g_min1})--(\ref{Helly_st3}), the maximum time headway corresponds to the synchronization time headway $\tau_{\rm G}$, which is larger than 
	the safe time headway $\tau_{\rm safe}$. From the mathematical simulations of the Helly's ACC model and the TPACC model, we may assume that if another
	single-leader deterministic car-following traffic flow model does not exhibit overdeceleration under free-flow conditions, then
	for the    model discontinuous overacceleration becomes evident within 
	some range of the flow rates at the bottleneck. In this case, the
	competition between this overacceleration and the speed adaptation effect ultimately 
	determines the nucleation characteristics of traffic breakdown (the F$\rightarrow$S transition) at the bottleneck.
	This assumption may be valid for the many classical single-leader deterministic car-following traffic flow models.
	The  justification of this assumption---which we have demonstrated above for Helly's ACC model and the TPACC model---remains outside the scope of this review. This represents an interesting avenue for future traffic research.

\subsection{Inconsistent Interpretation of The Concept ``Overacceleration'' \label{Term_OA_OA_S}}

Taken literally, the term ``overacceleration'' suggests ``excessive acceleration,'' which can easily introduce ambiguity. In particular, it might be mistakenly interpreted as describing situations where drivers accelerate more aggressively than necessary. Within the framework of three-phase traffic theory, however, the concept of vehicle overacceleration has a precise theoretical meaning that differs from common intuition, which can sometimes lead to inconsistent or erroneous interpretations.

For instance, one might incorrectly view discontinuous overacceleration as an independent, exogenous trigger that initiates traffic breakdown. Such a misconception inevitably invites a fundamental causality paradox---the classic ``chicken-and-egg dilemma.'' Under that flawed assumption, a sound analytical argument would indeed be required to prove that overacceleration chronologically and causally \textit{precedes} localized drops in speed or deceleration waves.

However, the mathematical modeling of the single-leader deterministic 
car-following traffic flow models in Secs.~\ref{Safety_OA_sec} and~\ref{TPACC_St_S} (see discussions in Sec.~\ref{Dis_OA_Feature_S}) directly refutes this mistaken interpretation. The modeling results yield the following rigorous conclusions regarding the nature of the phenomenon
\textit{discontinuous overacceleration}:
\begin{itemize}
\item[--] Discontinuous overacceleration is \textit{not} an independent, exogenous trigger introduced external to the traffic flow dynamics.
\item[--] Instead, it is an \textit{endogenous} traffic phenomenon that emerges from the complex spatiotemporal competition between vehicle overacceleration and speed adaptation.
\item[--] Within single-leader deterministic car-following models, discontinuous  overacceleration temporally and causally \textit{follows} initial local drops in speed or deceleration waves.
\item[--] Consequently, the fundamental causality paradox (the ``chicken-and-egg dilemma'') does not apply to this phenomenon. Overacceleration is neither a random exogenous catalyst nor a minor secondary symptom; rather, it is the decisive dynamic mechanism that dictates whether an existing deceleration wave is amplified into a full traffic breakdown (F$\rightarrow$S transition) or successfully dissipated.
\end{itemize}

\subsection{Discontinuous Overacceleration as a Critical Self-Organized Spatiotemporal Traffic Phenomenon} \label{Dis_OA_Sp-Tem_S}
	
The spatiotemporal interplay between speed adaptation and discontinuous overacceleration---as observed in the single-leader deterministic
 car-following models (Secs.~\ref{Safety_OA_sec} and~\ref{TPACC_St_S})---allows overacceleration to be understood as a \textit{critical self-organized} spatiotemporal traffic phenomenon. 

Indeed, vehicle braking induced by a slower vehicle merging at an on-ramp bottleneck generates a wave of localized speed decrease upstream of the bottleneck. The subsequent overacceleration of vehicles at the bottleneck acts to counteract and suppress the upstream propagation of this deceleration wave. Overacceleration successfully prevents this propagation as long as the initial speed drop within the wave remains below a critical threshold. In this scenario, due to this self-organized spatiotemporal process, only a localized speed decrease persists at the bottleneck, without triggering upstream wave propagation. However, when the amplitude of the speed decrease exceeds a critical threshold (serving as the nucleus for traffic breakdown), overacceleration can no longer arrest the upstream propagation of the deceleration wave, resulting in the F$\rightarrow$S transition at the bottleneck.

A considerably more complex manifestation of this critical 
self-organized spatiotemporal phenomenon occurs in multi-lane traffic (Sec.~\ref{Lane_OA_sec}). In this case---in addition to the baseline competition between the upstream-propagating deceleration wave and the dampening effect of overacceleration attempting to restore free-flow conditions---the overacceleration mechanism induced by lane-changing maneuvers contributes significantly to the emergent dynamics (Figs.~\ref{Free_Flow_Bottl_tr} and~\ref{Induced_F_S_Bottl_tr}).

Specifically, after vehicle 2 changes from the right to the left lane (denoted as ``2-left (overacceleration)'' 
in Fig.~\ref{Free_Flow_Bottl_tr}(d)), two competing effects emerge:
\begin{enumerate}
\item[(i)] The speed adaptation of vehicle 4 in the left lane as it follows vehicle 2-left (``4, speed adaptation'' in Fig.~\ref{Induced_F_S_Bottl_tr}(a)).
\item[(ii)] The overacceleration of vehicle 3 in the right lane (``3, overacceleration'' in Fig.~\ref{Induced_F_S_Bottl_tr}(b))~\footnote{To maintain a focused scope for this review, we omit further details of the self-organized spatiotemporal features of overacceleration on two-lane roads. Notably, 
it was shown in~\cite{KKl2025A} that both speed adaptation and overacceleration within a single lane can qualitatively alter the overacceleration
 mechanism triggered by lane-changing. In particular, when no overacceleration exists in a road lane,
there is also no overacceleration through lane-changing and,
respectively, traffic breakdown does not exhibit the nucleation
character.}.
\end{enumerate}
 
This interaction demonstrates that the outcome of the competition between the upstream propagation of a deceleration wave and the opposing stabilization effect of overacceleration depends, in the general case, on the collective overacceleration and deceleration behaviors of multiple vehicles both upstream and downstream of the bottleneck. This further supports the framework wherein discontinuous overacceleration is fundamentally classified as a critical, self-organizing spatiotemporal traffic phenomenon.

\subsection{Overacceleration versus Overdeceleration  -- \newline Theoretical  Highway Capacities \label{OA-OD_Cap_S} }

In   standard traffic models, in which traffic instability  
determines traffic dynamics (an overview of this {\it traffic flow dynamics},
derived from theoretical studies and simulations of standard traffic models, can be found in the book by Treiber and Kesting~\cite{Treiber-Kesting}), there are at least two characteristic 
flow rates in free flow: (i) A  maximum flow rate at which  traffic instability occurs spontaneously
 in free flow at a bottleneck\footnote{It must be emphasized  that a
 simulated value $q_{\rm max}$ can depend considerably on
the flow rates  at the bottleneck as well as on bottleneck parameters~\cite{Treiber-Kesting,Helbing}.}  denoted by $q_{\rm max}$.
(ii) A minimum flow rate denoted by $q_{\rm min}$  at which moving jams can still exist.
The  minimum  flow rate $q_{\rm min}$
can be considerably less than $q_{\rm max}$. The   flow rates $q_{\rm min}$ and
  $q_{\rm max}$
 determine a range of theoretical highway capacities of free flow at the bottleneck found     in
simulations of standard traffic models~\cite{Treiber-Kesting,Helbing}.
As mentioned  in Sec.~\ref{Overdeceleration_S},  the standard traffic models assume that traffic instability due to vehicle overdeceleration is
the cause for traffic breakdown; in turn, the traffic instability leads to the F$\rightarrow$J transition, i.e.,
moving jam emergence in free flow.  In other words, theoretical highway capacities of free flow at the bottleneck found     in
simulations of standard traffic models  are  related to the theoretical flow-rate range
 within which moving jams can occur or been induced in free flow at the bottleneck. 
 
 It must be emphasized that the theoretical highway capacities of free flow at the bottleneck     found in simulations of
  the standard traffic models~\cite{Treiber-Kesting,Helbing},
have {\it no} relation to the maximum and minimum 
highway capacities $C_{\rm min}$ and $C_{\rm max}$ of three-phase traffic theory~\cite{KernerBook1,KernerBook2,KernerBook3,KernerBook4}.
\begin{itemize}
\item [--] Features of  theoretical highway capacities of free flow at the bottleneck found in
the standard traffic models~\cite{Treiber-Kesting,Helbing} are determined by
   traffic instability 
caused by vehicle
{\it overdeceleration}.
\item [--] Contrarily,  in three-phase traffic theory
 the maximum and minimum 
highway capacities $C_{\rm min}$ and $C_{\rm max}$ of free flow at the bottleneck are determined   
by a competition between discontinuous vehicle 
{\it overacceleration} and speed adaptation (see Sec.~\ref{Z_highway_C_Sec}).
\end{itemize}
 As shown in this review, vehicle overdeceleration leading to traffic instability shows
totally  opposite effect on traffic flow dynamics in comparison with  vehicle overacceleration.
\begin{itemize}
\item   [--]
Theoretical highway capacities found
in the theory of traffic flow dynamics of the standard traffic models~\cite{Treiber-Kesting,Helbing},
have {\it no} relation to the maximum and minimum 
highway capacities $C_{\rm min}$ and $C_{\rm max}$ of three-phase traffic theory.
\end{itemize}

 \subsection{About Possible Contribution of
 Overdeceleration  To  Nucleation of F$\rightarrow$S Transition   at Bottleneck: Real-World Traffic \label{OA-OD_Real-World_S} }

The presented results of mathematical modeling demonstrate the sufficiency of the overacceleration mechanism within the proposed modeling framework. 
However, they do not fully establish the non-necessity of overdeceleration for
the nucleation of traffic breakdown in real traffic systems. Demonstrating that traffic breakdown can be reproduced without overdeceleration in a model is not equivalent to showing that overdeceleration does not participate in traffic breakdown in reality. In  real-world traffic, reaction delays, car-following errors, lane-changing disturbances, abrupt braking maneuvers, and ACC control imperfections may coexist. 	As a result, overdeceleration and overacceleration may interact rather than representing completely separable mechanisms. Indeed, theoretical studies~\cite{Kerner2019AA} show that overdeceleration and overacceleration interact strongly with one another \textit{in synchronized flow}:  this non-linear self-organizing effect leads to a spatiotemporal 
competition between S$\rightarrow$F  and S$\rightarrow$J  instabilities~\footnote{As noted in Sec.~\ref{Focus_S},
	this topic lies outside the scope of this review.}.

Therefore,   the   question arises as to whether  and how overdeceleration  \textit{in  free flow}  can  contribute
to the nucleation of the F$\rightarrow$S transition at a bottleneck in real-world traffic.

 \subsubsection{Macroscopic real-world spatiotemporal traffic data \label{Mac_Real-World_S}}

Since 1995, the author had the opportunity to study a huge volume of  \textit{macroscopic} spatiotemporal traffic data. These data, measured by road detectors installed on many German and Dutch highways, continue to be collected today. 
In these data, we did not observe traffic instability occurring spontaneously (i.e., without an accident) in free flow at bottlenecks~\cite{KernerBook1}.
Thus, no contribution of overdeceleration to the nucleation of a spontaneous free-to-synchronized flow transition (F$\rightarrow$S transition)
  at bottlenecks is measurable in macroscopic data.
First, the F$\rightarrow$S transition has occurred. Only later, within the synchronized flow, could traffic instability be observed, which sometimes led to the formation of wide moving jams~\cite{KernerBook1}. This empirical result---that 
moving jams occur through a sequence of F$\rightarrow$S$\rightarrow$J transitions---initiated the development of the three-phase traffic theory
in which the concept ``discontinuous overacceleration'' was introduced (see Sec.~\ref{3_OA_Br_Sec}). 

\subsubsection{Microscopic   spatiotemporal traffic data of probe vehicles \label{Prob_Real-World_S}}

Since approximately 2008, another opportunity arose to study a vast amount of \textit{microscopic} spatiotemporal traffic data. These data are collected via probe vehicles across many industrialized regions (Europe and the USA) where traffic services have been developed.
Although the field measurements    revealed isolated instances where a few vehicles exhibited overdeceleration under free-flow bottleneck conditions,
 we could find no continuous development of traffic instability leading to the emergence of moving jams
  in free flow at the bottlenecks~\cite{KernerBook4}. 
	
	We can explain this empirical result as follows:
	At least some of the vehicles, which follow  the overdecelerating vehicle, exhibit
speed adaptation within the indifferent zone for car-following (Sec.~\ref{Cause_Adap_sec}). 
 The speed adaptation within the indifferent zone interrupts the instability 
development~\footnote{Furthermore, the interruption of the development of traffic instability 
can be caused by overacceleration during a lane change (Sec.~\ref{Lane_OA_sec}). However, the effects of lane changes cannot
 be usually resolved using data from probe vehicles.}; 
as a result, although a few vehicles can exhibit overdeceleration,  rather than traffic instability, the F$\rightarrow$S transition occurs.
This can explain why in real-world traffic measured via probe vehicle data
 moving jams consistently occur spontaneously through a sequence of F$\rightarrow$S$\rightarrow$J transitions.

It should be emphasized that although the share of probe vehicle data is only about 4--8\%, nevertheless, due to the
study a huge volume of   probe vehicle data measured during many years on various highways in many countries we can make the following 
conclusions~\cite{KernerBook4}:
\begin{itemize}
\item[--] \textit{In synchronized flow} at the bottleneck, vehicle overdeceleration is frequently observed. 
Vehicle overdeceleration of the following vehicles
 leads often to  traffic instability with moving jam emergence.
\item[--] \textit{In free flow}  at the bottleneck, vehicle overdeceleration   can  also occasionally be observed.
Contrary to synchronized flow at the bottleneck, in free flow no  continuous development of traffic instability and, consequently, no moving
 jams emergence is observed.
\end{itemize}

\subsubsection{Microscopic measurements of all vehicle trajectories}

Since the measurement of all vehicle trajectories in the NGSIM project~\cite{NGSIM,NGSIM2,NGSIM3}, 
several other datasets---such as I-24 MOTION~\cite{I-24 MOTION}, drone-based trajectory datasets HighD~\cite{HighD,HighD2,HighD3},  
   MiTra~\cite{MiTra}, SPT~\cite{SPT},  pNEUMA~\cite{pNEUMA} as well as~\cite{Kaufmann2018A}---have been realized for microscopic studies of phase transitions in real-world 
	traffic~\footnote{We do not consider here experiments with ACC/CACC platoons. This limitation is explained as follows:
	The most available real-world traffic data that are related to traffic flow of human-driven vehicles show that
the nucleation of traffic breakdown at bottlenecks is caused by the discontinuously in overacceleration,
 not due to vehicle overdeceleration and resulting traffic instability. However, the dynamics of automated vehicles
in the ACC/CACC platoons can be chosen totally different in comparison with the   dynamics of
	human-driven vehicles in real-world traffic. 
	Specifically, the choice of dynamic coefficients in ACC/CACC systems can lead to string instability, driven by the overdeceleration behavior of automated vehicles.
In experiments with 100\% string-unstable ACC/CACC, moving traffic jams can occur spontaneously at bottlenecks due to this string instability. As we have stated in   Sec.~\ref{Meth_Int_S},  in models of ACC- and TPACC-vehicles that we consider in this review  no string instability occurs. In other words, a consideration of string-unstable ACC/CACC systems  is out of the scope of the review.}.
The literature includes numerous studies  of moving jam emergence and propagation in synchronized flow in
 these datasets (see, e.g.,~\cite{Treiber-Kesting,KKl2009A,MiTra,Coifman2017,Punzo2011,Thiemann2008,Zhengbing-He2017,Li_Jiang2020,Leclercq2007A,MontaninoPunzo2013,Kesting2008AA,ThiemannKesting2008A,Li-Huang2020A,Yi-He-Bo2022A,RajuArkatkar2022A,Yang-Wan2024A,Sharma-Zheng2018A}).
In particular, the data confirm the conclusion made above from studies of probe vehicle data: \textit{In synchronized flow} at the bottleneck, vehicle overdeceleration is frequently observed; vehicle overdeceleration of the following vehicles
 leads often to  traffic instability with moving jam emergence.

However, we could not find any microscopic empirical studies of traffic breakdown \textit{in initial free flow} at bottlenecks that provide a scientifically sound conclusion regarding the potential contribution of vehicle overdeceleration to the F$\rightarrow$S transition. This lack of empirical evidence has a clear explanation: very few microscopic data are available for studying traffic breakdown, 
because the nucleation of the transition from free flow to synchronized flow at a bottleneck lasts only about 0.3--1 min.
In contrast, a large amount of microscopic data is available for studying the emergence and propagation of moving jams in synchronized flow, even when measured on a single highway section on a specific day; these data can be observed throughout the entire duration of bottleneck congestion, which typically lasts from 30 to 150 minutes or longer.
Thus, deriving scientifically sound statistical conclusions from these microscopic data regarding the potential contribution of vehicle overdeceleration to the nucleation of the  F$\rightarrow$S transition (traffic breakdown) in initial free flow at bottlenecks---which is outside the scope of this review---remains an interesting task for future traffic research.
 
We can draw  the following conclusion from the studies of real-world spatiotemporal traffic data:
\begin{itemize}
\item[--] Regardless of whether vehicle overdeceleration contributes   to traffic breakdown nucleation, most available
 real-world traffic data show that the nucleation of traffic breakdown (the F$\rightarrow$S transition) at bottlenecks is caused by a discontinuity in overacceleration rather than overdeceleration. That is, the F$\rightarrow$S transition is \textit{not} triggered by the traffic instability incorporated into most standard traffic flow models.
\end{itemize}

\subsection{About Simulations of Possible Contribution of
 Overdeceleration  To  Nucleation of F$\rightarrow$S Transition   at Bottleneck  \label{OA-OD_Modeling_S} }

In the traffic flow model for human-driven vehicles---defined
 by Eqs.~(\ref{g_v_g_min1_ad})--(\ref{Helly_st2_ad}), (\ref{pinch-formula}), (\ref{g_safe-min}), and (\ref{G-min}) 
in Sec.~\ref{Gen_Model_Sec}---we have incorporated a fundamental empirical observation: moving jams can occur spontaneously only within synchronized flow, rather than in free flow (see Sec.~\ref{OA-OD_Real-World_S}). 
To mathematically realize this model feature,   the speed threshold $v_{\rm pinch}$  is introduced
in model (\ref{g_v_g_min1_ad})--(\ref{Helly_st2_ad}), (\ref{pinch-formula}), (\ref{g_safe-min}), and (\ref{G-min}). 
Condition (\ref{pinch-formula}) restricts the choice of the model parameter $v_{\rm pinch}$:
we have found that  as long as condition (\ref{pinch-formula}) holds,
 increasing $v_{\rm pinch}$ does not qualitatively change the GP formation (Fig.~\ref{overreaction-OA}(b)). Consequently, this model for human-driving vehicles
 cannot be used to study the potential effect of overdeceleration on the nucleation of traffic breakdown. At any permissible value of $v_{\rm pinch}$, no overdeceleration occurs in free flow at the bottleneck. This explains why the time-delay $T^{\mathrm{(B)}}$ of traffic breakdown remains 
robust against variations of $v_{\rm pinch}$ in model (\ref{g_v_g_min1_ad})--(\ref{Helly_st2_ad}), (\ref{pinch-formula}), (\ref{g_safe-min}), and (\ref{G-min}).

Contrary to model (\ref{g_v_g_min1_ad})--(\ref{Helly_st2_ad}), (\ref{pinch-formula}), (\ref{g_safe-min}), and (\ref{G-min}), 
both overacceleration and overdeceleration exist at free-flow speed in the Kerner-Klenov stochastic three-phase traffic 
flow model~\cite{KKl,KKl2003A,KKl2009A}. In simulations of this model, overdeceleration can occur in free flow when the space-gap $g$ of a vehicle at the bottleneck becomes considerably smaller than the safe space-gap $g_{\mathrm{safe}}$. Nevertheless, we do  \textit{not}  observe a continuous growth of the deceleration wave upstream of the bottleneck in free flow; that is, no traffic instability caused by overdeceleration is found. This can be explained as follows: at least some of the vehicles following the overdecelerating vehicle exhibit speed adaptation within the indifferent zone for car-following (Sec.~\ref{Cause_Adap_sec}). This speed adaptation within the indifferent zone interrupts the development of the instability. This mechanism is intended to reproduce the empirical finding that, in the data analyzed so far, no continuous development of traffic instability has been observed in free flow at bottlenecks (Sec.~\ref{OA-OD_Real-World_S}).

	Unfortunately, the Kerner-Klenov model does not permit a scientifically rigorous study of
	whether overdeceleration contributes  to the nucleation of the F$\rightarrow$S transition (traffic breakdown) at a bottleneck.
	This limitation arises because both overacceleration and overdeceleration are incorporated via model fluctuations. 
	The parameters governing these fluctuations influence all simulation outcomes---including the emergence of moving jams in synchronized flow---rather 
	than specifically isolating the features of the F$\rightarrow$S transition.
	
	Consequently, we can conclude that no empirical or theoretical traffic studies to date have rigorously 
	investigated the potential contribution of overdeceleration to the nucleation of the F$\rightarrow$S transition at a bottleneck. This subject, which is out of the scope of this review, remains a compelling objective for future traffic research.
	However, as we have stressed in Sec.~\ref{OA-OD_Real-World_S},
	regardless of whether vehicle overdeceleration contributes  to traffic breakdown nucleation, most available
 real-world traffic data show that the nucleation of traffic breakdown (the F$\rightarrow$S transition) at bottlenecks is caused by a discontinuity in overacceleration rather than overdeceleration; that is, the F$\rightarrow$S transition is  not  initiated by the traffic instability incorporated into most standard traffic flow models.

\subsection{Commonality and Individuality of Different Overacceleration Mechanisms \label{OA-Commonality_S}}

We have emphasized that there is a cooperation of different overacceleration mechanisms     (Secs.~\ref{Adv_TPACC_model_sec}
and~\ref{Lane_OA_sec}). The different overacceleration mechanisms
exhibit both a commonality and an individuality.

The different overacceleration mechanisms share the following commonalities: (i) They satisfy the definition of overacceleration (Sec.~\ref{Definition_OA_sec}); specifically, an overacceleration mechanism represents vehicle acceleration behavior   that  causes 
the free flow
metastability with respect to the F$\rightarrow$S transition at the
bottleneck. (ii) Each overacceleration mechanism aims to recover the maximum speed allowed under free-flow conditions.

The  individuality of 
 the overacceleration mechanism is associated with different physical/behavioral effects in vehicular traffic:
 (i) Safety acceleration   tries to suppress the upstream propagation of
a speed deceleration wave occurring
at the bottleneck. (ii)  A lane change out of the right lane  causes vehicles following in that lane to accelerate; this promotes free traffic
 flow and limits the upstream propagation of the   wave of deceleration originating at the bottleneck.

 We must emphasize that overacceleration mechanisms---via both safety acceleration and lane-changing---emerge
 naturally without being explicitly introduced into the models of
Secs.~\ref{Safety_OA_sec} and~\ref{TPACC_St_S}. 
There is  \textit{no}  stronger tendency to accelerate than to decelerate among drivers or automated systems in response to upstream speed disturbances.
Moreover, the models do  \textit{not}  assume a greater preference for free-flow conditions over compliance with safe spacing. 
We account for  \textit{no}  information from vehicles further ahead of the preceding vehicle. Furthermore, \textit{no} driver psychological speed thresholds 
	and \textit{no} non-linear bounded rationality are applied.  
Finally, \textit{no} other model extensions---such as heterogeneous driving behavior, randomness, or multi-phase velocity-spacing relationships---are incorporated.
\begin{itemize}
\item[--] These overacceleration mechanisms are inherent model properties: they result
 directly from mathematical car-following traffic flow models for both human-driven and automated vehicles.
\item[--] Overacceleration mechanisms driven by safety acceleration and lane-changing 
become  evident  when vehicle overdeceleration is not incorporated into the model
 \textit{under free-flow conditions}. This aligns with real-world traffic observations
 (Sec.~\ref{OA-OD_Real-World_S}) and follows the approaches taken in~\cite{Kerner2023B,KKl2025A} as well as in this review. 
\end{itemize}

We assume that there can be different mechanisms of overacceleration 
that  are associated with  other  physical/behavioral effects in vehicular traffic that are not incorporated in elementary 
mathematical models considered in this review.   
Analysis of such mechanisms of overacceleration that is out of the scope of this review
can be an interesting task for future traffic research.

	  \subsection{Why Do Empirical Microscopic Validations of Standard Models Fail to  Uncover Overacceleration Mechanisms?
		\label{Valid_OA_Model_S}}

	In this review, we have argued that vehicle overacceleration is a fundamental microscopic characteristic governing traffic breakdown. A widely used methodology for validating assumptions about driver behavior relies on empirical vehicle trajectory data
	(see, e.g.,~\cite{Treiber-Kesting,Brockfeld2003,NaLinZong2014A,Xianda_Chen,Montanino-Punzo2015A}). However, when such assumptions are implemented in standard traffic flow models, 
	they fail to reproduce the empirically observed nucleation nature of traffic breakdown (F$\rightarrow$S transition) 
	at bottlenecks (Fig.~\ref{20041998_MSP}(a)).

One possible explanation is that discontinuous vehicle overacceleration, as introduced in three-phase traffic theory, makes no sense within the framework of standard traffic theories and models. This raises a fundamental question: which aspects of real traffic dynamics are not captured by the prevailing methodology of validating driver behavior solely through empirical trajectory data?

A qualitative explanation can be proposed as follows. Traffic is inherently a spatiotemporal phenomenon. Therefore, prior to validating a traffic model using microscopic trajectory data, it is essential to verify whether the model can reproduce empirical spatiotemporal traffic patterns. In particular, the model should be capable of simulating the nucleation nature of traffic breakdown (F$\rightarrow$S transition) at bottlenecks. Only after such macroscopic and spatiotemporal consistency is established does it become meaningful to use microscopic trajectory data for parameter calibration.
A comprehensive investigation of this issue lies beyond the scope of this review but represents an important direction for future traffic research.

\section{Some Ideas to Microscopic  Overacceleration Management  Through     Automated Vehicles  and AI \label{S-F_gen_sec}}

 \subsection{About Strategies for Traffic Breakdown Prevention 
\label{Ad_Abs_Sec}}

In empirical traffic data measured in real-world traffic, traffic breakdown at a bottleneck is the F$\rightarrow$S transition that exhibits the nucleation nature.
Real-world traffic data show that the nucleation of the F$\rightarrow$S transition is not caused by traffic instability---i.e., a continuously growing local wave of speed decrease propagating upstream of a bottleneck
(Sec.~\ref{OA-OD_Real-World_S}). 
Instead of the growing moving jam that results from traffic instability, real-world traffic data reveal  synchronized flow in which moving jams do not necessarily emerge (Fig.~\ref{20041998_MSP}). 

This explains the criticism raised in Sec.~\ref{Overdeceleration_S} regarding the application of ``jam absorption strategies'' for traffic breakdown control. Here, a question arises: Within the framework of three-phase traffic theory, what fundamental adjustments are needed for traffic breakdown absorption strategies?

To address this question, we should first recall that three-phase traffic theory 
explains the nucleation nature of traffic breakdown at bottlenecks through a spatiotemporal 
competition between discontinuous overacceleration and the opposing effect of speed adaptation. 
While overacceleration drives a tendency toward free flow at the bottleneck, 
speed adaptation promotes a transition toward synchronized flow (Sec.~\ref{Definition_OA_sec}). 
When overacceleration outweighs speed adaptation on average in the immediate vicinity of the bottleneck, 
traffic breakdown is prevented. Conversely, if speed adaptation dominates, traffic breakdown occurs.
 
Consequently, traffic breakdown can potentially be prevented using congestion management strategies implemented via automated vehicles and AI. 
To achieve this, in accordance with three-phase traffic theory, these management strategies should aim to:
\begin{itemize}  
\item [--] Increase vehicle overacceleration   in the immediate vicinity upstream and downstream of the bottleneck.  
 \end{itemize}
Below we explore some ideas for such management strategies~\footnote{To our knowledge, no  studies have yet explored using automated vehicles and AI to increase vehicle overacceleration   near bottlenecks. This remains an important area for future traffic research.
}.

\subsection{Microscopic  Overacceleration Management 
\label{S-F_sec}}

As shown in this review, for the   control of traffic breakdown at the bottleneck,
automated vehicles should control the competition between discontinuous vehicle overacceleration and speed adaptation
 at the bottleneck, {\it not} traffic instability. We   define $\lq\lq$microscopic overacceleration management" as follows:
 \begin{itemize}
\item [--]
   Microscopic overacceleration management is the 
	control of the competition between discontinuous overacceleration and speed adaptation together with
	the control of speed disturbances at the bottleneck.
\item [--] The objective
of this microscopic overacceleration management depends on the current traffic state
 at the bottleneck: 
\begin{itemize}
\item [(i)] If free flow still exists at   the bottleneck, then  
 traffic breakdown (F$\rightarrow$S transition) should be prevented.
\item [(ii)] If synchronized flow   exists at   the bottleneck, then   a return S$\rightarrow$F transition should be forced.
\end{itemize}
 \end{itemize}
Microscopic overacceleration management can be achieved through 
 cooperative driving of automated vehicles near the bottleneck using AI models.
 By incorporating
extensive, real-world datasets previously measured at this bottleneck,
AI models can identify the current phase of data traffic using real-time data.

If the free traffic flow phase is currently located at the bottleneck, AI models can suggest which behavior (e.g., lane changing, accelerating, braking, or maintaining the current speed) should be performed by each automated vehicle near the bottleneck. This can increase overacceleration, whereas the speed adaptation effect does not increase significantly within a local speed decrease at the bottleneck. AI models can also suggest the timing of the appropriate action for the automated vehicle, as well as the magnitude and duration of the acceleration or braking.

 For example, based on the results in Sec.~\ref{Lane_OA_sec}, it can be assumed that if a sufficiently large gap in the left lane is detected, an automated vehicle changing from the right to the left lane can increase overacceleration in the right lane without causing significant speed adaptation in the left lane.
However, we are currently unaware of any studies that have investigated these and other possible methods for controlling microscopic overacceleration.
In other words, the aforementioned proposals for controlling the free-flow phase at bottlenecks could represent very interesting tasks for future traffic research.

If synchronized flow exists at the bottleneck, a return S$\rightarrow$F transition should be forced through
microscopic overacceleration management.
Simulations of the return S$\rightarrow$F transition, shown below in Sec.~\ref{S-F1_sec}, can
illustrate how microscopic overacceleration management can be performed.

\subsection{Potential Cooperation of Overacceleration Management and Jam Absorption Driving to Restore  
 Free Flow     
\label{S-C-F_sec}} 

As explained above, rather than traffic instability, the cause of
 traffic breakdown is associated with a competition between discontinuous overacceleration with speed adaptation.
For this reason,  approaches such as $\lq\lq$jam absorption driving" (also referred to as dissipation of stop-and-go waves, stop-and-go wave suppression, or shockwave damping), which try to prevent
 traffic instability, are not capable for controlling  traffic breakdown.

Nevertheless, we can assume that overacceleration management in cooperation with  jam absorption driving could be
useful for the prevention of GP formation (Fig.~\ref{not-driver-overreaction_short}(b)) as follows. We assume that
the use of overacceleration management has failed to prevent traffic breakdown (F$\rightarrow$S transition) at the bottleneck. 
Then, as explained in Sec.~\ref{GP_Common_in_theories}, moving jams emerge in synchronized flow, i.e., the GP appears (Fig.~\ref{not-driver-overreaction_short}(b)). 
Now, through  the application of
	jam absorption driving of automated vehicles, we try to 
	  dissolve moving jams in the synchronized flow.
	In this case, the GP in Fig.~\ref{not-driver-overreaction_short}(b)  transforms into the WSP shown in Fig.~\ref{not-driver-overreaction_short}(a).
	Later, overacceleration management can be applied with the aim of the initiating of
	the  S$\rightarrow$F instability at the bottleneck. Such a cooperation of jam absorption driving with
	the following overacceleration management	 can be used to initiate a return  
S$\rightarrow$F transition at the bottleneck.

\begin{figure} 
 \begin{center}
\includegraphics[width = 8 cm]{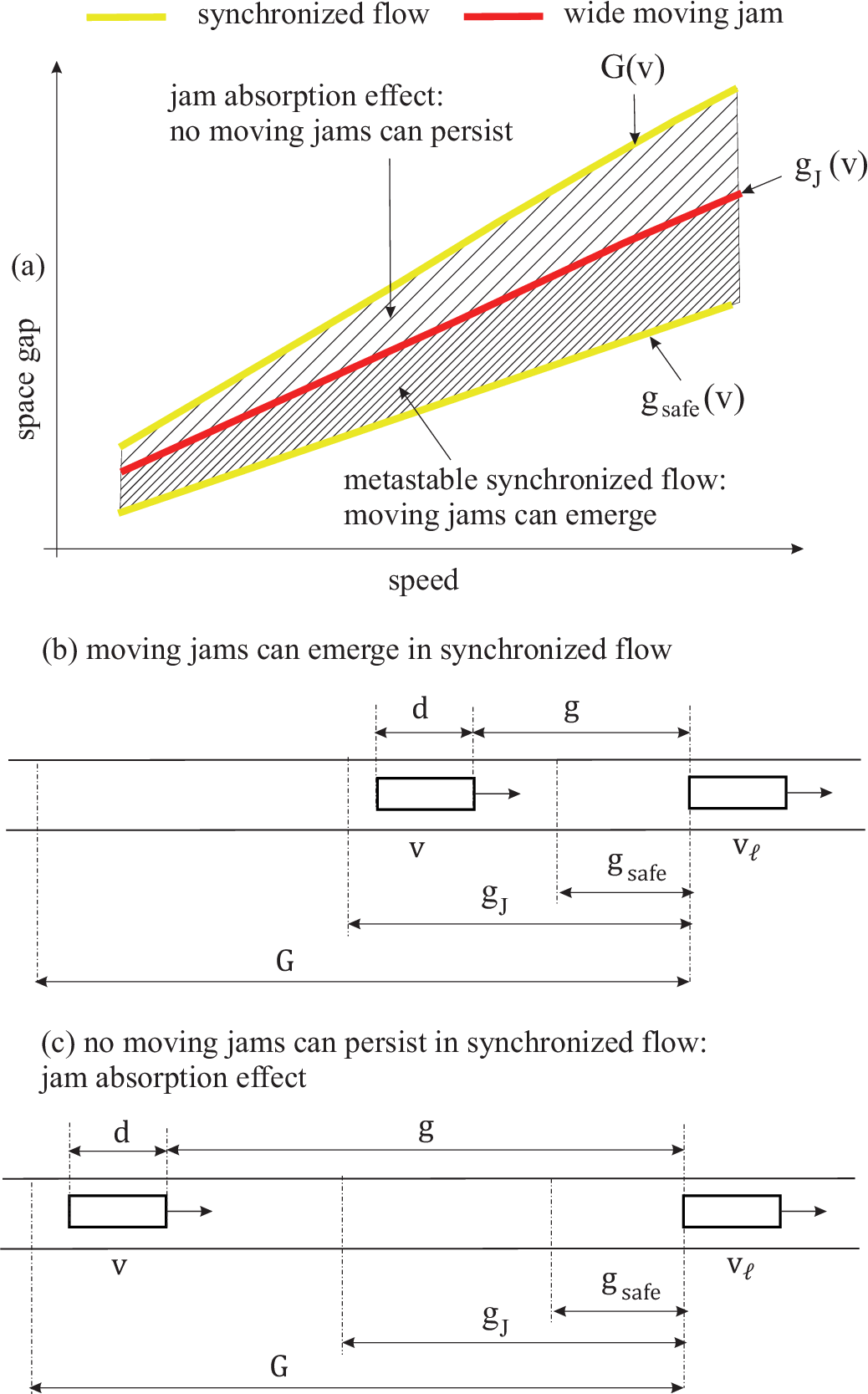}
\caption{Qualitative explanation of jam absorption effect in the three-phase traffic theory~\cite{KernerBook1,Kerner2012A_Jam,Kernerhem2013A_Jam}. (a) Two classes of steady 
states of synchronized flow
 (dashed 2D-regions) in the  space-gap--speed plane; red curve $g_{\rm J}(v)$  presents characteristic parameters of wide moving jam propagation in the  space-gap--speed plane;
yellow curves  $G(v)$ and $g_{\rm safe}(v)$ are related to speed-dependencies of synchronization   and safe space gaps, respectively.
(b, c)  Schemes of car-following within
indifferent zone  (\ref{indif_g}) under condition $g<g_{\rm J}$ (b) and under condition  $g>g_{\rm J}$ (c) 
for steady states of synchronized flow. $g$ is the space gap between two vehicles. 
}
\label{Line_J_g}
\end{center}
\end{figure}

To explain how the application of jam absorption driving can be used for the transformation of  
the GP (Fig.~\ref{not-driver-overreaction_short}(b))   into the WSP (Fig.~\ref{not-driver-overreaction_short}(a)),
we consider briefly the jam absorption effect in the framework of three-phase traffic theory (Fig.~\ref{Line_J_g}).
 As known~\cite{KernerBook1}, in synchronized flow states that are
  above the curve $g_{\rm J}(v)$  (Fig.~\ref{Line_J_g}(a)) no wide moving jams can be induced or persist over time.
	Therefore, if vehicles maintain on average a large enough
 space-gap
that satisfies conditions (\ref{indif_g}) and $g (v)>g_{\rm J}(v)$, all moving jams dissolve
	and only synchronized flow remains.  
  Through this jam absorption effect, the GP (Fig.~\ref{not-driver-overreaction_short}(b))   transforms into
 the WSP (Fig.~\ref{not-driver-overreaction_short}(a)).  

 The fact that the classical traffic flow instability
occurs only when the average space gap in traffic flow is small enough (the vehicle
 density is large enough) was well-known from the classical papers of
 Herman, Gazis, Rothery, Montroll,  
Chandler, and Potts~\cite{GM_Com1,GM_Com2,GM_Com3} as well as
Kometani and Sasaki~\cite{KS,KS1,KS2,KS4}. In other words, it was clear
that no traffic flow instability in traffic flow, in which the average space gap between vehicles is  large enough, can occur.
Therefore, the consideration of the jam absorption effect made here
can be considered the application of this well-known feature of the classical traffic flow
 instability~\cite{GM_Com1,GM_Com2,GM_Com3,KS,KS1,KS2,KS4,Nagatani_R,Gazis,Treiber-Kesting,Schadschneider,Chowdhury,Helbing,Bando_2,KK1994,Helbing1999A,Helbing2001A,Helbing2002A,SternCui2018,WangJin2023,HanWang2021,NishiTomoeda2013,TaniguchiNishi2015,Nishi2020A,HeZheng2017,LiYanagisawa2024,LiuZheng2025,ZhengZhang2020,WangLi2022,SuzukiNishi2026,ZhengbingHe}
for 2D-region of synchronized flow states of the three-phase traffic theory~\cite{KernerBook1,Kerner2012A_Jam}.

As explained above, a cooperation of overacceleration management and jam absorption driving is a possible option
for congestion mitigation. However, it must be emphasized that when standard traffic models are used for simulations, then incorrect
conclusions   about state of traffic flow resulting from the application of the jam absorption driving could be expected.
To explain this statement, we should mention that
 synchronized flow  resulting from the F$\rightarrow$S transition exhibits very complex   spatiotemporal  traffic dynamics.
In this traffic dynamics, a competition between the S$\rightarrow$F instability and the S$\rightarrow$J instability 
has a great importance~\cite{Kerner2019AA}.  
	In the three-phase traffic theory,	the S$\rightarrow$F instability is  explained by   discontinuous
	vehicle overacceleration. However,  discontinuous
  overacceleration makes no sense for    the standard traffic theories and models.  For this reason,
	simulations of the proposed above cooperation of overacceleration management and jam absorption driving
	should be made with a microscopic model in the framework of the three-phase traffic theory.
However, we are currently unaware of 
  studies that have investigated these and other possible methods for the dissolving of the GP with
	initiating of
 the  S$\rightarrow$F instability at the bottleneck.
In other words, the aforementioned proposals for the recovering of free flow
 could represent very interesting tasks for future traffic research.

\subsection{Control of a Single Automated Vehicle Initiating S$\rightarrow$F Instability 
\label{S-F1_sec}}

\begin{figure}
\begin{center}
\includegraphics[width = 8 cm]{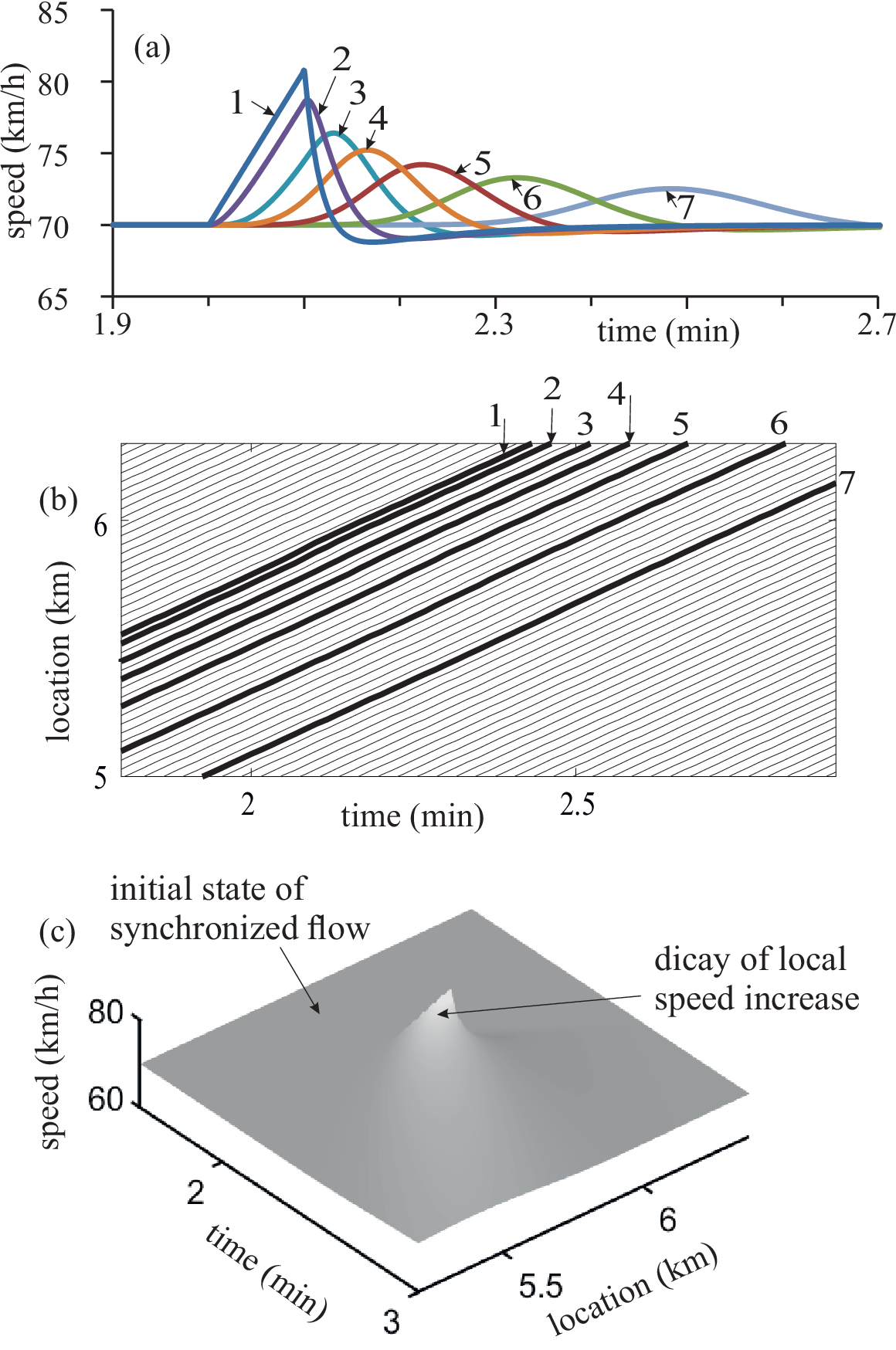}
\end{center}
\caption[]{Nucleation character of S$\rightarrow$F instability: No S$\rightarrow$F instability occurs. Simulations  
with model  (\ref{g_v_g_min1_ad})--(\ref{Helly_st2_ad}) made
on single-lane  road of length 8 km    without bottlenecks with initial steady synchronized flow state  
 at $v=$ 70 km/h and  $g=$ 27.5 m:
(a, b)  
time-development of speeds (a) and trajectories (b) of  vehicles 1--7   caused by
  initial local speed increase of vehicle 1   simulated through 
	short-time acceleration of vehicle 1 with $a=$ 0.5
 $\rm m/s^{2}$ during   6.5 s. (c) Spatiotemporal development of the vehicle speed   during
the decay of the initial local increase in the speed of vehicle 1   in (a).
Other model parameters can be found in~\cite{Kerner2023B}.
Adapted from~\cite{Kerner2023B}.
}
\label{SF-instability_1}
\end{figure}

\begin{figure}
\begin{center}
\includegraphics[width = 8 cm]{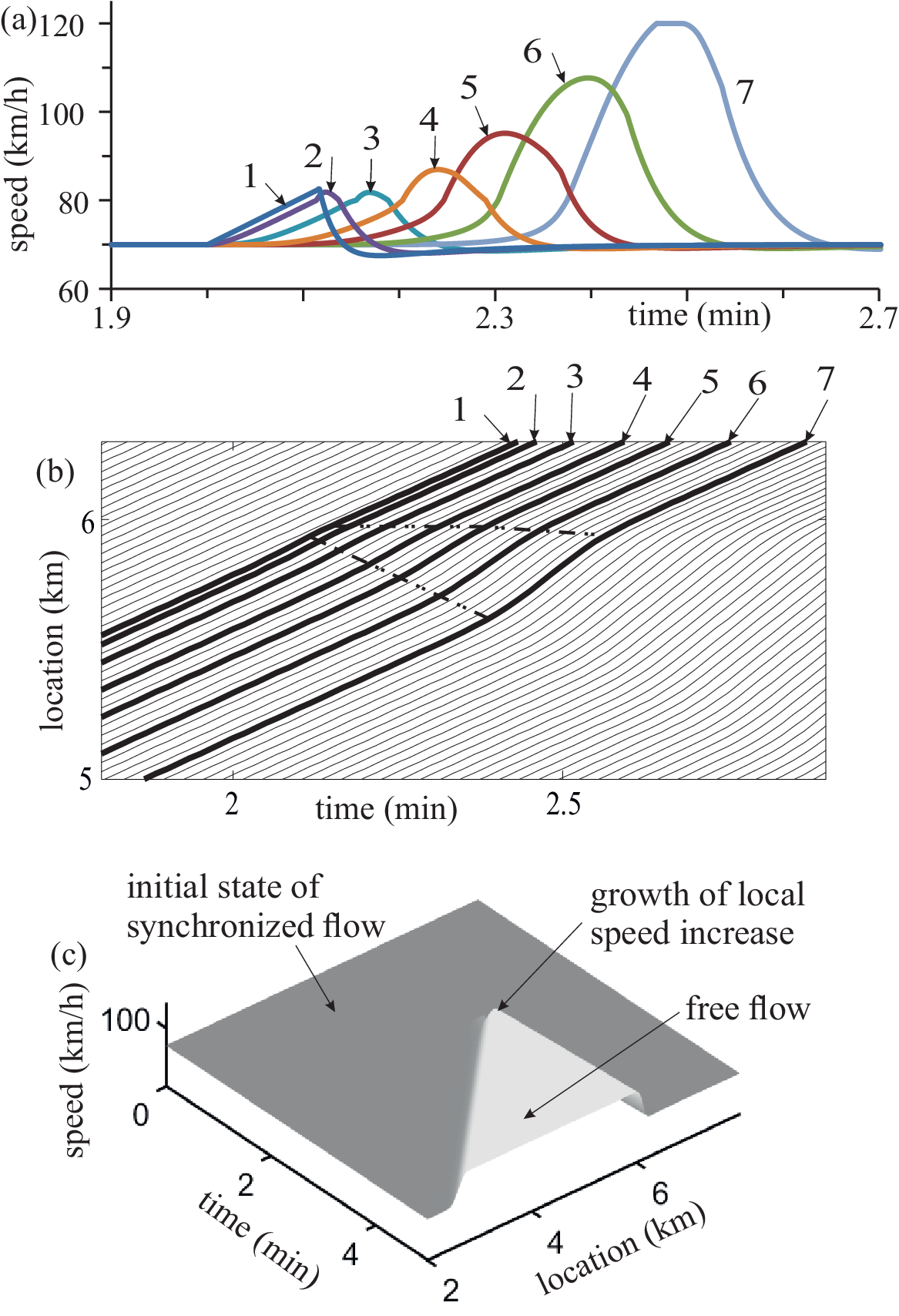}
\end{center}
\caption[]{Nucleation character of S$\rightarrow$F instability. Simulations  
  made with the same model and  model parameters as those in Fig.~\ref{SF-instability_1}, however,
with the only one exception:  The duration of short-time acceleration of  vehicle 1  with $a=$ 0.5 $\rm m/s^{2}$ is equal to 7 s,  i.e., the initial    acceleration
	of vehicle 1, which causes the local speed increase in synchronized flow, is  0.5 s longer than
	that of vehicle 1  in Fig.~\ref{SF-instability_1}.
 In (a, b),
time-development of speeds (a) and trajectories (b) of  vehicles   1--7. (c) Spatiotemporal development of speed   during
 S$\rightarrow$F instability shown in (a, b).
Dashed-dotted curves in (b) denote
the development of S$\rightarrow$F instability in synchronized flow leading
to the S$\rightarrow$F transition. Other model parameters can be found in~\cite{Kerner2023B}.
Adapted from~\cite{Kerner2023B}. 
}
\label{SF-instability_2}
\end{figure}

The S$\rightarrow$F instability caused by a competition between discontinous overacceleration with speed adaptation
 has been predicted in~\cite{Kerner2015C}. Additionally, it has been found that in synchronized   flow there can be a spatiotemporal competition
between S$\rightarrow$F and S$\rightarrow$J instabilities~\cite{Kerner2019AA}. Contrary to~\cite{Kerner2015C,Kerner2019AA}, where
stochastic three-phase traffic flow models have been applied,
 we consider simulations of the S$\rightarrow$F instability with
the  TPACC-model (\ref{g_v_g_min1_ad})--(\ref{Helly_st2_ad}) (Figs.~\ref{SF-instability_1} and~\ref{SF-instability_2}).
We show that  a single automated vehicle can initiate the S$\rightarrow$F instability.
In this case, through the control of the single automated vehicle
a return S$\rightarrow$F transition could occur leading to the recovering of free flow.

We assume that the initial traffic state     is homogeneous
 synchronized flow with the speed 70 km/h  (Fig.~\ref{SF-instability_1}). If vehicle 1
accelerates with $a=$ 0.5
 $\rm m/s^{2}$ during   6.5 s, a local speed increase is realized in the synchronized flow state.
The  local speed increase  decays over time and the initial homogeneous
 synchronized flow recovers   (Fig.~\ref{SF-instability_1}).

 However, if in the same initial state of synchronized flow
vehicle 1 accelerates only 0.5 s longer than vehicle 1 in Fig.~\ref{SF-instability_1}, the local speed increase    initiated by
 vehicle 1
grows  over time  (vehicles 2--7 in Figs.~\ref{SF-instability_2}(a, b)) leading to  
  the  S$\rightarrow$F instability  
 (Fig.~\ref{SF-instability_2}).   The time-development of
the S$\rightarrow$F instability leads
to a local S$\rightarrow$F transition  (Fig.~\ref{SF-instability_2}(c)).

Thus, there is a critical maximum speed within the local speed increase in synchronized flow for
the  S$\rightarrow$F instability: (i) When the maximum
speed is less than the critical maximum speed, no S$\rightarrow$F instability occurs  (Fig.~\ref{SF-instability_1}). (ii)
 Contrarily, when the maximum
speed is higher than the critical maximum speed, the S$\rightarrow$F instability
does occur  (Fig.~\ref{SF-instability_2}).  In other words,
the  S$\rightarrow$F instability exhibits the nucleation nature.
As found in~\cite{Kerner2019AA,Wiering}, the S$\rightarrow$F instability exhibits also the nucleation nature in synchronized flow occurring   due to 
traffic breakdown (F$\rightarrow$S transition)
 at a road or moving bottleneck. 

The nucleation nature of the  S$\rightarrow$F instability  can be used to initiate a return  
S$\rightarrow$F transition with the use of automated vehicles and AI.  In fact, we can assume that AI models based on historical  data  and real-time data can   detect how the S$\rightarrow$F instability can be initiated.
Then, this information can be used by one or several
automated vehicles for the initiating a local speed increase in synchronized flow that is larger than the critical
local speed increase. This causes the  S$\rightarrow$F instability that development leads to the recovering of free flow.
However, we are currently unaware of   studies that have investigated these and other possible methods for the initiating of
 the  S$\rightarrow$F instability with the use of automated vehicles and AI.
In other words, the aforementioned proposals for the recovering of free flow
on the road could represent very interesting tasks for future traffic research.

 \section{Conclusions 
\label{Conl_S}}

1. The essential contribution of this review article is that, by employing microscopic  deterministic three-phase traffic flow models, we are able to demonstrate that the key element in controlling traffic breakdown lies in the control of vehicle overacceleration---and not, as assumed in the most standard traffic theories and models, in the control of traffic instability. The control of vehicle overacceleration can be realized through the individual control of vehicle motion. For this reason, vehicle overacceleration constitutes a fundamental microscopic feature for the control of traffic breakdown with the use of automated vehicles and AI.

2.    We have shown that the empirical  nucleation nature  of traffic breakdown is indeed caused by
 the competition of discontinous overacceleration with speed adaptation, not by vehicle overdeceleration, i.e., not by traffic instability.

3. Vehicle overdeceleration  leading to    traffic instability does indeed cause moving traffic jams
  (S$\rightarrow$J transition) within synchronized flow, but it is not the cause of the initial traffic breakdown at the bottleneck.
	In other words, overdeceleration  that leads to traffic instability generates moving traffic jams within synchronized traffic flow; traffic instability is thus a consequence of traffic breakdown, not its cause.
 
4. The interplay of   discontinuous vehicle overacceleration and vehicle speed adaptation explains  the empirical
 nucleation nature of traffic breakdown (F$\rightarrow$S transition) with the use of \textit{elementary} 
  single-leader deterministic car-following traffic flow models, in which the leading vehicle follows only the motion of the preceding vehicle.  
	Discontinuous vehicle overacceleration does occur when:
	\begin{itemize}
	\item[--]
{\it No} additionally accounts for information from vehicles further ahead of the preceding vehicle,
and \textit{no}   situation-dependent safety margins have been used.
\item[--]   
  {\it No}  driver/vehicle psychological speed thresholds, and {\it no} non-linear bounded rationality   have been applied.
	\item[--]
	There is \textit{no}  stronger tendency to accelerate than to decelerate among drivers or automated systems in response to upstream speed disturbances. Additionally, \textit{no} greater preference for free-flow conditions over compliance with safe spacing has been incorporated.
	\item[--]  {\it No}   other model extensions, such as heterogeneous driving behavior, randomness, or multi-phase velocity-spacing relationships have been included.
 \end{itemize}

5. Discontinuous overacceleration   occurs   in   single-leader deterministic car-following traffic flow models   without explicitly introducing
it in the model.

6. We have considered several microscopic three-phase traffic flow models for automated and human-driving vehicles,  in which 
 traffic breakdown (F$\rightarrow$S transition) at the bottleneck is realized exclusively through vehicle overacceleration: In the models, neither classical traffic instability nor string instability can occur  under free-flow conditions.

7. The nucleation nature of traffic breakdown
(F$\rightarrow$S transition) at the bottleneck is qualitatively  the same in traffic with  human-driving vehicles or in traffic with
automated vehicles: The nucleation nature of traffic breakdown
is caused by the interplay of   discontinuous vehicle overacceleration and vehicle speed adaptation.

8. Simulations  of traffic consisting of automated vehicles
 can reproduce 
  empirical  data of traffic with human-driving vehicles,
	in which the MSP while propagating upstream has induced the F$\rightarrow$S transition at the bottleneck
with the LSP emergence: As found in the empirical data (Fig.~\ref{20041998_MSP}(a)), in simulations of automated vehicles 
no moving jams have been observed during the MSP propagation as well as
 during and after the traffic breakdown (Figs.~\ref{MSP_Breakdown} and~\ref{MSP_color}).

9. The following mechanisms of vehicle overacceleration can be distinguished: (i) overacceleration due to safety acceleration; (ii) overacceleration resulting from a lane change; (iii) explicit overacceleration $a_{\rm OA}$. It is expected that further investigations into the discontinuous acceleration behavior of vehicles will enable the identification of additional mechanisms of vehicle overacceleration.

10. To emphasize the fundamental importance of  
   discontinuous vehicle overacceleration for traffic theory, in Secs.~\ref{Safety_OA_sec},~\ref{TPACC_St_S}, and~\ref{Lane_OA_sec}  we have shown that:
	\begin{itemize}
\item[--] The  overacceleration mechanisms    due to safety  acceleration
and resulting from a lane change are inherent model properties: they result
 directly from mathematical deterministic car-following traffic flow models for both human-driven and automated vehicles.	 
  \item[--] 
	The nucleation behavior of the  F$\rightarrow$S transition caused by overacceleration mechanisms (driven by safety acceleration and lane-changing) 
	becomes evident when vehicle overdeceleration is not incorporated into the model \textit{under free-flow conditions}.
 \end{itemize}
 This aligns with real-world traffic observations
 (Sec.~\ref{OA-OD_Real-World_S}).

11. Contrary to three-phase traffic flow models, most standard traffic flow models---where traffic breakdown
 at a bottleneck is driven by either classical traffic instability or string instability---fail to replicate 
the empirical data shown in Fig.~\ref{20041998_MSP}(a). This limitation demonstrates that standard theoretical approaches developed for congestion mitigation, such as ``jam absorption driving'' (also termed stop-and-go wave dissipation, wave suppression, or shockwave damping), cannot prevent
  traffic breakdown (the F$\rightarrow$S transition) in free flow at bottlenecks. Ultimately, because these standard approaches target the wrong underlying mechanism, they are fundamentally incapable of preventing breakdown.
	
12. 	Nevertheless, integrating overacceleration management with jam absorption driving could potentially both prevent and dissipate traffic congestion in the future. Exploring this combined approach for free-flow recovery presents a highly promising avenue for future traffic research.

13.  AI-equipped automated vehicles can effectively maintain free-flow traffic conditions through microscopic overacceleration management. Specifically, vehicle overacceleration should be increased  in the immediate vicinity upstream and downstream of the bottleneck. Consequently, using AI-equipped automated to individually control 
the competition between discontinuous overacceleration and speed adaptation---an approach that remains unexplored---represents a crucial direction for the future development of microscopic mixed traffic theory.

\appendix

\section{Choice of Dynamic Coefficient in Helly's Model of ACC-Vehicles \label{ACC_safety_S}}

 To avoid vehicle collisions,   
   for safety  
				acceleration $a_{\rm safety}(g, v, v_{\ell})$  in  (\ref{ACC_Cl})  we   apply ideas of
				  dynamic breaking strategies of General Motors car-following model
					 of Herman, Gazis, Montroll, Potts,
Rothery, and Chandler~\cite{GM_Com1,GM_Com2,GM_Com3,Gazis}, in which at $\Delta v<0$ vehicle deceleration is proportional to the term
	\begin{eqnarray}
\frac{v\tau_{\rm safe}}{g}\Delta v.  
\label{GM_formula}
\end{eqnarray}
Respectively, we choose $K_{\rm ACC,2}=K_{\rm ACC,2}(g, v, \Delta v)$: 
 \begin{equation}
K_{\rm ACC,2}(g, v, \Delta v)=\left\{\begin{array}{ll} 
K^{(0)}_{\rm ACC,2} \frac{v\tau_{\rm d}}{g} \ \textrm{at $ \Delta v \leq 0$ and $g< v\tau_{\rm d}$}, \\  
K^{(0)}_{\rm ACC,2}    \ \textrm{otherwise}, \\
\end{array} \right.
\label{Helly_ACC_K}
\end{equation}
$K^{(0)}_{\rm ACC,2}$ is a constant. In simulations (Fig.~\ref{ACC_safety_acc}),  we have chosen in (\ref{ACC_Cl})
$K_{\rm ACC,1}=  0.3 \ s^{-2}$ and $K^{(0)}_{\rm ACC,2}= 0.9 \ s^{-1}$. 

We have also simulated the ACC-model
(\ref{ACC_Cl}) with   the same constant values
	$K_{\rm ACC,1}=  0.3 \ s^{-2}$ and $K_{\rm ACC,2}=
0.9 \ s^{-1}$ as those in~\cite{Kerner2023A}
	at which condition for string stability (\ref{ACC_stability}) is satisfied. We have found that  
at chosen model parameters no moving jams occur in synchronized flow
 and the term $v\tau_{\rm d}/g$  in (\ref{Helly_ACC_K})
does not influence results noticeably. For this reason, 
   simulations of the  ACC-model
(\ref{ACC_Cl}), (\ref{Helly_ACC_K}) presented in the paper can be applied for a comparison with
those found in~\cite{Kerner2023A}   
for both single-lane and two-lane roads.

\section{Boundary Conditions and On-Ramp Model \label{Road_S}}
 
Open boundary conditions are applied. At the beginning of the   road $x=0$ vehicles  are generated one after another
in each of the lanes of the road at time instants
$t^{(k)}=k\tau_{\rm in}$, $k=1,2,\ldots$, 
where $\tau_{\rm in}=1/q_{\rm in}$, $q_{\rm in}$ is a given time-independent flow rate   per road lane.
The initial vehicle speed is equal to  
$v_{\rm free}$. After the vehicle has reached the end of the road $x=L$ it is removed.
Before this occurs, the farthest downstream vehicle  maintains its speed and lane.

The motion of a vehicle in a road lane is found under conditions $0 \leq v \leq v_{\rm free}$
from the solution of   equations
$dv/dt=a$,  $dx/dt=v$. These equations    are solved with the second-order Runge-Kutta method with time step 
$10^{-2}$ s.

In the on-ramp model, there is
a merging region of length $L_{\rm m}$ in the right   lane that begins at   location $x=x_{\rm on}$
within which  vehicles can merge from the on-ramp.  
  Vehicles
are generated at the on-ramp one after another at time instants
$t^{(m)}=m\tau_{\rm on}$, $m=1,2,\ldots$, 
where $\tau_{\rm on}=1/q_{\rm on}$, $q_{\rm on}$ is the on-ramp inflow rate.  To reduce
a  local  speed decrease occurring through the vehicle merging at the on-ramp bottleneck, vehicles merge   with the speed of the preceding vehicle $v^{+}$ at a middle location
$x=(x^{+}+x^{-})/2$ between  the preceding and following vehicles in the right lane, when the space gap between the  vehicles 
  exceeds some safety value $g^{\rm (min)}_{\rm target}=\lambda_{\rm b}v^{+}+ d$, i.e., some safety condition
$x^{+}-x^{-}-d>g^{\rm (min)}_{\rm target}$ should be
  satisfied, where $\lambda_{\rm b}$ is a constant. In accordance with these merging conditions,
 the space gap for a vehicle merging between each pair of consecutive vehicles in the right   lane is checked 
within merging region $L_{\rm m}$, starting from the upstream boundary of the merging region. If there is such a pair of consecutive vehicles, the vehicle merges onto the right   lane;
 if there is no pair of consecutive vehicles, for  which the safety condition   is satisfied at the current time step, the procedure is repeated at the next time step,
and so on.

\section{Choice of Dynamic Coefficients in Model of TPACC-Vehicles   \label{TPACC_safety_S}}

To make   vehicle deceleration without jumps,
when the space gap $g$ intersects boundaries $G$ and $g_{\rm safe}$ of the indifferent zone,
 in (\ref{g_v_g_min1}) we apply
\begin{equation}
K_{\Delta v}=\left\{\begin{array}{ll}
K_{2}   \ \textrm{at $\Delta v > 0$}, \\
K^{(1)}_{\Delta v}  \ \textrm{at $ \Delta v \leq 0$}, \\
\end{array} \right.
\label{K_Deltav}
\end{equation}
where $K_{2}=K^{(1)}_{4}$,
\begin{equation} 
K^{(1)}_{\Delta v}=\Biggl(K_{2}-K^{(2)}_{4} \frac{v\tau_{\rm safe}}{g}\Biggr) f(g, v) +  K^{(2)}_{4} \frac{v\tau_{\rm safe}}{g},
\label{K_Deltav3}
\end{equation}
\begin{equation} 
f(g, v)=\left\{\begin{array}{ll}
\frac{g-g_{\rm safe}(v)}{G(v)-g_{\rm safe}(v)}   \ \textrm{at $v > 0$}, \\
1 \ \textrm{at $ v=0$}. \\
\end{array} \right.
\label{K_Deltav4}
\end{equation}

\section{Speed Functions   of Synchronized   and Safety Space Gaps   \label{G-g-safe_Sec}}

 In generalized model (\ref{g_v_g_min1_ad})--(\ref{Helly_st2_ad}), (\ref{pinch-formula}) of Sec.~\ref{Gen_Model_Sec}, rather than    $g_{\rm safe}=v\tau_{\rm safe}$ and   $G=v\tau_{\rm G}$
 (\ref{G_g_safe_simple}), we use  the following formulations for the speed-functions $g_{\rm safe}(v)$ and $G(v)$:
		\begin{equation}
g_{\rm safe}=\left\{
\begin{array}{ll}
v\tau_{\rm safe} \ {\rm at} \ v \geq v_{\rm pinch}
 \\
g_{\rm min}+v\left(\tau_{\rm safe}
-\tau_{\rm min}\right) \ {\rm at} \ v < v_{\rm pinch}, \\
\end{array}\right.
  \label{g_safe-min}
  \end{equation}
\begin{equation}
G=\left\{
\begin{array}{ll}
v\tau_{\rm G} \ {\rm at} \ v \geq v_{\rm pinch}
 \\
g_{\rm min}+v\left(\tau_{\rm G}
-\tau_{\rm min}\right) \ {\rm at} \ v < v_{\rm pinch}, \\
\end{array}\right.
  \label{G-min}  
  \end{equation}
		where $\tau_{\rm min}=g_{\rm min}/v_{\rm pinch}$,   
		$g_{\rm min}$  is a parameter.

{\bf Acknowledgments:}

I would like to thank Sergey Klenov for   useful suggestions.

\end{document}